\documentclass[fleqn,usenatbib]{mnras}

\usepackage{newtxtext,newtxmath}

\usepackage[T1]{fontenc}
\pdfoutput=1
\DeclareRobustCommand{\VAN}[3]{#2}
\let\VANthebibliography\thebibliography
\def\thebibliography{\DeclareRobustCommand{\VAN}[3]{##3}\VANthebibliography}

\usepackage{graphicx}	
\usepackage{amsmath}	
\usepackage{amssymb}	
\usepackage{comment}
\usepackage{bm}
\usepackage{subcaption}
\usepackage{xr}

\title[Self-calibration of weak lensing systematics]{Self-calibration of weak lensing systematic effects using combined two- and three-point statistics}

\author[S. Pyne \& B. Joachimi]{
Susan Pyne$^{1}$\thanks{E-mail: ucapsep@ucl.ac.uk }
       and Benjamin Joachimi$^{1}$
\\
$^{1}$Department of Physics and Astronomy, University College London, Gower Street, London WC1E 6BT, UK\\
}

\date{Accepted XXX. Received YYY; in original form ZZZ}

\pubyear{2021}

\begin{document}
\label{firstpage}
\pagerange{\pageref{firstpage}--\pageref{lastpage}}
\maketitle

\begin{abstract}
We investigate the prospects for using the weak lensing bispectrum alongside the power spectrum to control systematic uncertainties in a  \textit{Euclid}-like survey.  Three systematic effects are considered: the intrinsic alignment of galaxies, uncertainties in the means of tomographic redshift distributions, and multiplicative bias in the measurement of the shear signal.  We find that the bispectrum is very effective in mitigating these systematic errors. Varying all three systematics simultaneously, a joint power spectrum and bispectrum analysis reduces the area of credible regions for the cosmological parameters  $\Omega_\mathrm{m}$ and $\sigma_8$ by a factor of 90 and for the two parameters of a time-varying dark energy equation of state by a factor of almost 20, compared with the baseline approach of using the power spectrum alone and of imposing priors consistent with the accuracy requirements specified for \textit{Euclid}. We also demonstrate  that including the bispectrum self-calibrates all three systematic effects to the stringent levels required by the forthcoming generation of weak lensing surveys, thereby reducing the need for external calibration data.
\end{abstract}

\begin{keywords}
gravitational lensing: weak -- cosmological parameters -- large-scale structure of Universe -- methods: analytical
\end{keywords}

\section{Introduction}
One of the primary aims of modern cosmology is to constrain cosmological parameters within the concordance cosmological model.   An increasingly reliable tool for this purpose is weak gravitational lensing. Recent galaxy surveys including
the Kilo-Degree Survey\footnote{http://kids.strw.leidenuniv.nl/index.php} (KiDS),
 the Dark Energy Survey\footnote{https://www.darkenergysurvey.org}  (DES) and the Hyper Suprime-Cam Subaru Strategic Survey\footnote{https://hsc.mtk.nao.ac.jp/ssp/} (HSC) have already produced strong constraints on parameters of structure growth \citep{troxel2018dark,hikage2019cosmology,asgari2021kids}. The next generation of  surveys such as \textit{Euclid}\footnote{http://sci.esa.int/euclid/} \citep{laureijs2011euclid} and the Vera C. Rubin Observatory Legacy Survey of Space and Time\footnote{https://www.lsst.org} (LSST) will represent a step change in the quantity and precision of weak lensing data and deliver even tighter parameter constraints.  Moreover, the increased volume and accuracy of the data will make it possible to use methods and statistics which are not feasible with current surveys.  

One possibility is to make more use of three-point weak lensing statistics.  These are inherently more difficult to measure and analyse than two-point statistics but nevertheless a three-point weak lensing signal was first detected as early as 2003 \citep{bernardeau2003patterns, pen2003detection}. Subsequently  \citet{semboloni2010weak} successfully used three-point aperture mass statistics from the Cosmic Evolution Survey \citep{scoville2007cosmic} to estimate cosmological parameters.  This work was an important proof of concept. Although the survey was small, with an area of only 1.64 deg$^2$, combining two-point and three-point statistics produced a modest improvement in parameter constraints.  More recently the feasibility and usefulness of  three-point measures were confirmed by \citet{ fu2014cfhtlens} using the larger Canada France Hawaii Telescope Lensing Survey (CFHTLenS; \citealt{heymans2012cfhtlens}).  

Several theoretical studies have investigated the weak lensing bispectrum from the point of view of reducing statistical uncertainties \citep{takada2004cosmological,kayo2012information,kayo2013cosmological,coulton2019constraining,rizzato2019tomographic}.  All these authors concluded that in principle the bispectrum can provide worthwhile additional information and thus improve cosmological parameter constraints, with \citet{coulton2019constraining} additionally showing improved constraints on the sum of the neutrino mass.  However these investigations did not take account of systematic uncertainties and their conclusions must be considered optimistic. If anything, the results reinforce the need to control systematic uncertainties.   Currently systematic and statistical errors are of similar size but future surveys will drastically reduce statistical uncertainties, making control of systematic effects a priority.    Accounting for systematic effects can also shed light on tension  between recent results from weak lensing \citep{troxel2018dark,abbott2019dark,hikage2019cosmology,asgari2020kids+,heymans2021kids,joudaki2020kids+} and the latest \textit{Planck} analyses of the cosmic microwave background \citep{aghanim2018planck}. In particular there are discrepant results for the value of the structure growth parameter \mbox{$S_8=\sigma_8\sqrt{\Omega_\mathrm{m}/0.3}$}, derived from the matter fluctuation amplitude parameter $\sigma_8$  and the matter density parameter $\Omega_\mathrm{m}$.    The possibility that this apparent tension between results from different probes stems from uncontrolled systematic effects has not been ruled out.

In the light of this we investigate the feasibility of using three-point statistics to control some of the major systematic uncertainties which beset  weak lensing.  
Our work is partly motivated by existing evidence that some systematics affect two-point and three-point statistics in different ways, for example \citet{semboloni2008sources} for weak lensing, \citet{foreman2020baryonic} for the matter bispectrum.  For tomographic weak lensing we might expect these differences to be substantial because the weak lensing power spectrum and bispectrum are differently-weighted projections of their matter counterparts.   We also explore the potential for using the combined bispectrum and power spectrum to enable self-calibration -- mitigating systematic effects using only data from the survey itself.

We focus on three major sources of systematic error: intrinsic alignments of galaxies, residual uncertainty in the shape of tomographic redshift distributions expressed through potential shifts in their means, and multiplicative bias in  shear estimation. 
The effects of  these (and other) systematic errors on two-point weak lensing statistics have been studied extensively and there is a significant literature discussing specific types of uncertainty and presenting general approaches to estimating and controlling systematics \citep{huterer2005calibrating,huterer2006systematic,ma2006effects,bridle2007dark,kitching2008systematic,bernstein2009comprehensive,hearin2012general,kirk2012cosmological,massey2012origins,cropper2013defining,joachimi2015galaxy,kirk2015galaxy,troxel2015intrinsic,mandelbaum2018weak,schaan2020photo}.  
The resulting methods have been implemented  in the analysis of two-point statistics from recent weak lensing surveys \citep{hoyle2018dark,zuntz2018dark,hikage2019cosmology,samuroff2019dark,giblin2021kids,joachimi2020kids}. 

In contrast, relatively little attention has been paid to the effect of systematics on  three-point weak lensing statistics such as the bispectrum  even though many of the concepts developed for the power spectrum can readily be adapted.  Of the few studies which  did consider systematics in three-point statistics, \citet{huterer2006systematic} investigated generic multiplicative and additive biases in future surveys such as LSST and found that using the bispectrum as well as the power spectrum could increase the scope for  self-calibration without undue degradation of parameter constraints, and   \citet{semboloni2013effect} showed that combining two- and three-point statistics 
can largely remove systematics due to baryonic feedback.  For intrinsic alignments, \citet{shi2010controlling} extended a nulling method from two-point to three-point statistics, which mitigated the effects of intrinsic alignments but at the expense of loss of constraining power, and
\citet{troxel2011self,troxel2012self} used the redshift dependency within single redshift bins to inform a self-calibration method.  
Theoretical explorations of three-point intrinsic alignment statistics have been presented by \citet{semboloni2008sources} based on simulations and \citet{merkel2014theoretical} using a tidal alignment model. 
 
In this work we provide a more complete assessment of the value of the bispectrum to mitigate weak lensing systematics through self-calibration in a tomographic survey. 
We model the effect of each systematic on the weak lensing power spectrum and bispectrum and use Fisher matrix analysis to forecast the potential for self-calibration in a \textit{Euclid}-like survey.   

In Sect. \ref{sec:WL}  we summarise  the tomographic weak lensing power spectrum and bispectrum and the structure of their covariance matrices. Section \ref{sec:cosmoparams} records our survey and modelling assumptions.  In Sect. \ref{sec:sysmodelling} we describe our parameterization of the three systematic effects, for both the power spectrum and bispectrum, and in Sect. \ref{sec:inf} we describe our inference methodology.  In Sect. \ref{sec:sysresults} we present our results. Our conclusions are in Sect. \ref{sec:sysconc}. 
Appendices  \ref{sec:covterms} to  \ref{sec:WLcov} give details of our power spectrum and bispectrum covariance methodology.  
Appendix  \ref{sec:additional_plots} contains supplementary plots demonstrating self-calibration.

\section{Weak gravitational lensing}\label{sec:WL}
\subsection{Tomographic weak lensing power spectrum and bispectrum}\label{sec:WLBS}
Throughout this work we assume a flat universe.  With this assumption the convergence field $\kappa^{(i)}$ for the $i $\thinspace th tomographic bin at angular position $\boldsymbol{\theta}$ is 
\begin{align} 
\kappa^{(i)}(\boldsymbol{\theta})&= \int_0^{\chi_\mathrm{lim}} \mathrm{d}\chi \ q^{(i)}(\chi)\,\delta(\chi,\chi\boldsymbol{\theta})\ ,
\end{align}
where $\chi_\mathrm{lim}$ is the maximum comoving distance  of the survey, $\delta$ is the matter density contrast, and  the weight $q^{(i)}(\chi)$ is defined  as 
\begin{align}
	q^{(i)}(\chi)&= \frac{3H_0^2\Omega_\mathrm{m}}{2c^2}\frac{\chi}{a(\chi)} \int _\chi^{\chi_{\mathrm{lim}}}\mathrm{d}\chi^\prime \ p^{(i)}(\chi^\prime)\frac{\left(\chi^\prime-\chi\right)}{\chi^\prime} \ .\label{eq:q}
\end{align}
Here $a(\chi)$ is the scale factor,
  $ p^{(i)}(\chi)$ is the line-of-sight distribution of galaxies in the $i$\thinspace th tomographic bin, $H_0$ is the Hubble constant and $\Omega_\mathrm{m}$ is the matter density parameter.

Assuming that the Limber and flat-sky approximations are valid, the tomographic weak lensing power spectrum at angular multipole $\ell$ between redshift bins $i$ and $j$ is 
\begin{align}
 	C^{(ij)}(\ell) 	&= \int_0^{\chi_\mathrm{lim}}\mathrm{d}\chi \ q^{(i)}(\chi)\,q^{(j)}(\chi)\, \chi^{-2} P_\delta\left(k; \chi\right)\ ,\label{eq:WLPS}
\end{align}
where $P_\delta$ is the matter power spectrum, $k \,\chi(z)=\ell +1/2$,  and we use the more accurate extended Limber approximation \citep{loverde2008extended} which includes higher-order terms from a series expansion of $(\ell+1/2)^{-1}$. 

The corresponding bispectrum is \citep{takada2004cosmological,kayo2013cosmological}
\begin{align}	
	B^{(ijk)}(\boldsymbol{\ell}_1,\boldsymbol{\ell}_2,\boldsymbol{\ell}_3) &= \int_0^{\chi_\mathrm{lim}} \mathrm{d}\chi  \ q^{(i)}(\chi)\, q^{(j)}(\chi)\,q^{(k)}(\chi)\, \chi^{-4} \notag\\
	& \hspace{2.8cm} \times B_\delta\left(\bm{k}_1,\bm{k}_2, \bm{k}_3;\chi\right)\  ,\label{eq:WLBS}
\end{align}
where  $B_\delta$ is the matter bispectrum and we again use the extended Limber approximation  \citep{munshi2011higher}. The vectors $\bm{k}_i$ form a  triangle so that \mbox{$\bm{k}_1+\bm{k}_2+\bm{k}_3=0$}.  Thus the bispectrum has only three degrees of freedom, two from the triangle condition and one from the orientation of the triangle in space.  
\subsection{Summary statistics}
In this work we treat the weak lensing power spectrum and bispectrum as observables, even though they are not directly measurable in practice because of complications such as incomplete sky coverage of surveys.  Nevertheless we expect our results  to be valid for alternative more practical Fourier-space summary statistics. In the case of  two-point analyses such alternatives include band powers \citep{van2018kids,joachimi2020kids} and pseudo-$C_\ell$ estimators \citep{hikage2011shear,asgari2018flat,alonso2019unified}, both of which contain essentially the same information as the power spectrum. 

Three-point summary statistics are less well-developed and it is less clear how they relate  to the underlying bispectrum. The most recent three-point analyses of survey data  \citep{semboloni2010weak,fu2014cfhtlens}  have used aperture mass statistics,  which can be estimated from correlation functions or modelled from the bispectrum \citep{schneider1998new}.  One advantage is that third-order aperture mass statistics separate E- and B-modes of the shear signal well \citep{shi2014well}. This is desirable
since the detection of B-modes can indicate the presence of systematics. 
Other three-point Fourier-space estimators have also been suggested, for example the integrated bispectrum which is sensitive mainly to squeezed triangles  \citep{munshi2020estimating}. This has recently been generalised to other bispectrum configurations through a \lq pseudo\rq \ estimator \citep{munshi2020weak}. So far these new  statistics have not been used to analyse survey data and their practicality and realism are unknown.
\subsection{Weak lensing covariance}\label{sec:WLC}
Both the matter and weak lensing covariance matrices have the general form  
\begin{align}
	\mathbfss{Cov}_{\mathrm{full}} &= 
	\mathbfss{Cov}_{\mathrm{G}}+\mathbfss{Cov}_{\mathrm{NG}}	+\mathbfss{Cov}_{\mathrm{SSC}}\ ,	\label{eq:cov}
\end{align}
where the subscripts denote \lq Gaussian\rq, \lq in-survey non-Gaussian\rq \ and \lq supersample covariance\rq \ \citep{takada2009impact,kayo2012information,takada2013power}. 
Appendix \ref{sec:covterms} summarises the origin of these terms in the matter power spectrum and bispectrum covariance.

To calculate the matter covariance  we use  the halo model formalism, following  \citet{takada2013power} for the power spectrum covariance  and  \citet{chan2017assessment} and \citet{chan2018bispectrum} for the  bispectrum covariance.  Appendix \ref{sec:SSC} gives further details of the bispectrum supersample covariance, since the full expression has not been widely used, and also discusses conflicting results in the literature, justifying our choice of model.  

For ease of computation we consider only equilateral triangles when calculating the bispectrum and its covariance, but recognize that in doing this we are discarding potentially valuable information from other triangle shapes. For example \citet{barreira2019squeezed} found that squeezed triangles with two large and one small side provided useful information.  This reduction in information means that our conclusions about the efficiency of self-calibration are likely to be conservative.

To estimate the weak lensing power spectrum and bispectrum covariances and their cross-covariance, we follow the methods in \citet{takada2004cosmological}  and \citet{kayo2012information}.  In \mbox{App. \ref{sec:WLcov}} we give expressions for all the components of the weak lensing power spectrum and bispectrum covariances for a single tomographic bin.
Similar results, including for the power spectrum-bispectrum cross-covariance,  can be found in \citet{kayo2012information} and \citet{rizzato2019tomographic}.

Appendix \ref{sec:WLcov} also illustrates the relative sizes of terms in the weak lensing covariance matrices.With our assumptions the in-survey non-Gaussian terms of both the power spectrum and bispectrum covariance are sub-dominant.
Consequently, to simplify calculation, in our analysis we  include only the Gaussian  and supersample terms.  Over most of the relevant angular scales the power spectrum and bispectrum supersample covariance are both dominated by the one-halo terms, but we nevertheless retain all terms apart from small dilation terms.
\section{Cosmological parameters, survey characteristics and modelling assumptions}\label{sec:cosmoparams}
We  assume a spatially flat wCDM model and consider six cosmological parameters with fiducial values as shown in \mbox{Table \ref{tab:params}}. 
We model the evolving dark energy equation of state parameter by \mbox{$w(a) = w_0 + (1-a)\, w_a$} where $a$ is the cosmological scale factor, which introduces two further parameters: $w_0$, the value of $w$ at the present day, with fiducial value -1, and $w_a$ which has fiducial value 0. 
\begin{table}
    \begin{tabular}{  lcc }
    \hline
	Parameter&Symbol &Fiducial value \\\hline
    Matter density parameter& $\Omega_\mathrm{m}$ &0.27\\
    Baryon density parameter&$\Omega_\mathrm{b}$&0.05\\
    Density fluctuation amplitude&$\sigma_8$&0.81\\
    Hubble constant (scaled)& $h$&0.71\\
    Scalar spectral index& $n_\mathrm{s}$&0.96\\
    Dark energy equation of state&w&-1.0\\[1ex]
    \hline
    \end{tabular}
    \caption{Fiducial cosmological parameters.}
    \label{tab:params}
\end{table}
We assume a \textit{Euclid}-like survey with area 15\,000 deg$^2$, total galaxy density 30 arcmin$^{-2}$ and redshift range $0.0 \le z \le 2.5$. The assumed overall redshift probability distribution of source galaxies is 
\begin{align}
p(z)&\propto z^\alpha \exp\left[-\left(\frac{z}{z_0}\right)^\beta\right]\ ,\label{eq:pz}
\end{align}
with $\alpha = 2.0$, $\beta = 1.5$, $z_0 =z_\mathrm{med}/ \!\sqrt{\mathstrut 2}$, $z_\mathrm{med}=0.8$.
We model statistical uncertainty in photometric redshift values by assuming that the redshift distribution within each tomographic bin is Gaussian with a dispersion $\sigma_\mathrm{ph}$. Thus the conditional probability of obtaining a photometric redshift $z_\mathrm{ph}$ given the true redshift $z$ has the form
\begin{align}
p(z_\mathrm{ph}|z)\propto \exp\left[-\frac{\left(z_\mathrm{ph}-z\right)^2}{2\sigma^2_\mathrm{ph}\left(1+z\right)^2}\right]\ .
\end{align}
We take  $\sigma_\mathrm{ph}$ to be 0.05. 

With these assumptions,
we divide the redshift distribution into five bins, each containing the same number of galaxies.    Because of uncertainties in photometric measurements, this results in a narrower redshift range with photometric redshift bin boundaries [0.20,0.51], [0.51,0.71], [0.71,0.91], [0.91,1.17] and [1.17,2.00]. 
Future surveys such as \textit{Euclid} will allow much finer bin division than this. However  we find that,  in the absence of systematic uncertainties, increasing the number of bins beyond five for either the power spectrum or the bispectrum provides little extra information. This is consistent with results in  \citet{ma2006effects}, \citet{joachimi2010simultaneous} and \citet{rizzato2019tomographic}.  In fact we find that, considering statistical uncertainties only, if five or more bins are used for the power spectrum, there is little to be gained from using more than two bins for the bispectrum.  This will not necessarily be true if systematic uncertainties are considered.  For example, using only the power spectrum,  if intrinsic alignments are present the information content does not level off until up to 20 bins are used \citep{bridle2007dark,joachimi2010simultaneous}.  Nevertheless we restrict the main self-calibration analysis to five bins to reduce the complexity of the bispectrum and its covariance. It is reasonable to expect that the self-calibration power would increase if more than five bins were used. In Sect. \ref{sec:10bins} we briefly discuss results from using ten bins for the power spectrum.

We use 20 angular bins equally logarithmically spaced from \mbox{$\ell_\mathrm{min} =30$} to \mbox{$\ell_\mathrm{max} =3000$}.  This range avoids large scales where the Limber approximation breaks down and in any case little information is available, and also small scales where the modelling of non-linear effects on the matter distribution becomes very uncertain. 
This maximum angular scale is conservative compared to \mbox{$\ell_\mathrm{max} =5000$}  used for most  \textit{Euclid} analysis.

We model the non-linear matter power spectrum with the fitting formula from \citet{takahashi2012revising}. For the three-dimensional matter bispectrum we use the well-established formula from \citet{gil2012improved}, recognising that this was calibrated over a relatively narrow range with \mbox{$k< 0.4 \, h \, \mathrm{Mpc}^{-1}$} and so could be unreliable at the smallest angular scales which we use.
Recently, \citet{takahashi2020fitting} derived a more accurate prescription for the matter bispectrum, especially at highly non-linear scales \mbox{$k< 10 \, h \, \mathrm{Mpc}^{-1}$}, which is also the first such formula to  include the impact of baryonic feedback. This new formula is likely to be more suitable for weak lensing studies and opens up the possibility of additional self-calibration. This will be the subject of further work, including an assessment of the consistency of the new formula with established feedback approaches for the power spectrum \citep{mead2021hmcode}. 

We employ the transfer function from \citet{eisenstein1998baryonic}.

\section{Modelling of systematics}\label{sec:sysmodelling}
In this section we discuss our parameterization of the systematic effects, in each case starting from methods which have been shown to work well for the power spectrum and extending them to the bispectrum.

\subsection{Intrinsic alignment of galaxies}\label{sec:IA}  
The observed (lensed) ellipticity of a galaxy, $\epsilon_\mathrm{obs}$, is related to its intrinsic ellipticity, $\epsilon_\mathrm{I}$, by \citep{seitz1997steps}
\begin{align}
\epsilon_{\mathrm{obs}}&=\frac{ \epsilon_\mathrm{I} +g}{1+g^\ast\epsilon_\mathrm{I}}\ ,
\end{align}
where $g = \gamma/(1-\kappa)$ is the reduced shear.  In this equation all variables are complex numbers and $g^\ast$ is the complex conjugate of $g$. In the weak lensing regime $\kappa \ll 1, \gamma \ll 1$, so $g \approx \gamma$ and 
\begin{align}
\epsilon_{\mathrm{obs}}\approx\gamma_\mathrm{G}+\epsilon_\mathrm{I}\ .\label{eq:IAobs}
\end{align}
(See  \citet{deshpande2020post} for a discussion of the validity of the reduced shear  assumption for  a \textit{Euclid}-like survey.)

Assuming that intrinsic ellipticities are random so that \mbox{$\langle\epsilon_\mathrm{I}\rangle=0$}, the average ellipiticity over a large number of galaxies provides an estimate of the true gravitational shear $\gamma_\mathrm{G}$.  

From Eq. (\ref{eq:IAobs}) we can construct a correlator of the ellipticities of two galaxy samples, labelled $i$ and $j$, as
\begin{align}
\left\langle\epsilon_{\mathrm{obs}}^{(i)}\epsilon_{\mathrm{obs}}^{(j)}\right\rangle&= \left\langle\gamma_\mathrm{G}^{(i)}\gamma_\mathrm{G}^{(j)}\right\rangle +  \left\langle\gamma_\mathrm{G}^{(i)}\epsilon_\mathrm{I}^{(j)}\right\rangle +  \left\langle\gamma_\mathrm{G}^{(j)}\epsilon_\mathrm{I}^{(i)}\right\rangle+ \left\langle\epsilon_\mathrm{I}^{(i)}\epsilon_\mathrm{I}^{(j)}\right\rangle\label{eq:corr2}\\
&\equiv \mathrm{GG} + \mathrm{GI} +\mathrm{II}\ .
\end{align}
Note that the correlators on the right-hand side of Eq. (\ref{eq:corr2}) are illustrative and not true correlation functions because they do not explicitly take account of the fact that shear is a spin-2 quantity.

The first term of the right-hand side of Eq. (\ref{eq:corr2}) is the lensing signal, GG, and the fourth represents intrinsic alignment auto-correlation, II.  There are two GI terms representing cross-correlations between shear and intrinsic alignment.  Although we model both of these, the first will be small if $z_i<z_j$ unless the two redshift distributions overlap substantially, because intrinsically-aligned galaxies at higher redshift cannot affect the lensing of galaxies at lower redshift. In a tomographic analysis we associate the labels $i$ and $j$  with different redshift bins.

 Analogues of Eq. (\ref{eq:WLPS}) can be used to calculate the two intrinsic alignment power spectra in the extended Limber approximation:
\begin{align}
 C_\mathrm{GI}^{(ij)}(\ell) 	&= \int_0^{\chi_\mathrm{lim}}\mathrm{d}\chi \ q^{(i)}(\chi)\,p^{(j)}(\chi)\, \chi^{-2} P_{\delta\delta_\mathrm{I}}\left(k; \chi\right)\\
C_\mathrm{II}^{(ij)}(\ell) 	&= \int_0^{\chi_\mathrm{lim}}\mathrm{d}\chi \ p^{(i)}(\chi)\,p^{(j)}(\chi)\, \chi^{-2} P_{\delta_\mathrm{I}\delta_\mathrm{I}}\left(k; \chi\right)	\ ,
\end{align}
where, as in Sect. \ref{sec:WLBS}, $q^{(i)}(\chi)$ is defined by Eq. (\ref{eq:q}), $p^{(i)}(\chi)$ is the distribution of galaxies in the $i$\thinspace th tomographic bin, and \mbox{$ k \,\chi(z)=\ell +1/2$}.  The power spectra $P_{\delta\delta_\mathrm{I}}$ and $P_{\delta_\mathrm{I}\delta_\mathrm{I}}$ are defined by
\begin{align}
\left\langle \tilde{\delta}_\mathrm{G}(\bm{k}_1;\chi)\tilde{\delta}_\mathrm{I}(\bm{k}_2;\chi) \right\rangle &= (2\mathrm{\pi})^3 \delta_\mathrm{D}(\bm{k}_1+\bm{k}_2)
 P_{\delta\delta_\mathrm{I}}(k_1;\chi)\\
\left\langle \tilde{\delta}_\mathrm{I}(\bm{k}_1;\chi)\tilde{\delta}_\mathrm{I}(\bm{k}_2;\chi) \right\rangle &= (2\mathrm{\pi})^3 \delta_\mathrm{D}(\bm{k}_1+\bm{k}_2)
 P_{\delta_\mathrm{I}\delta_\mathrm{I}}(k_1;\chi) \ ,
\end{align}
where $\tilde{\delta}_\mathrm{I}$ denotes the Fourier transform of the density contrast of the field which produces the intrinsic alignment.

This formalism can be extended to three-point statistics.
In analogy to Eq. (\ref{eq:corr2})  we construct a three-point correlator 
\begin{align}
\left\langle \epsilon_{\mathrm{obs}}^{(i)}\epsilon_{\mathrm{obs}}^{(j)}\epsilon_{\mathrm{obs}}^{(k)}\right\rangle &= \mathrm{GGG} +\mathrm{GGI}+\mathrm{GII}+\mathrm{III}\ ,\label{eq:corr3}
\end{align}
where again $i$, $j$ and $k$ denote galaxy samples.
The four terms on the right-hand side of this equation are given by
 \begin{align}
  \mathrm{GGG} &=\left\langle\gamma_\mathrm{G}^{(i)}\gamma_\mathrm{G}^{(j)}\gamma_\mathrm{G}^{(k)}\right\rangle \\
   \mathrm{GGI}&= \left\langle\gamma_\mathrm{G}^{(i)}\gamma_\mathrm{G}^{(j)}\epsilon_\mathrm{I}^{(k)}\right\rangle + \left\langle\gamma_\mathrm{G}^{(j)}\gamma_\mathrm{G}^{(k)}\epsilon_\mathrm{I}^{(i)}\right\rangle +\left\langle\gamma_\mathrm{G}^{(k)}\gamma_\mathrm{G}^{(i)}\epsilon_\mathrm{I}^{(j)}\right\rangle\\
   \mathrm{GII} &= \left\langle\gamma_\mathrm{G}^{(i)}\epsilon_\mathrm{I}^{(j)}\epsilon_\mathrm{I}^{(k)}\right\rangle + \left\langle\gamma_\mathrm{G}^{(j)}\epsilon_\mathrm{I}^{(k)}\epsilon_\mathrm{I}^{(i)}\right\rangle +\left\langle\gamma_\mathrm{G}^{(k)}\epsilon_\mathrm{I}^{(i)}\epsilon_\mathrm{I}^{(j)}\right\rangle\\
   \mathrm{III}&=\left\langle\epsilon_\mathrm{I}^{(i)}\epsilon_\mathrm{I}^{(j)}\epsilon_\mathrm{I}^{(k)}\right\rangle\ .
\end{align}
As before these are simplified illustrative correlators.

In a similar way, we can split the observed bispectrum $B_{\mathrm{obs}}$ into four terms
\begin{align}
B_{\mathrm{obs}}^{(ijk)}&= B_{\mathrm{GGG}}^{(ijk)}+B_{\mathrm{GGI}}^{(ijk)}+B_{\mathrm{GII}}^{(ijk)}+B_{\mathrm{III}}^{(ijk)}\ .\label{eq:Bobs}
\end{align}
 The term $B_{\mathrm{GGG}}^{(ijk)}$ is the lensing bispectrum defined by
\begin{align}
\left\langle \tilde{\kappa}_\mathrm{G}^{(i)} (\boldsymbol{\ell}_1) \tilde{\kappa}_\mathrm{G}^{(j )}(\boldsymbol{\ell}_2)\tilde{\kappa}_\mathrm{G}^{(k)}(\boldsymbol{\ell}_3) \right\rangle 
&=(2\mathrm{\pi})^2 \delta_\mathrm{D}(\boldsymbol{\ell}_1+\boldsymbol{\ell}_2+\boldsymbol{\ell}_3)B_{\mathrm{GGG}}^{(ijk)}(\boldsymbol{\ell_1},\boldsymbol{\ell}_2,\boldsymbol{\ell}_3)\ ,
\end{align}
where $\tilde{\kappa}_\mathrm{G}$ is the Fourier transform of the convergence   and only unique combinations of $i$,  $j$ and  $k$ are included.  The bispectrum $B_{\mathrm{GGG}}^{(ijk)}(\boldsymbol{\ell_1},\boldsymbol{\ell}_2,\boldsymbol{\ell}_3)$  is given by Eq. (\ref{eq:WLBS}).

The other three terms on the right-hand side of Eq. (\ref{eq:Bobs}) can be defined similarly, replacing $\tilde{\kappa}_\mathrm{G}$ by $\tilde{\kappa}_\mathrm{I}$ as appropriate.
For example for $\mathrm{GGI}$
\begin{align}
\left\langle \tilde{\kappa}_\mathrm{G}^{(i)} (\boldsymbol{\ell}_1) \tilde{\kappa}_\mathrm{G}^{(j)} (\boldsymbol{\ell}_2)\tilde{\kappa}_\mathrm{I}^{(k)}(\boldsymbol{\ell}_3) \right\rangle 
&=(2\mathrm{\pi})^2 \delta_\mathrm{D}(\boldsymbol{\ell}_1+\boldsymbol{\ell}_2+\boldsymbol{\ell}_3)B_{\mathrm{GGI}}^{(ijk)}(\boldsymbol{\ell}_1,\boldsymbol{\ell}_2,\boldsymbol{\ell}_3)\ ,
\end{align}
and, in analogy to Eq. ( \ref{eq:WLBS}),
\begin{align}
B_{\mathrm{GGI}}^{(ijk)}(\boldsymbol{\ell}_1,\boldsymbol{\ell}_2,\boldsymbol{\ell}_3)&=
\int_0^{\chi_{\mathrm{lim}} } \mathrm{d}\chi  \ q^{(i)}(\chi)\, q^{(j)}(\chi)\, p^{(k)} (\chi) \ \chi^{-4}\notag\\
&\hspace{2cm} \times B_{\delta\delta\delta_\mathrm{I}}\left(\bm{k}_1,\bm{k}_2, \bm{k}_3;\chi\right)\ , \label{eq:BGGI}\end{align}
where again the constituents of the equation are  defined in Sect. \ref{sec:WLBS}, with $B_{\delta\delta\delta_\mathrm{I}}$  defined by
\begin{align}
\left\langle \tilde{\delta}_\mathrm{G}(\bm{k}_1;\chi)\tilde{\delta}_\mathrm{G}(\bm{k}_2;\chi)\tilde{\delta}_\mathrm{I}(\bm{k}_3;\chi) \right\rangle &= (2\mathrm{\pi})^3 \delta_\mathrm{D}(\bm{k}_1+\bm{k}_2+\bm{k}_3)\\\notag
&\quad\quad\quad \times B_{\delta\delta\delta_\mathrm{I}}\left(\bm{k}_1,\bm{k}_2, \bm{k}_3;\chi\right)\ .
\end{align}
 
With these ingredients we can evaluate the full observed lensing bispectrum $B_\mathrm{obs}$ in terms of $B_{\delta\delta\delta}$, $B_{\delta\delta\delta_\mathrm{I}}$,  $B_{\delta\delta_\mathrm{I}\delta_\mathrm{I}}$ and $B_{\delta_\mathrm{I}\delta_\mathrm{I}\delta_\mathrm{I}}$.  
Our method is similar to that in \citet{troxel2012self}, 
 \citet{merkel2014theoretical} and \citet{deshpande2020euclid}.
 
The matter bispectrum $B_{\delta\delta\delta}$ is determined straightforwardly from the fitting function in \citet{gil2012improved}, which has the form
\begin{align}
  B_{\delta\delta\delta}(\bm{k}_1,\bm{k}_2,\bm{k}_3)&= 2 F_2^{\mathrm{eff}}(\bm{k}_1,\bm{k}_2) P_\mathrm{NL}(k_1)P_\mathrm{NL}(k_2)+ \text {2 perms.}\ ,\label{eq:GM}
 \end{align}
 where $F_2^{\mathrm{eff}}$ are modifications of the normal perturbation theory kernels $F_2$ \citep{bernardeau2002large} and $P_\mathrm{NL}(k)$ is the non-linear matter power spectrum.
 
To obtain expressions for $B_{\delta\delta\delta_\mathrm{I}}$,  $B_{\delta\delta_\mathrm{I}\delta_\mathrm{I}}$ and $B_{\delta_\mathrm{I}\delta_\mathrm{I}\delta_\mathrm{I}}$ we adapt the linear alignment model developed by \citet{hirata2004intrinsic} for intrinsic alignment power spectra.  This model assumes that the ellipticity of a galaxy is linearly related to the local quadrupole of the gravitational potential at the time the galaxy formed. 
The model is well-established for two-point statistics \citep{bridle2007dark, kirk2012cosmological} and has been found to be a good fit to direct measurements of intrinsic alignments \citep{singh2015intrinsic,singh2016intrinsic,johnston2019kids+}. We adopt the so-called non-linear alignment
model  introduced by \citet{bridle2007dark} which replaces the linear power spectrum used in \citet{hirata2004intrinsic} with the non-linear matter power spectrum $P_\mathrm{NL}(k)$.

Based on this approach, we relate $\tilde{\delta}_\mathrm{I}$ to the Fourier transform of the matter density contrast, $ \tilde{\delta}_\mathrm{G}$,  by \mbox{$  \tilde{\delta}_\mathrm{I} = f_{\mathrm{IA}} \tilde{\delta}_\mathrm{G} $}, where the factor $ f_{\mathrm{IA}}$ has the form
\begin{align}
f_{\mathrm{IA}}&=-A_{\mathrm{IA}}\frac{C_1\Omega_\mathrm{m}\rho_\mathrm{cr}}{(1+z)D(z)}\left(\frac{1+z}{1+z_0}\right)^{\eta_\mathrm{IA} }\ .\label{eq:IAmodel}
\end{align}
Here $\rho_\mathrm{cr}$ is the critical density and $D(z)$ is the growth factor normalised to unity at the present day. The parameter $C_1$ is a normalisation factor  which in principle can be determined from observations or simulations. We use the value derived by \citet{bridle2007dark} which is $5\times 10^{-14} \,  h^{-2} \mathrm{M}_{\odot}^{-1}\mathrm{Mpc}^3$, leading to \mbox{$C_1\rho_\mathrm{cr} = 0.0134$} \citep{joachimi2011constraints}.

Our parameterization of $f_\mathrm{IA}$ allows for uncertainty in the intrinsic alignment amplitude and possible redshift dependence through the free parameters $A_{\mathrm{IA}}$  and  $\eta_{\mathrm{IA}}$ respectively. 
We do not model luminosity dependence but $\eta_\mathrm{IA}$ acts as a proxy for any indirect redshift dependence through luminosity \citep{troxel2018dark}.   
We set  the fiducial value of $\eta_\mathrm{IA}$ to be zero 
and take the fiducial value of $A_{\mathrm{IA}}$
to be 1, consistent with recent survey results.
The quantity $z_0$ is an arbitrary pivot value which we set to 0.3 in line with previous work \citep{joachimi2011constraints, joudaki2016cfhtlens}.

In the two-point case the two intrinsic alignment power spectra  are given by
\begin{align}
P_{\delta\delta_ \mathrm{I}}(k)&= f_{\mathrm{IA}}P_\mathrm{NL}(k)\ ,\\
P_{\delta_ \mathrm{I}\delta_ \mathrm{I}}(k)&=f_{\mathrm{IA}}^2P_\mathrm{NL}(k)\ .
\end{align}
 
\begin{figure*}
\centering
\begin{subfigure}[b]{1\linewidth}
\includegraphics[width=18cm]{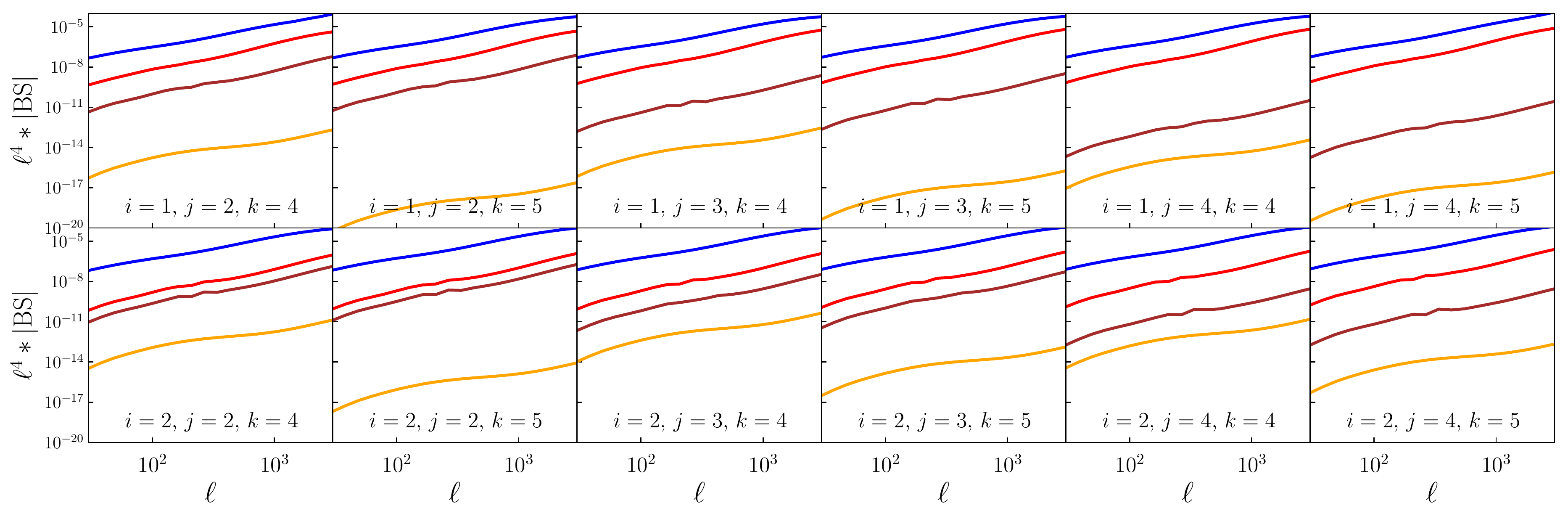}
\vspace{0.2cm}
\end{subfigure}
     \begin{subfigure}[b]{0.5\linewidth}
        \includegraphics[width=1\linewidth]{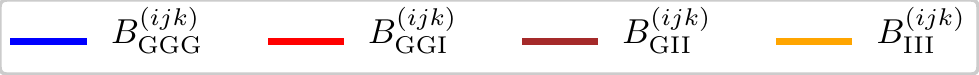}  
    \end{subfigure}  
   \vspace{3pt}
  
\caption{Absolute values of the weak lensing bispectrum, $B_{\mathrm{GGG}}^{(ijk)}$, and the three intrinsic alignment bispectra, $B_{\mathrm{GGI}}^{(ijk)}$, $B_{\mathrm{GII}}^{(ijk)}$ and $B_{\mathrm{III}}^{(ijk)}$, for illustrative tomographic bin combinations $i,j,k$.  Results are for equilateral triangle configurations using five redshift bins, assuming the fiducial values of unity for the  intrinsic alignment amplitude  $A_{\mathrm{IA}}$, and zero for the redshift exponent $\eta_{\mathrm{IA}}$ in Eq. (\ref{eq:IAmodel}). }\label{fig:IABS} 
\end{figure*}

We extend this  to the three-point case using tree-level perturbation theory  and the fitting function from Eq. (\ref{eq:GM})  to get
\begin{align}
B_{\delta\delta\delta_\mathrm{I}}(\bm{k}_1,\bm{k}_2,\bm{k}_3)&=
 2\, \Big[ f_{\mathrm{IA}}^2F_2^{\mathrm{eff}}(\bm{k}_1,\bm{k}_2) P_\mathrm{NL}(k_1)P_\mathrm{NL}(k_2)\\\notag
 &\hspace{0.4cm}+ f_{\mathrm{IA}}F_2^{\mathrm{eff}}(\bm{k}_2,\bm{k}_3) P_\mathrm{NL}(k_2)P_\mathrm{NL}(k_3)\\\notag
&\hspace{0.4cm} +f_{\mathrm{IA}}F_2^{\mathrm{eff}}(\bm{k}_3,\bm{k}_1) P_\mathrm{NL}(k_3)P_\mathrm{NL}(k_1)\big]\ ,\\ 
 B_{\delta\delta_\mathrm{I}\delta_\mathrm{I}}(\bm{k}_1,\bm{k}_2,\bm{k}_3)
 &=2\, \Big[ f_{\mathrm{IA}}^3F_2^{\mathrm{eff}}(\bm{k}_1,\bm{k}_2) P_\mathrm{NL}(k_1)P_\mathrm{NL}(k_2)\\\notag
 &\hspace{0.4cm}+ f_{\mathrm{IA}}^2F_2^{\mathrm{eff}}(\bm{k}_2,\bm{k}_3) P_\mathrm{NL}(k_2)P_\mathrm{NL}(k_3)\\\notag
 &\hspace{0.4cm}+f_{\mathrm{IA}}^3F_2^{\mathrm{eff}}(\bm{k}_3,\bm{k}_1) P_\mathrm{NL}(k_3)P_\mathrm{NL}(k_1)\big] \ ,\\ 
B_{\delta_\mathrm{I}\delta_\mathrm{I}\delta_\mathrm{I}}(\bm{k}_1,\bm{k}_2,\bm{k}_3)&= f_{\mathrm{IA}}^4 B_{\delta\delta\delta}(\bm{k}_1,\bm{k}_2,\bm{k}_3)\ .
 \end{align}
 Then integrating as in Eq. (\ref{eq:BGGI}) gives expressions for the weak lensing intrinsic alignment bispectra.  Fig. \ref{fig:IABS} shows examples of resulting bispectra for some illustrative tomographic bin combinations.  This figure shows equilateral triangle bispectra obtained with five redshift bins, assuming fiducial values of the intrinsic alignment parameters $A_{\mathrm{IA}}$ and $\eta_{\mathrm{IA}}$. The GGI bispectrum is negative and its magnitude can be almost as large as the GGG signal. The other bispectra are positive. The GII bispectrum is generally several orders of magnitude less than the GGI bispectrum, but in some bin combinations it is as much as 20\% of the GGI bispectrum.  The III bispectrum is always sub-dominant, which is consistent with the findings in  \citet{semboloni2008sources} from simulations of a survey similar to CFHTLenS.  

\begin{figure*}
\includegraphics[width=8.5cm]{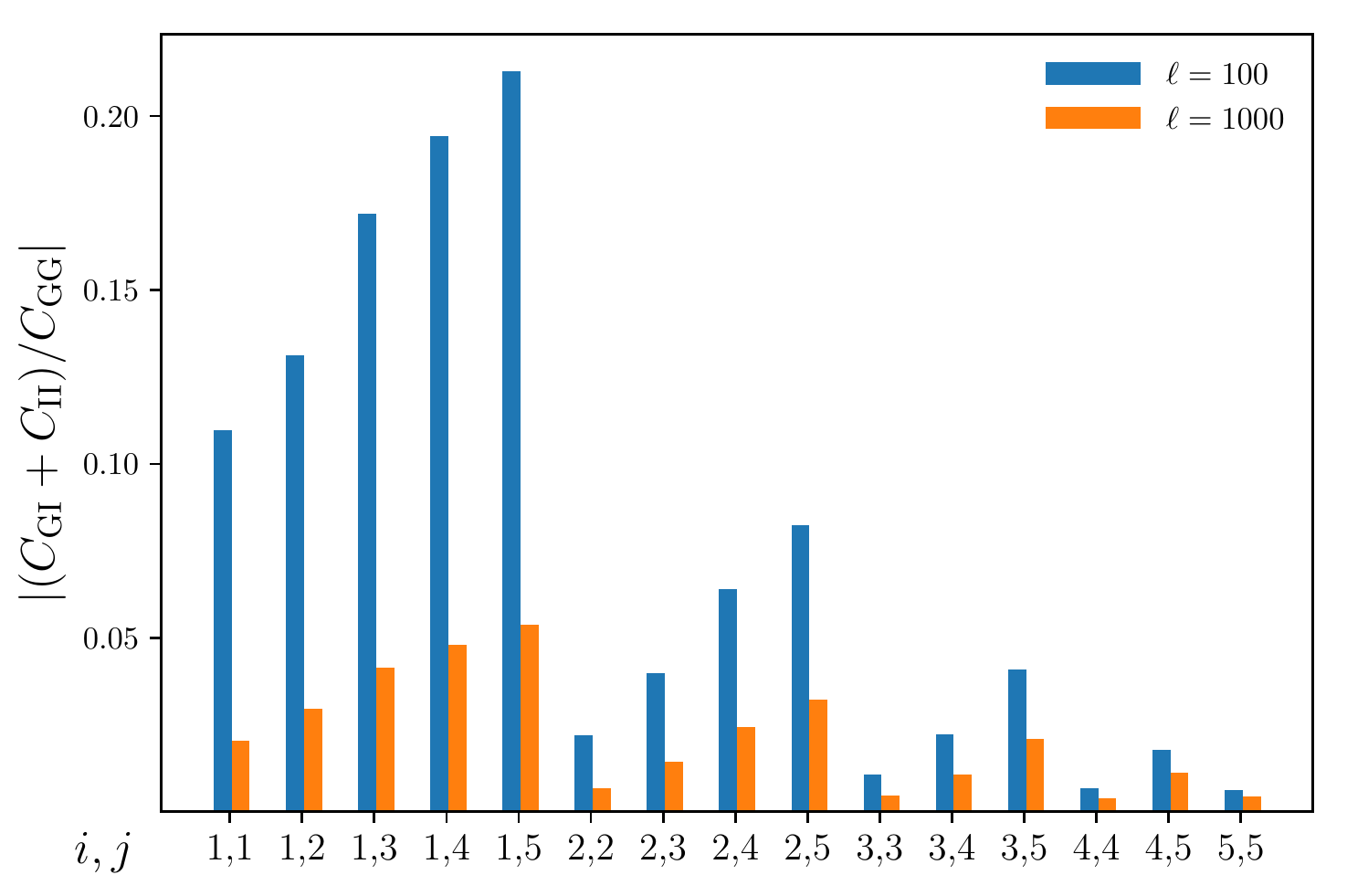}
\includegraphics[width=8.5cm]{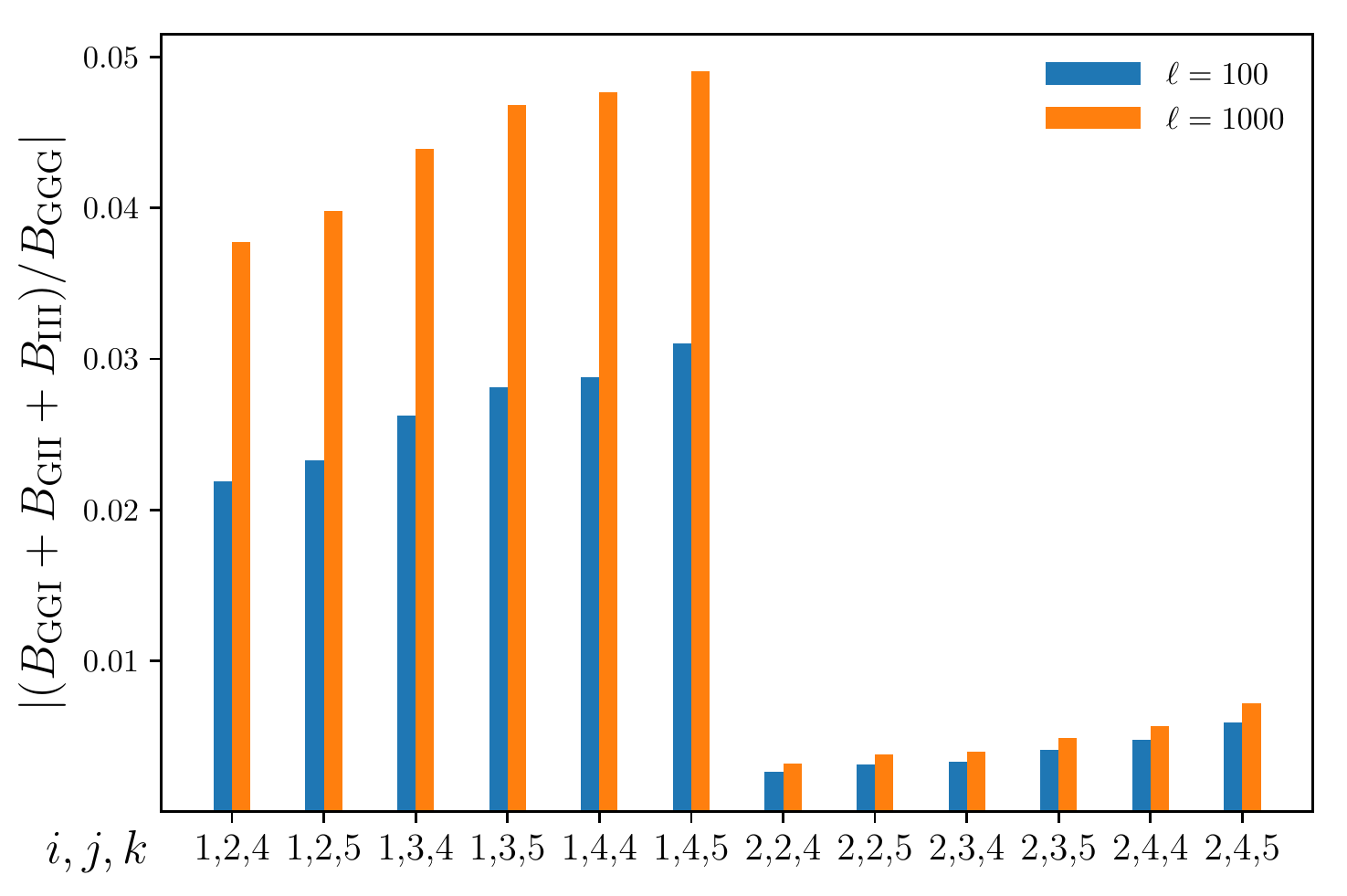}
\caption{  \textit{Left}: Absolute values of the total intrinsic alignment power spectrum relative to the lensing power spectrum for all tomographic bin combinations. \textit{Right}: Absolute values of the total intrinsic alignment bispectrum relative to the lensing bispectrum  for the same tomographic bin combinations $i,j,k$ as in Fig.  \ref{fig:IABS}, for equilateral triangles.  In both panels results are shown for two illustrative angular scales, $\ell=100$ and $\ell=1000$.  Results are for five redshift bins, assuming the fiducial values of unity for the  intrinsic alignment amplitude  $A_{\mathrm{IA}}$, and zero for the redshift exponent $\eta_{\mathrm{IA}}$ in Eq. (\ref{eq:IAmodel}).  Note the different scales on the vertical axes.}\label{fig:IAratios} 
\end{figure*}

In Fig. \ref{fig:IAratios}  we show the relative importance of the intrinsic alignment terms compared with the pure lensing signal, plotted at two representative angular scales, \mbox{$\ell=100$} and \mbox{$\ell=1000$}, for both the power spectrum and bispectrum.  For the power spectrum all redshift bin combinations are plotted whereas for the bispectrum we show  the  same subset as in Fig. \ref{fig:IABS}. Noting the different vertical scales in these two figures, we find that intrinsic alignment affects the power spectrum more than the bispectrum. This contrasts with the findings from simulations in \citet{semboloni2008sources}.   These authors measured three-point aperture mass statistics  and concluded that  the III/GGG ratio was generally higher than the II/GG ratio. They also found that the III signal is negative whereas we find it is positive.    Despite these disagreements we see no obvious reason to discard the linear alignment model which is well-established and robust for two-point statistics. 
The key point for our analysis is that intrinsic alignments affect the power spectrum and bispectrum differently.  Even if  our model is not entirely accurate in detail, conclusions based on it will still hold.  We plan to revisit and validate the modelling assumptions in future work. 

\subsection{Redshift uncertainties}\label{sec:zuncert}
Another source of systematic uncertainty is the calibration of tomographic redshift distributions.
Here we consider a single source of uncertainty due to the use of photometric redshift measurements: bias in the mean redshift of each tomographic bin. Thus we consider the effect of shifting the whole distribution of galaxies in a bin  to a higher or lower redshift, without changing the shape of the distribution.  This has been found to be a good proxy for the uncertainty in the distribution within a bin \citep{hikage2019cosmology,hildebrandt2020kids}. We allow for different uncertainty and hence different shifts  in each bin so that the redshift distribution, $p^{(i)}$, in bin $i$ is modelled as
\begin{align}
p^{(i)}(z) = p^{(i)}_{\mathrm{obs}}(z-\Delta z_i)\ ,
\end{align}
where $p^{(i)}_{\mathrm{obs}}(z)$ is the observed redshift distribution. The shifts in the mean, $\Delta z_i$, are treated as free parameters.
A similar method for forecasting  redshift uncertainties was used by \citet{huterer2006systematic}. It is also the standard approach used for current surveys  \citep{joudaki2016cfhtlens,  
abbott2018dark, hoyle2018dark,hikage2019cosmology, hildebrandt2020kids+}. 
\subsection{Multiplicative shear bias}
The final type of systematic error which we consider is multiplicative shear bias which alters the amplitude of the weak lensing signal. We ignore additive bias which can be calibrated directly on the data.

 Multiplicative biases can have several quite distinct origins \citep{massey2012origins, cropper2013defining, kitching2019propagating}. For example they can arise from incorrect modelling of the point spread function, especially of its size \citep{cropper2013defining,mandelbaum2018weak,giblin2020kids}, or from an inappropriate galaxy surface brightness model \citep{miller2013bayesian}. A more pervasive source of multiplicative bias is noise bias.  This is an unavoidable consequence of the non-linear transformation from image pixels to ellipticity measurements and would be present even if the galaxy profile was known perfectly \citep{melchior2012means,viola2014probability}.  

Simple models of multiplicative bias have been developed by several authors \citep{heymans2006shear, huterer2006systematic,kacprzak2012measurement,massey2012origins}.  We follow   \citet{huterer2006systematic} who assumed that multiplicative biases in different redshift bins of a tomographic survey are independent and uncorrelated. Thus the measured shear $\hat{\gamma}^{(i)}$ in bin $i$  is 
\begin{align}
\hat{\gamma}^{(i)}&= (1+m_i)\, \gamma^{(i)}_\mathrm{true}\ ,\label{eq:multbias}
\end{align}
where $\gamma_\mathrm{true}$ is the true (but unmeasurable) shear.  We assume this equation holds for both components of the shear and that  $m_i$ is  a scalar which is the same for both components. 

From Eq. (\ref{eq:multbias}) we can construct the two-point correlator
\begin{align}
\hat{\xi}^{(ij)}_+(\theta) &= \left\langle \hat{\boldsymbol{\gamma}}^{(i)} \hat{\boldsymbol{\gamma}}^{\ast(j) }\right\rangle \\
&\approx (1+ m_i +m_j) \ \xi^{(ij)}_+(\theta)\ ,
\end{align}
where $\theta$ is the angle on the sky between a pair of galaxies, and in the final line we have dropped terms of order $m_i^2$ \citep{huterer2006systematic,massey2012origins}. An analogous expression can also be defined for  $\hat{\xi}^{(ij)}_-(\theta)$.  Note that these correlators are again simplifications which ignore the spin-2 nature of the shear.
Taking the Fourier transform leads to a similar expression for the  power spectrum 
\begin{align}
\hat{C}^{(ij)}(\ell)&\approx (1+ m_i +m_j) \ C^{(ij)}(\ell)\ .
\end{align}
Similarly for three-point statistics we can write a generic  correlator as
\begin{align}
\hat{\zeta}^{(ijk)}(\theta_1, \theta_2,\theta_3)&\approx (1+ m_i +m_j+m_k) \ \zeta^{(ijk)}(\theta_1, \theta_2,\theta_3)\ ,
\end{align}
where we use the facts that the  multiplicative factors are real and the same for both shear components. Once again this expression is a simplification which ignores the fact that shear is a spin-2 quantity. The shear three-point correlation function in fact has eight components, or four if considered as a complex quantity \citep{takada2002three,schneider2003three,zaldarriaga2003higher}.

The bispectrum is then modelled as \citep{huterer2006systematic,massey2012origins}
\begin{align}
\hat{B}^{(ijk)}(\boldsymbol{\ell}_1,\boldsymbol{\ell}_2,\boldsymbol{\ell}_3)&\approx (1+ m_i +m_j+m_k) \ B^{(ijk)}(\boldsymbol{\ell}_1,\boldsymbol{\ell}_2,\boldsymbol{\ell}_3) \ .
\end{align}
 
 This method of calibrating multiplicative shear bias by treating the multiplicative factors as nuisance parameters was used in two-point analyses of data from CFHTLenS \citep{kilbinger2013cfhtlens,miller2013bayesian} and DES  \citep{abbott2018dark}.   \citet{fu2014cfhtlens} extended the method in \cite{kilbinger2013cfhtlens} to analysis of three-point aperture mass statistics.  
\section{Inference methodology}\label{sec:inf}
\subsection{Fisher matrices and figures of merit}\label{sec:FM}
To investigate the impact of systematics we  use the  Fisher matrix \citep{tegmark1997karhunen}. In simplified notation the elements of the Fisher matrix are defined  by
\begin{align}
F_{\alpha\beta} &= \frac{\partial \bm{D}^\mathrm{T}}{\partial p_\alpha}\mathbfss{Cov}_\mathrm{D}^{-1}\frac{\partial \bm{D}}{\partial p_\beta}\label{eq:Fisher}\ ,
\end{align}
where $\bm{D}$ is the data vector, $\mathbfss{Cov}_\mathrm{D}$ is the corresponding covariance matrix, and $p_\alpha$ and $p_\beta$ are parameters which may be the cosmological parameters which we want to estimate or nuisance parameters associated with systematic uncertainties. 
In detail the matrix multiplication in Eq. (\ref{eq:Fisher}) is  a sum over all combinations of angular frequencies and tomographic bins. 

 Eq. (\ref{eq:Fisher}) assumes a Gaussian likelihood and  that the covariance is independent of the cosmological parameters.  As discussed by \citet{carron2013assumption}, using a parameter-dependent covariance matrix with a Gaussian likelihood would introduce a spurious term into the Fisher matrix. 
 
We consider two different data vectors, firstly the power spectrum and then the power spectrum and bispectrum combined.  In the second case  the covariance matrices, including their cross-covariance, are also combined \citep{kayo2012information}. We do not consider the bispectrum alone since if the bispectrum is available then we can assume that a two-point statistic has already been measured.

The diagonal element $(\bm{F}^{-1})_{\alpha\alpha}$ of the inverse Fisher matrix provides a lower bound for the variance of parameter $p_\alpha$ after marginalising over all other parameters.  Thus higher values in the Fisher matrix, or equivalently lower values in its inverse, correspond to lower uncertainty.  
In this work we are interested in  understanding how well we must constrain nuisance parameters in order to improve estimates of the cosmological parameters.  To do this we consider the effect of imposing priors on the nuisance parameters. To add a Gaussian prior with width $\Delta p_\alpha$ to parameter $p_\alpha$ we add $1/\Delta p_\alpha^2$ to $F_{\alpha\alpha}$. We then use the inverse of the updated Fisher matrix to determine revised constraints on the other parameters \citep{tegmark1997karhunen}.  
We use the inverse of the area of the Fisher ellipse  as a figure of merit (FoM), as defined by the Dark Energy Task Force \citep{albrecht2006report}.  This provides a single figure which quantifies how tightly the parameters  are constrained.  In the plane of the parameters $p_\alpha$ and $p_\beta$ the FoM is defined as 
\begin{align}
 \mathrm{FoM}_{\alpha\beta} &=\left\{ (\mathbfss{F}^{-1})_{\alpha\alpha}(\mathbfss{F}^{-1})_{\beta\beta}-\left[( \mathbfss{F}^{-1})_{\alpha\beta}\right]^2\right\}^{-1/2}\ .\label{eq:FoM}
 \end{align}
We focus on FoMs in the $\Omega_\mathrm{m}$ -- $\sigma_8$ and  $w_0$ -- $w_a$ planes which are most relevant for weak lensing.
\begin{figure}
      \begin{subfigure}[b]{0.25\linewidth}     
                  \includegraphics[width=8.5cm]{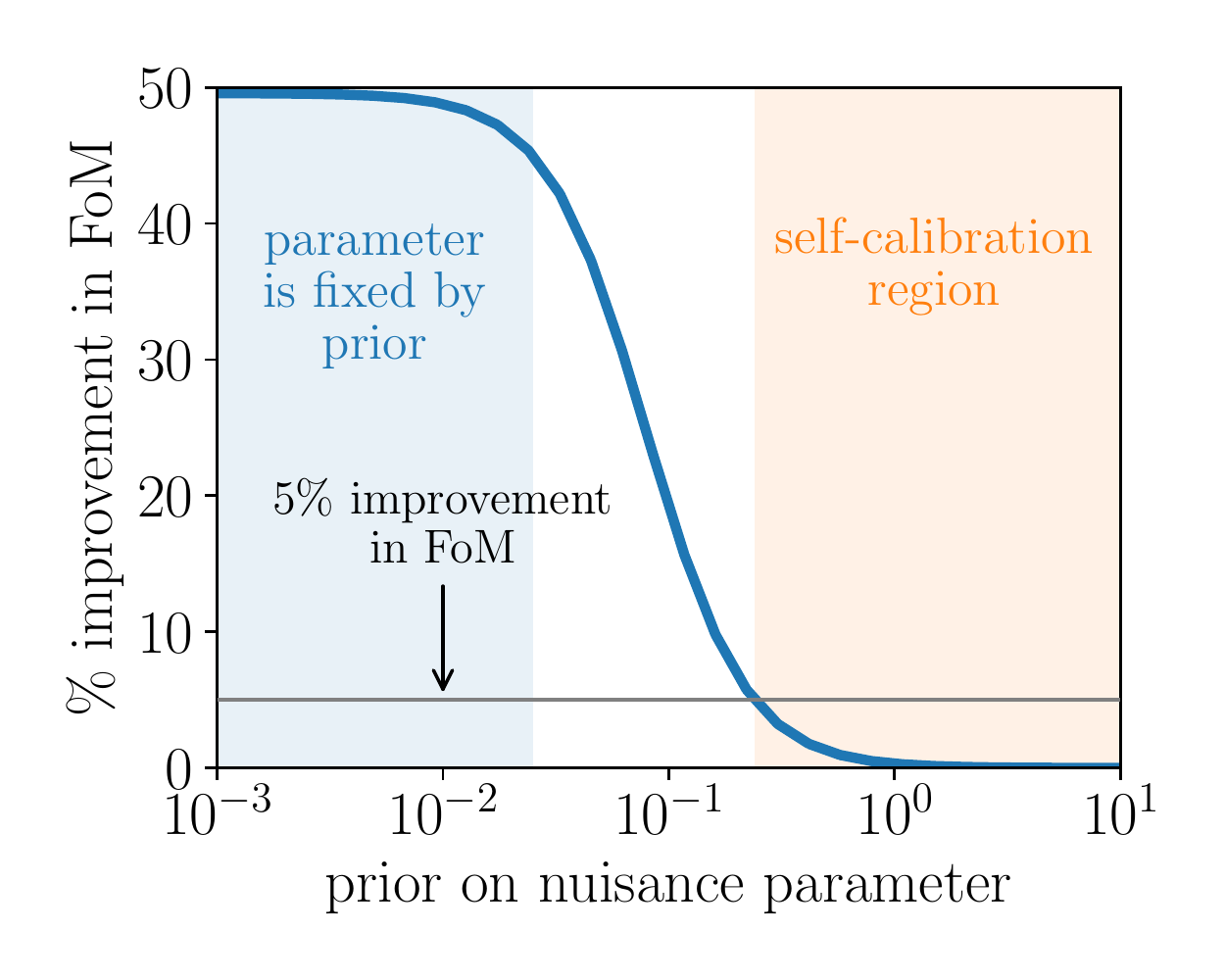}                    
    \end{subfigure}
 \caption{ Schematic diagram showing self-calibration. The blue line shows the typical shape of the relationship between a FoM and a prior on a nuisance parameter.  The vertical axis range is illustrative of the percentage improvement in the FoM compared with the FoM with a wide prior. We define the self-calibration regime as the region where the improvement in the FoM is less than five percent, shown by  the horizontal grey line.   }\label{fig:selfcal}
\end{figure}
The Fisher matrices and FoMs take account of the  cosmological parameters defined in Sect. \ref{sec:cosmoparams}  together with the nuisance parameters defined in Sect. \ref{sec:sysmodelling}: the parameters $A_{\mathrm{IA}}$ and $\eta_{\mathrm{IA}}$ from Eq. (\ref{eq:IAmodel}), five nuisance parameters $\Delta z_i$ denoting the shift in the mean value of the redshift bin centered on $z_i$, and five parameters $m_i$ representing the multiplicative bias in each tomographic bin.
To calculate the Fisher matrices we need the derivatives of the power spectrum and bispectrum with respect to the parameters.  The derivatives with respect to the intrinsic alignment and multiplicative bias parameters can be evaluated analytically 
but the cosmological parameters and redshift shifts require numerical derivatives for which we use a standard five-point stencil.  We confirmed the accuracy of the derivative calculations by verifying that, for each parameter $p_\alpha$, a Gaussian distribution centred on the fiducial parameter value with variance $(F_{\alpha\alpha})^{-1}$ matches the one-dimensional posterior for $p_\alpha$.  
\subsection{Interpretation of figures of merit}\label{sec:interpret}
  We use figures of merit in  several ways.  Firstly, FoMs in the presence of systematics can be compared to their values when there are no systematics. This quantifies the loss of information due to the systematic uncertainties.  This is particularly useful for comparing the relative importance of two different systematic effects. Secondly, we can quantify the extra information provided by the bispectrum (with or without systematics) by comparing the  FoMs obtained with the power spectrum only with those obtained with the combined power spectrum and bispectrum. 
Finally, we can consider how the FoMs change when we alter the priors on nuisance parameters.  This gives insight into the self-calibration regime where a nuisance parameter can be constrained purely from information in the survey without the need for external information to set priors, although at the expense of some loss of overall constraining power.  
\begin{figure*}
      \begin{subfigure}[b]{1\linewidth}     
                  \includegraphics[width=17cm]{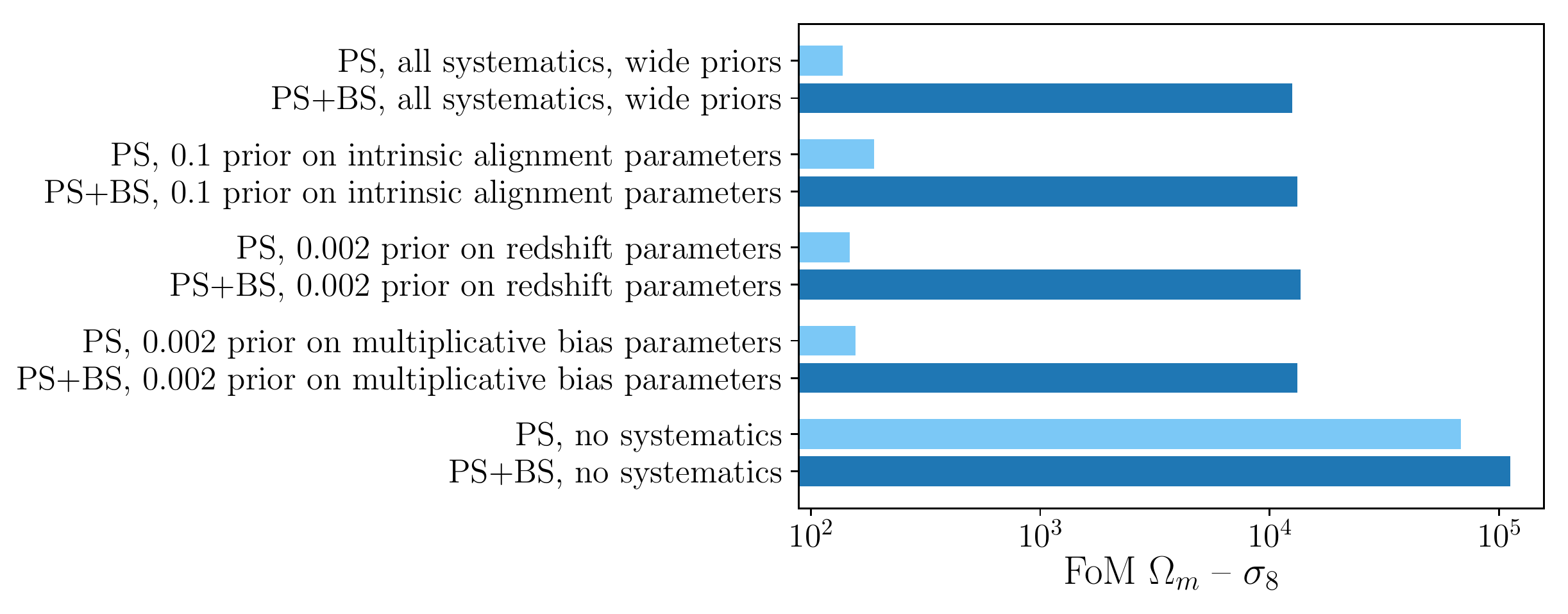}      
    \end{subfigure}
    
 \vspace{0.3cm} 
     \begin{subfigure}[b]{1.\linewidth}     
                  \includegraphics[width=17cm]{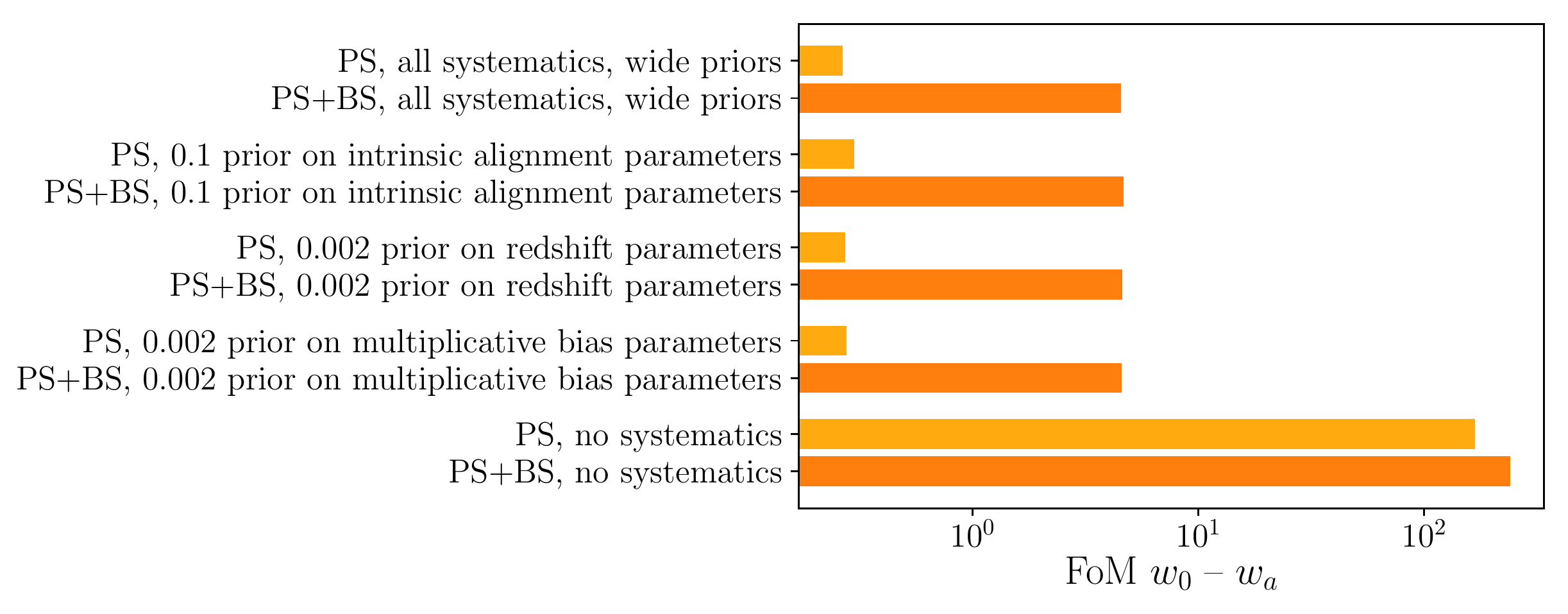}      
    \end{subfigure}
 \caption { The effects on the $\Omega_\mathrm{m}$--$\sigma_8$ and $w_0$ -- $w_a$ figures of merit of using the bispectrum as well as the power spectrum and the further effects of priors on the nuisance parameters.}\label{fig:summary}
\end{figure*}

Figure \ref{fig:selfcal} shows schematically how a figure of merit changes as the prior on a parameter is changed.  The values in this figure are purely for illustration.  The prior values considered are typical of those in our later analysis but the vertical axis values are simply illustrative of possible improvements in the FoM compared to the FoM with a wide prior of ten.  If the prior value is small (blue region in Fig. \ref{fig:selfcal}), the parameter is tightly constrained by the prior and further tightening of the prior does not affect it.  Conversely, in the orange region the parameter is independent of the prior; this is the self-calibration regime where the FoM can be determined purely by data from the survey.  
Between these two regimes, in the white area, the FoM rises rapidly as the prior is tightened. We choose to define the self-calibration regime as the region where the FoM is improved by less than five percent of the value it has with a wide prior, although this definition is somewhat arbitrary.   This level is indicated by the horizontal grey line in Fig.  \ref{fig:selfcal}.  The size of the step between the orange and blue regions indicates how strongly the FoM relies on priors outside the self-calibration region. A small step is desirable. 

In Sect. \ref{sec:sysresults} we consistently present results in the format of Fig. \ref{fig:selfcal}.
Throughout we show the percentage improvement in the FoMs compared with  \lq base\rq \ values obtained with wide priors.  In each case the self-calibration  regime, as we define it,  can be read off as the region where the FoM is improved by less than five percent.  The boundary of this region varies from case to case.
\subsection{Default priors}\label{sec:defpriors}
For our analysis we define a set of default priors to represent the baseline accuracy possible with \textit{Euclid}.
For redshift bin means and multiplicative bias we take the default priors to be the accuracy requirements specified in the Euclid Definition Study Report   \citep{laureijs2011euclid}.  This sets a requirement that the mean redshift should be known to at least an accuracy of $0.002\ (1+z)$ for each redshift bin. The corresponding  accuracy requirement for multiplicative bias is also 0.002, based on shear simulations in \citet{kitching2008bayesian}. 
There is no specified \textit{Euclid} accuracy requirement for intrinsic alignment parameters so we take as our default a conservative value of  0.1 for both parameters.

\section{Results}\label{sec:sysresults}
\subsection{Overview}
Our main results are summarised in Fig. \ref{fig:summary}.  This shows the FoMs obtained in three situations:  when all systematics are present but wide priors of ten are imposed on all nuisance parameters; when  default priors are imposed on each type of nuisance parameter in turn, but wide priors are imposed on the remaining parameters; and when no systematics are present -- this can be considered as  a baseline which exemplifies the maximum attainable information content. Table \ref{tab:FoMtable2} provides the numerical results behind Fig.  \ref{fig:summary}.   
Using the bispectrum can be much more beneficial than the alternative of using the power spectrum alone and imposing tight priors on the nuisance parameters.  When all systematics are taken together, combining the  power spectrum and bispectrum produces a 90-fold increase  in  the 
 \mbox{$\Omega_\mathrm{m}$ -- $\sigma_8$} FoM and a nearly 20-fold increase in the 
\mbox{$w_0$ -- $w_a$} FoM, compared with using the power spectrum alone, even when priors on all nuisance parameters are wide.  This improvement can be compared with the factor of 1.6 gain obtained from the bispectrum when only statistical uncertainties are considered.  
 
The default prior values used in Fig. \ref{fig:summary} are mainly in the self-calibration regions where the FoMs are insensitive to the prior. This explains why the FoMs are similar regardless of which systematic we consider here. This is especially true for the combined power spectrum and bispectrum, and less so for the power spectrum alone.   We discuss this further in Sect. \ref{sec:selfcal}. 

One important caveat in interpreting our results is that we have undoubtedly underestimated the constraining and self-calibration power of the power spectrum because we use only five tomographic  bins throughout, as discussed in Sect. \ref{sec:cosmoparams}. We return to this in Sect. \ref{sec:10bins}.

\subsection{Effect of the bispectrum - statistical errors}\label{sec:stat}
 The first line of each panel in Table \ref{tab:FoMtable1} shows FoMs (or ratios of FoMs) obtained when systematic uncertainties are ignored, so only statistical errors are present.  This situation has been investigated by several other authors \citep{kayo2012information,kayo2013cosmological,rizzato2019tomographic}. All  found that the bispectrum could improve cosmological parameter constraints: \citet{kayo2012information} estimated a 20-40\% improvement in the signal to noise ratio from using the bispectrum, \citet{kayo2013cosmological} forecast a 60\% improvement in the dark energy figure of merit and \citet{rizzato2019tomographic} forecast an improvement in the signal to noise ratio of around 10\%.  In comparison we find that including the bispectrum as well as the power spectrum increases the \mbox{$\Omega_\mathrm{m}$ -- $\sigma_8$} FoM by around 60\% and the \mbox{$w_0$ -- $w_a$} FoM by around 40\%.  The differences in the results can be attributed at least partly to different survey specifications and tomographic set-ups.
 \subsection{Effect of the bispectrum - systematic errors}
The remainder of Table \ref{tab:FoMtable1} shows the impact of systematic uncertainties on the two FoMs, assuming wide priors on all the nuisance parameters.  Intrinsic alignments have the most deleterious effect.  With the power spectrum only, the presence of intrinsic alignment nuisance parameters reduces  
 the \mbox{$\Omega_\mathrm{m}$ -- $\sigma_8$} FoM by a factor of more than 300, and the \mbox{$w_0$ -- $w_a$} FoM by a factor of 400.    Multiplicative bias is relatively harmless, although certainly not negligible. Again considering the power spectrum only, multiplicative bias causes both FoMs to fall by around $20 - 25 \%$.  The effect of redshift uncertainty is intermediate, reducing  the 
 \mbox{$\Omega_\mathrm{m}$ -- $\sigma_8$} FoM by a factor of around ten and the \mbox{$w_0$ -- $w_a$} FoM by a factor of 35.
Even with wide priors on all the nuisance parameters the bispectrum is hugely helpful in counteracting the effect systematic uncertainties.  This is largely because the bispectrum reduces the impact of intrinsic alignments, which affect the power spectrum and bispectrum very differently as seen in Sect. \ref{sec:IA}. However the bispectrum also considerably offsets the effect of uncertainty in redshift bin means.  
  \begin{table*}
\vspace{0.9cm}
\caption{Figures of merit obtained with the power spectrum only and with the power spectrum and bispectrum together, when tight priors are imposed on nuisance parameters. In each case  wide priors are assumed for all parameters which do not have priors explicitly imposed.}
    \label{tab:FoMtable2}
  \centering
    \begin{tabular}{lrr}
\hline
\multicolumn{1}{c}{Analysis type}& \multicolumn{2}{c} {Figure of merit/ratio}\\
\multicolumn{1}{c}{} &\multicolumn{1}{r} {$\Omega_\mathrm{m}$ -- $\sigma_8$} & \multicolumn{1}{r}{ $w_0$ -- $w_a$}\\
\hline
PS, wide priors on all nuisance parameters&138 &0.27 \\
PS, 0.1 prior on IA parameters & 188& 0.30 \\
  PS, 0.002 prior on redshift parameters & 148 & 0.27 \\
  PS, 0.002 prior on multiplicative bias parameters & 156 & 0.28\\[1ex]
 \hline  
PS+BS, wide priors on all nuisance parameters& 12\,557 &4.54\\
  PS+BS, 0.1 prior on IA parameters& 13\,182 & 4.69 \\
   PS+BS, 0.002 prior on redshift parameters & 13\,657 & 4.63 \\
   PS+BS, 0.002 prior on multiplicative bias parameters & 13\,183 & 4.59\\[1ex]
\hline(PS+BS)/PS, wide priors on all nuisance parameters&90.9&17.0\\
  (PS+BS)/PS, 0.1 prior on IA parameters & 70.0& 15.6\\
  (PS+BS)/PS,  0.002 prior on redshift parameters & 92.4 & 16.9\\
    (PS+BS)/PS,  0.002 prior on multiplicative bias parameters & 84.4 & 16.6\\[1ex]
  \hline
    \end{tabular}
    \end{table*}
 
\begin{table*}
 \caption{Figures of merit obtained with the power spectrum only, and with the power spectrum and bispectrum together, when wide priors are imposed on all nuisance parameters. Since wide priors have been imposed on the nuisance parameters, in this table it is not assumed that the redshift and multiplicative bias parameters meet the \textit{Euclid} accuracy requirements.}
    \label{tab:FoMtable1}
  \centering
    \begin{tabular}{lrr}
    \hline
\multicolumn{1}{c}{Spectrum type}& \multicolumn{2}{c} {Figure of merit/ratio}\\
\multicolumn{1}{c}{} &\multicolumn{1}{r} { $\Omega_\mathrm{m}$ -- $\sigma_8$} & \multicolumn{1}{r}{$w_0$ -- $w_a$}\\
\hline
  PS, no systematics & 68\,029 & 169 \\
  PS, intrinsic alignments only& 229 &0.4 \\
  PS, redshift bin shifts only& 7\,808 & 4.6 \\
  PS, multiplicative bias only& 53\,000 & 117 \\
  PS, all systematics & 138 & 0.3 \\[1ex]
\hline  
  PS+BS, no systematics & 111\,834 & 241 \\
  PS+BS, intrinsic alignments only& 16\,199&5.2\\
  PS+BS, redshift bin shifts only& 65\,972 & 34.0 \\
  PS+BS, multiplicative bias only& 97\,796 & 188 \\
  PS+BS, all systematics & 12\,557 & 4.5 \\[1ex]
  \hline
  (PS + BS)/PS, no systematics & 1.64 & 1.43\\
  (PS + BS)/PS, intrinsic alignments only& 70.7&13.9\\
  (PS + BS)/PS, redshift bin shifts only& 8.4 & 7.4 \\
  (PS + BS)/PS, multiplicative bias only& 1.8 & 1.6 \\
  (PS + BS)/PS, all systematics & 90.9 & 17.0 \\[1ex]
 \hline
    \end{tabular}
    \end{table*}
  \subsection{Number of tomographic bins}\label{sec:10bins}
An alternative to using the bispectrum is to use the power spectrum only but with more tomographic bins.  To investigate this we recalculate the figures of merit using the power spectrum only with ten bins.  The results are shown in Table \ref{tab:PS10bins}.  For redshift uncertainties and multiplicative bias this automatically increases the number of nuisance parameters so the figures of merit also increase. However for intrinsic alignments even when ten bins are used for the power spectrum the figures of merit do not approach those those shown in Table  \ref{tab:FoMtable1}.  Using the power spectrum with ten bins  reducess the \mbox{$\Omega_\mathrm{m}$ -- $\sigma_8$} figure of merit by a factor of 20 and the  \mbox{$w_0$ -- $w_a$} figure of merit by a factor of five compared with using the power spectrum and bispectrum combined but only five bins. 

  \begin{table*}
 \caption{ Figures of merit obtained with the power spectrum only with 10 tomographic bins, and with the power spectrum (10 bins)  and bispectrum (5 bins) together, when wide priors are imposed on all nuisance parameters. }\label{tab:PS10bins}
    \centering
    \begin{tabular}{lrr}
    \hline
\multicolumn{1}{c}{Analysis type}& \multicolumn{2}{c} {Figure of merit/ratio}\\
\multicolumn{1}{c}{} &\multicolumn{1}{r} { $\Omega_\mathrm{m}$ -- $\sigma_8$} & \multicolumn{1}{r}{$w_0$ -- $w_a$}\\
\hline
  PS, 10 bins, no systematics & 111\,892 & 306 \\
  PS, 10 bins, intrinsic alignments only&720&1.1 \\
  PS, 10 bins, redshift bin shifts only& 12\,024 & 7.7 \\
  PS, 10 bins, multiplicative bias only& 101\,056 & 257 \\
  PS, 10 bins, all systematics & 531 & 0.8 \\[1ex]
\hline  
  PS (10 bins)+BS (5 bins), no systematics & 160\,982 & 392 \\
  PS (10 bins)+BS (5 bins), intrinsic alignments only& 27\,140&8.9\\[1ex]
  \hline
  (PS+BS)/PS, 10+5 bins, no systematics & 1.43 & 1.28\\
  (PS+BS)/PS, 10+5 bins, intrinsic alignments only& 37.7&8.4\\[1ex]
 \hline
    \end{tabular}
    
 \end{table*}
 
\subsection{Self-calibration}\label{sec:selfcal}
\begin{figure*}
\hspace{-2.4cm}
     \begin{subfigure}[b]{0.25\linewidth}
          \includegraphics[width=8cm]{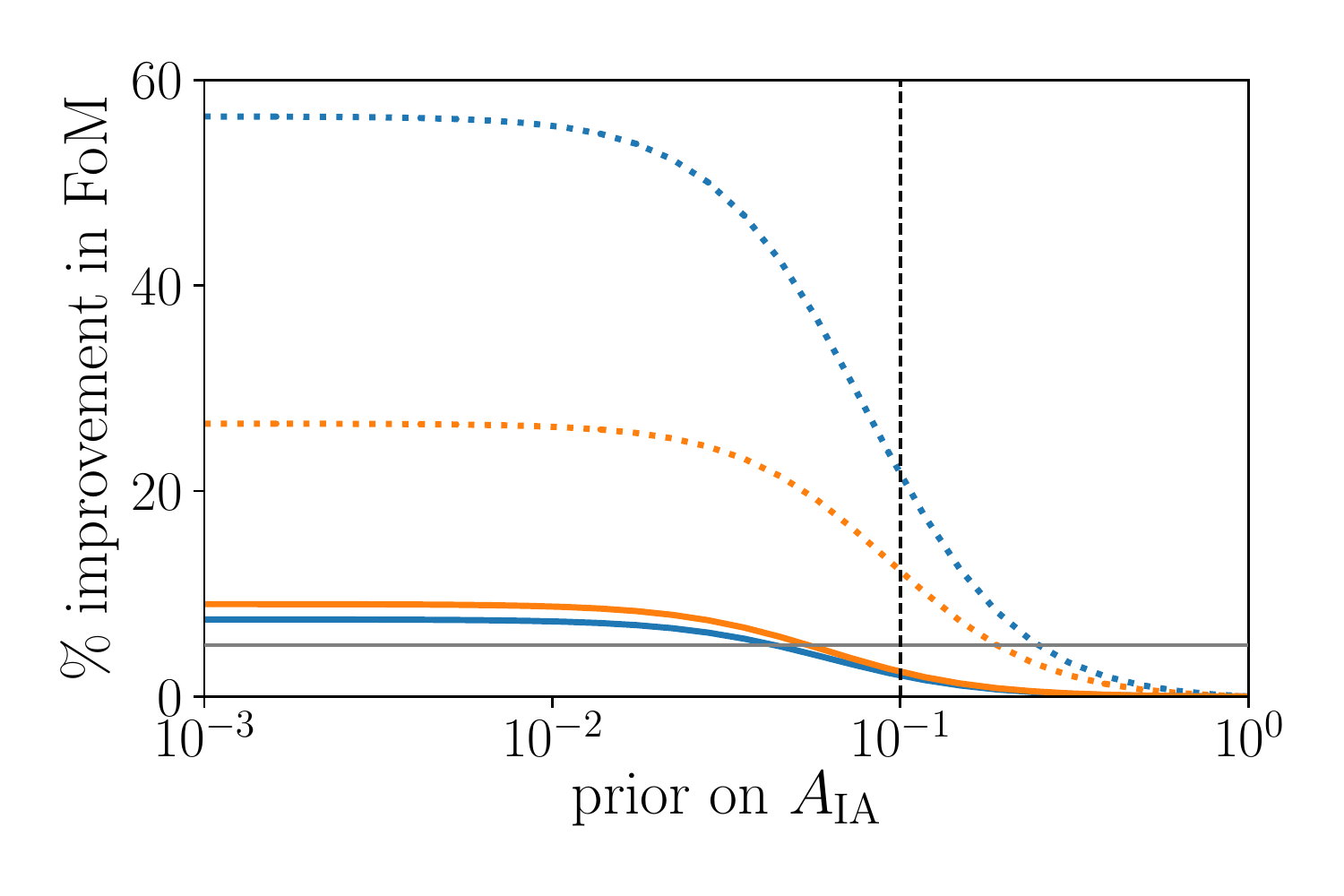}      
    \end{subfigure}
    \hspace{4cm} 
    \begin{subfigure}[b]{0.25\linewidth} 
    \includegraphics[width=8cm]{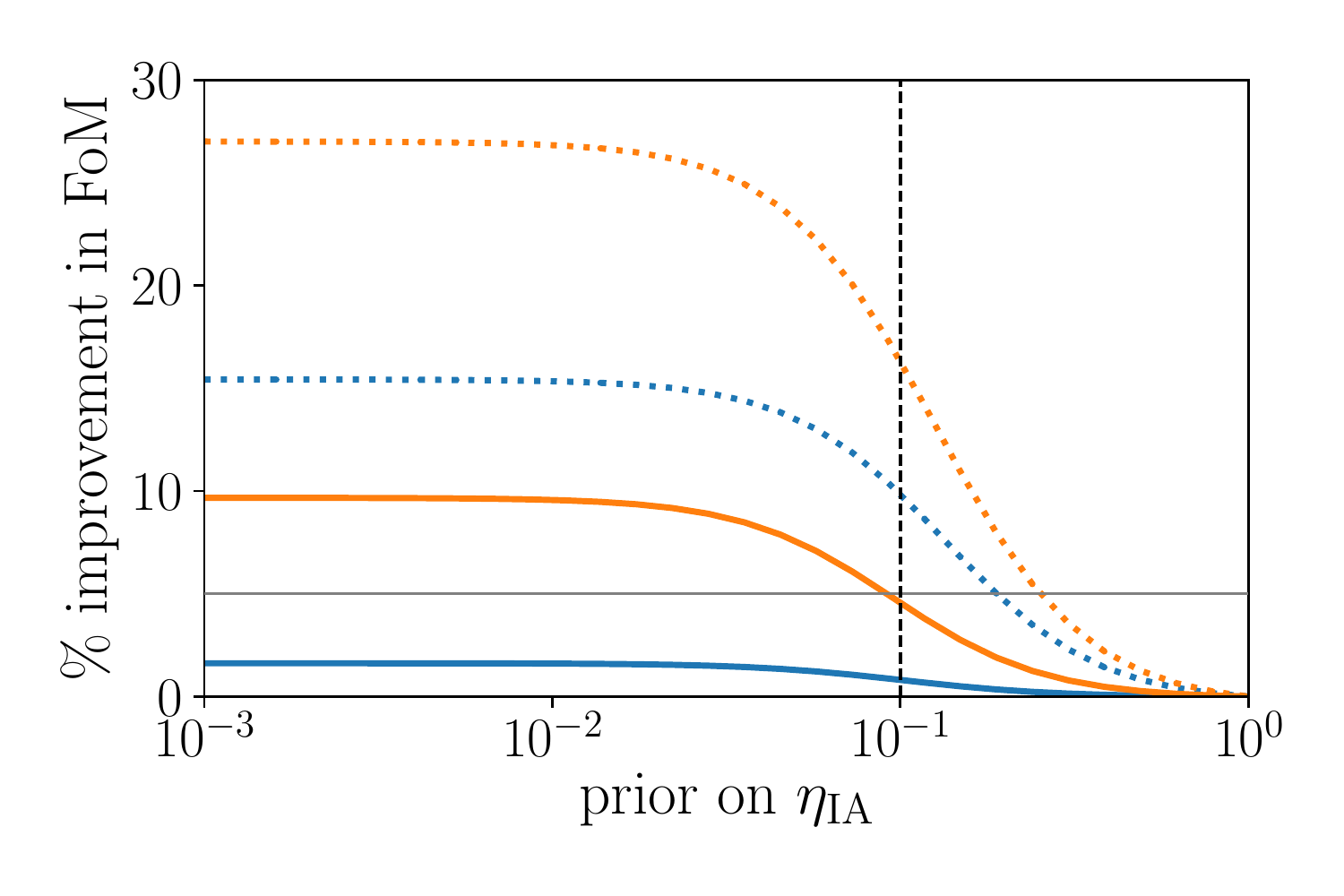}   
    \end{subfigure}
   
  \vspace{-0.6cm} 
  \hspace{-2.2cm}
     \begin{subfigure}[t]{0.25\linewidth}                   
   \vskip 0pt   \includegraphics[width=8cm]{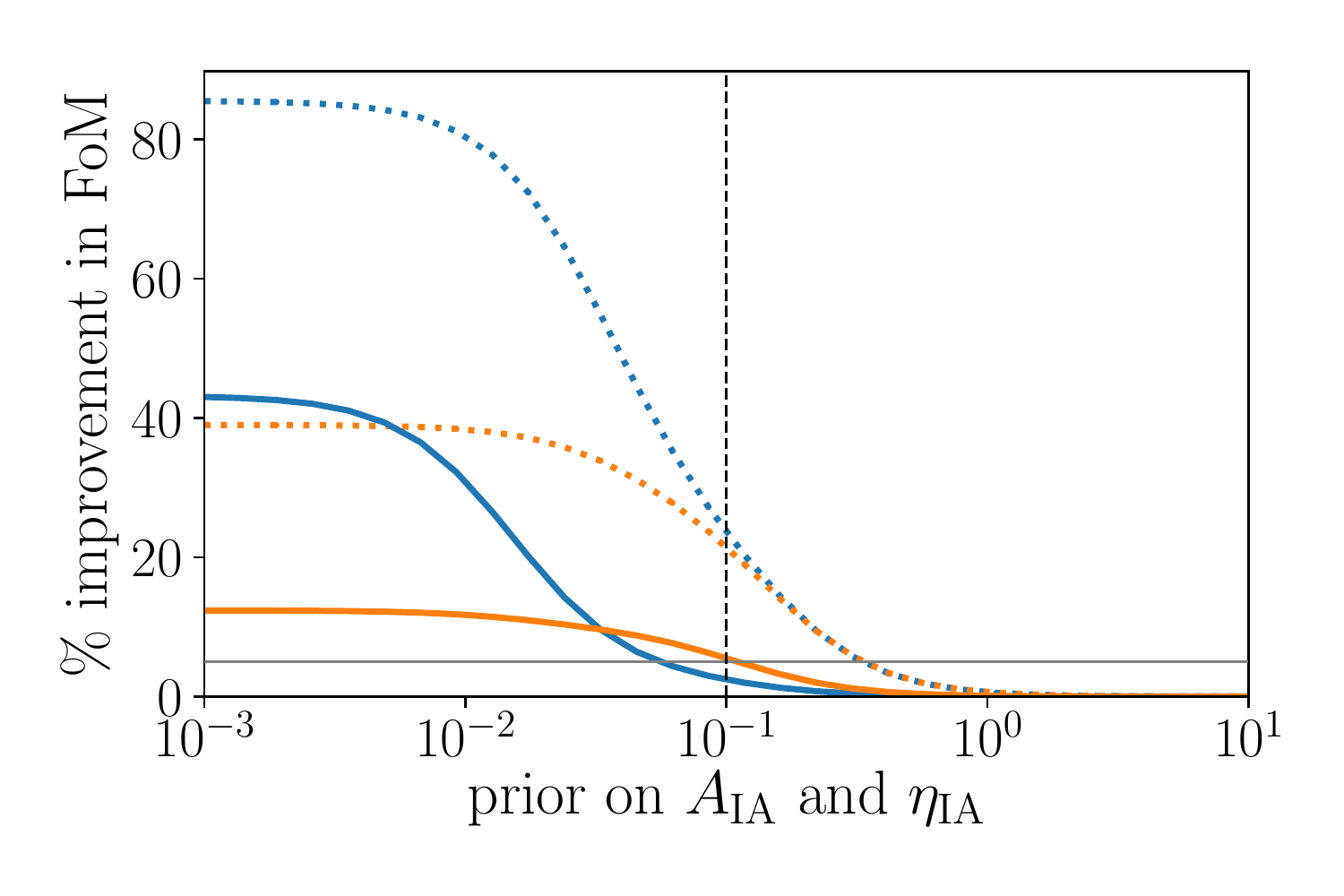}      
    \end{subfigure}
   \hspace{4.2cm}
   \begin{subfigure}[t]{0.25\linewidth}
     \vskip 1.6cm    \includegraphics[width=7cm]{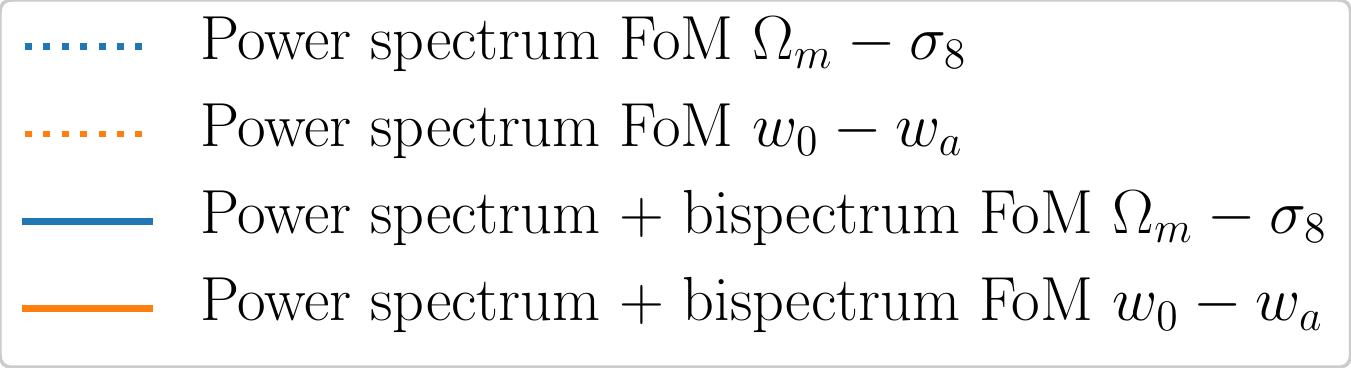}  
    \end{subfigure}
   \caption{ Percentage increase in figures of merit when the  priors on the parameters $A_{\mathrm{IA}}$ and $\eta_{\mathrm{IA}}$ are tightened, compared to wide priors of 10. Priors on all other nuisance parameters are set to their default values - see Sect. \ref{sec:defpriors}.  \textit{Top left}: Effect of tightening prior on $A_\mathrm{IA}$ only. \textit{Top right}: Effect of tightening prior on $\eta_{\mathrm{IA}}$ only. \textit{Bottom left}: Effect of tightening both priors simultaneously. The vertical dashed lines indicate our default prior of 0.1. The horizontal grey lines indicate a 5\% improvement in the FoM. An improvement less than this is our criterion for self-calibration. Note different vertical scales in each panel.}\label{fig:IAFoM1}
\end{figure*}

We next investigate the effect on the FoMs of tightening or relaxing the priors on the nuisance parameters and  hence the potential for self-calibration. 
Figure \ref{fig:IAFoM1} shows the effect of varying priors on the intrinsic alignment parameters, with fixed priors equal to the \textit{Euclid} requirements imposed on all other parameters.   The horizontal grey lines in each panel indicate our definition of self-calibration discussed in Sect. \ref{sec:interpret}.
The self-calibration regime is the region to the right of the point where these lines cross the orange or blue lines.
\begin{figure*}
\hspace{-2.4cm}
      \begin{subfigure}[b]{0.25\linewidth}     
                  \includegraphics[width=8cm]{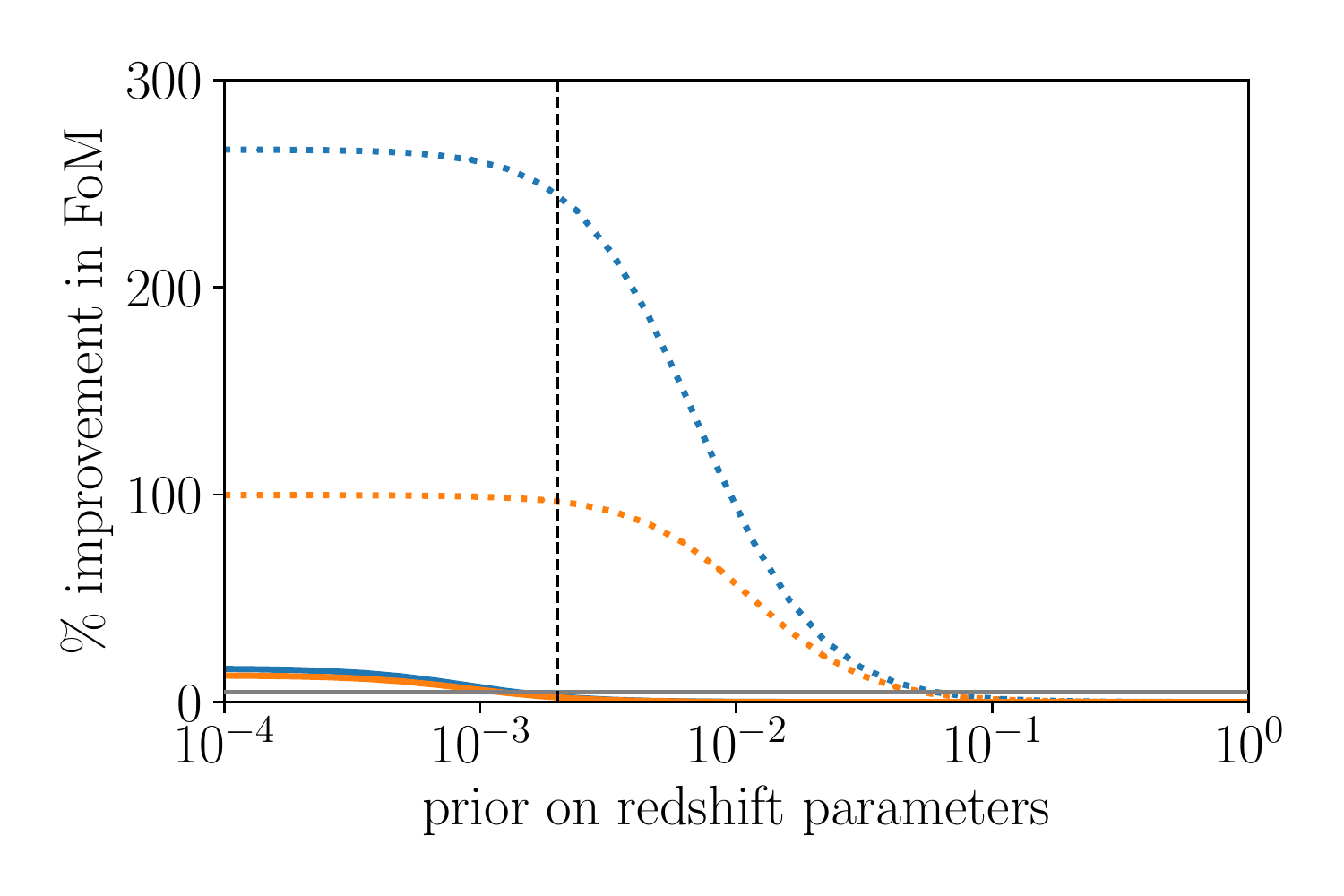}                    
    \end{subfigure}
 \hspace{4cm}
      \begin{subfigure}[b]{0.25\linewidth}     
                  \includegraphics[width=8cm]{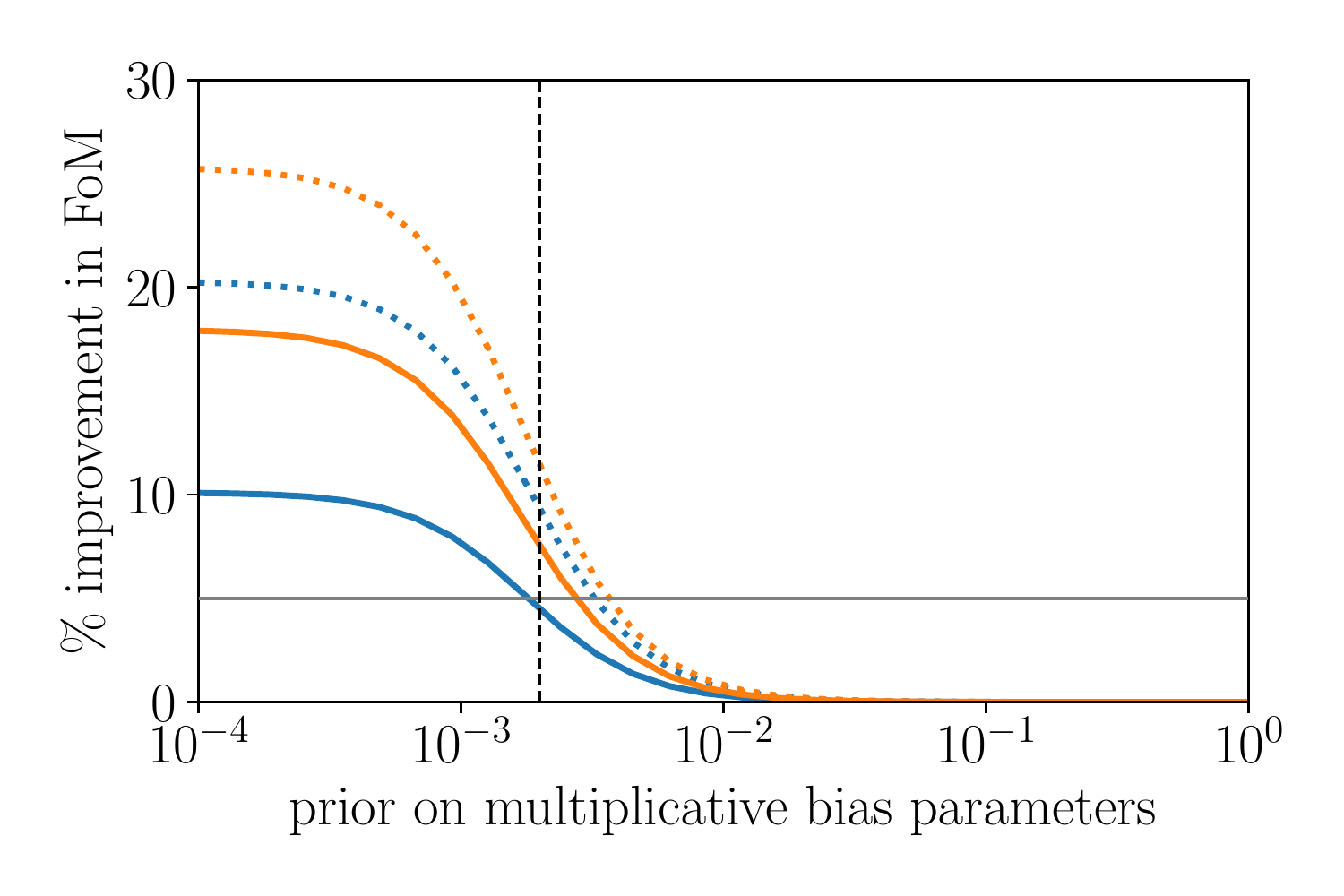}                    
    \end{subfigure}
    
        \vspace{0.1cm}
         \hspace{5.5cm}
    \begin{subfigure}[b]{1.\linewidth}
    \hspace{6cm}
          \includegraphics[width=7cm]{legendFoM.pdf}  
              \end{subfigure}
 \caption{Percentage increase in figures of merit when priors are tightened simultaneously. In each case the same prior is applied to every parameter. Priors on all other nuisance parameters are set to their default values - see Sect. \ref{sec:defpriors}.  \textit{Left}: Redshift parameters   \textit{Right}: Multiplicative bias parameters. The vertical dashed lines indicate the  accuracy requirements from the Euclid Definition Study Report  \citep{laureijs2011euclid}. The horizontal grey lines indicate a 5\% improvement in the FoM. An improvement less than this is our criterion for self-calibration. Note different vertical scales in each panel.}\label{fig:multFoM2}
\end{figure*}
\begin{figure*}
\hspace{-2.5cm}
      \begin{subfigure}[b]{0.25\linewidth}     
                  \includegraphics[width=8cm]{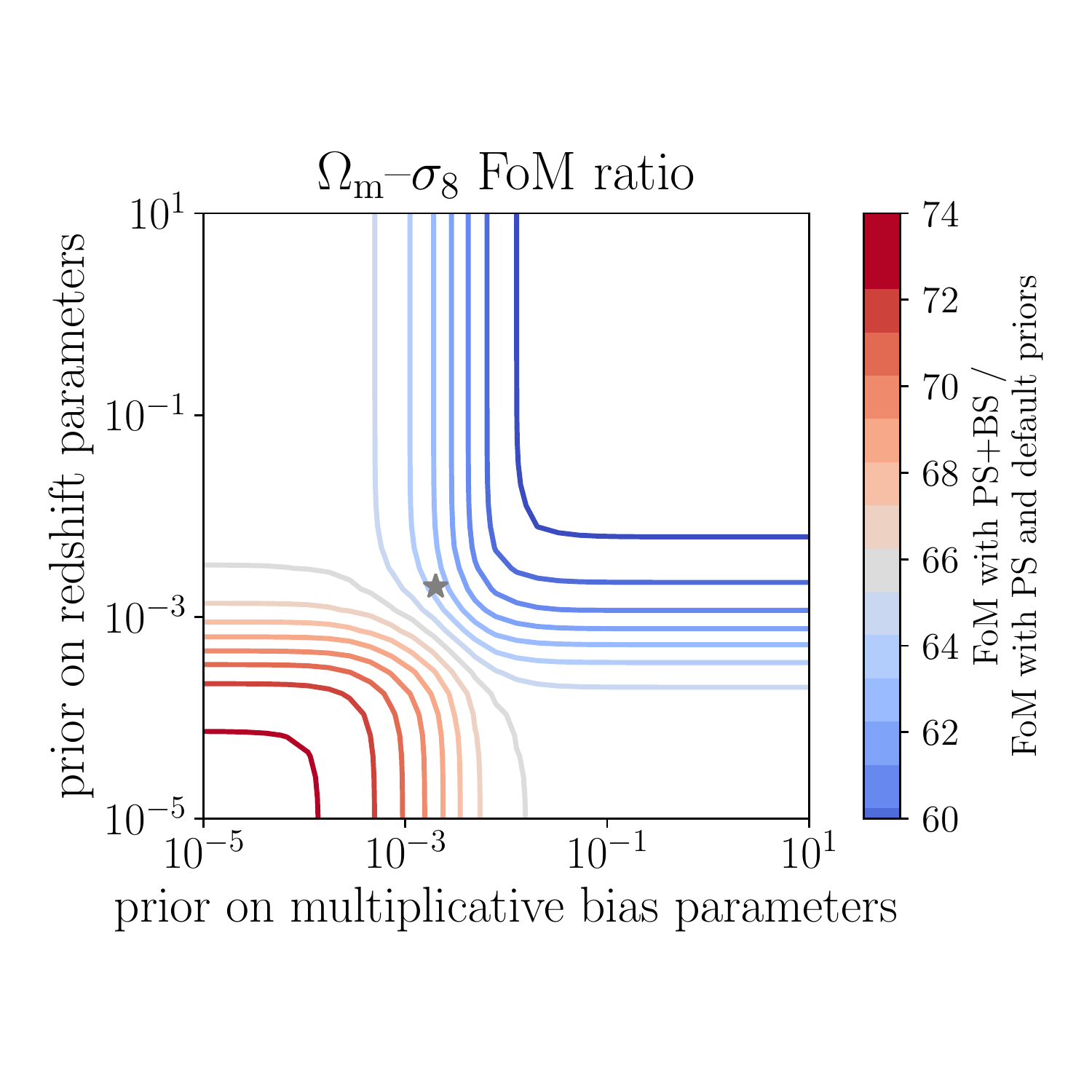}                    
    \end{subfigure}
\hspace{4cm}
    \begin{subfigure}[b]{0.25\linewidth}     
                  \includegraphics[width=8cm]{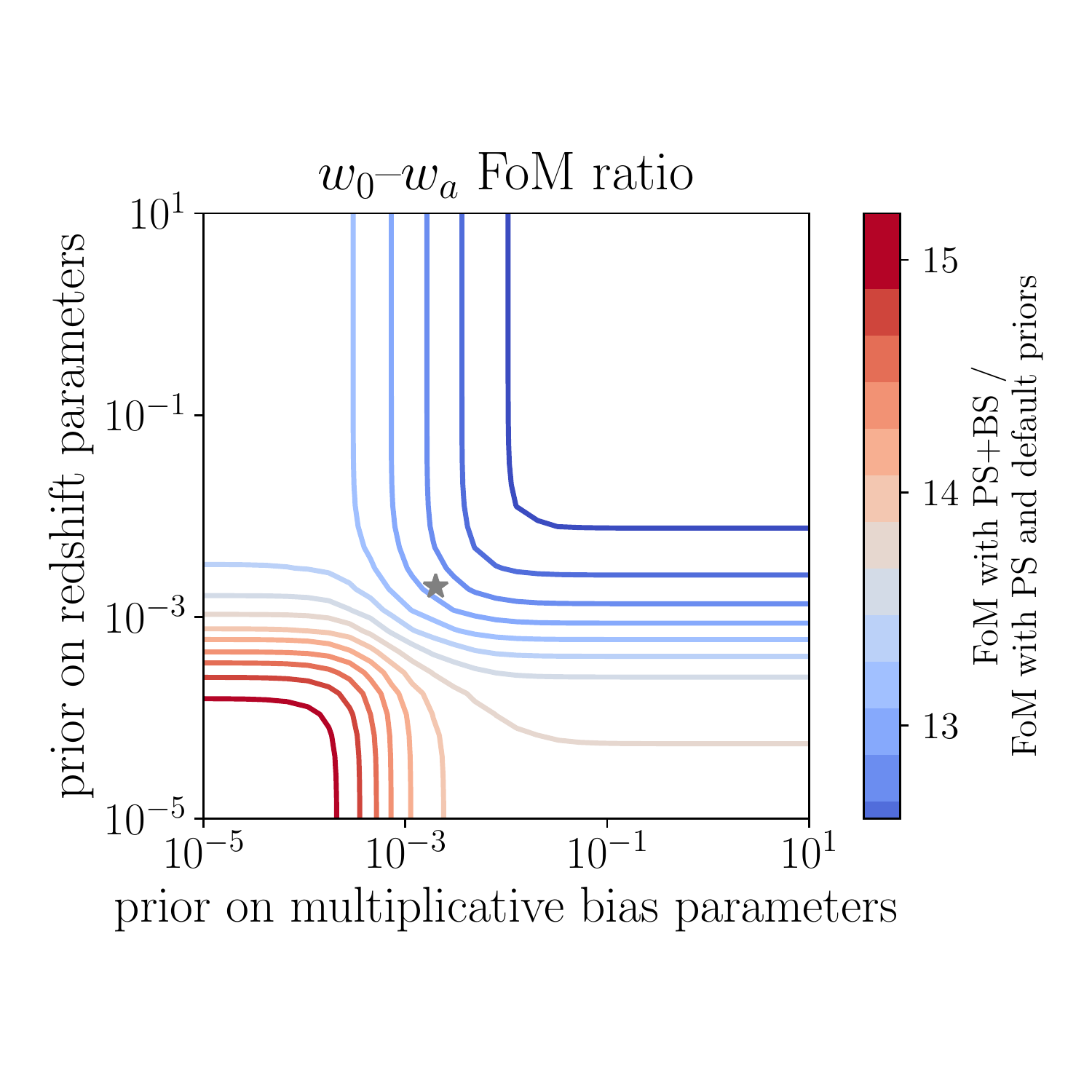}                    
    \end{subfigure}
    \vspace{-1cm} 
 \caption{ Ratio of FoM with power spectrum and bispectrum to FoM with power spectrum and default priors: 0.1 for the intrinsic alignment parameters and  \textit{Euclid} accuracy requirement of 0.002 for redshift and multiplicative bias  parameters \citep{laureijs2011euclid}. The grey stars indicate the default prior values for redshift bin mean and multiplicative bias parameters.  \textit{Left}: $\Omega_\mathrm{m}$ -- $\sigma_8$ FoM.   \textit{Right}: $w_0$ -- $w_a$ FoM. Note different scales in the two panels. Priors on the intrinsic alignment parameters are set to their default values - see Sect. \ref{sec:defpriors}.  
  }\label{fig:contour1}
\end{figure*}

When only the power spectrum is considered the self-calibration regime starts when the priors on $A_{\mathrm{IA}}$ and $\eta_{\mathrm{IA}}$ are about 0.5.
Using the bispectrum as well,  the self-calibration regime extends to a prior value of about 0.1 for $A_{\mathrm{IA}}$, but does not change for $\eta_{\mathrm{IA}}$.   
The bottom panel of  Fig. \ref{fig:IAFoM1} shows that if priors on both intrinsic alignment parameters are tightened simultaneously, the self-calibration requirements are around 0.5 if only the power spectrum is considered.  In contrast, when the power spectrum and bispectrum are combined, the self-calibration point is around 0.05, within our default prior of 0.1.

In all three panels of  Fig. \ref{fig:IAFoM1} the size of the step between the self-calibration regime and the regime where the FoM is controlled by the priors is much smaller when  the  bispectrum and power spectrum are combined.  This means that even outside the self-calibration regime the bispectrum massively reduces the requirement for tight external priors and the degradation of the FoM within the self-calibration is much less than for the power spectrum only.

For the power spectrum,  the \mbox{$\Omega_\mathrm{m}$ -- $\sigma_8$} figure of merit is most sensitive to the amplitude parameter $A_\mathrm{IA}$ and the \mbox{$w_0$ -- $w_a$} figure of merit is most sensitive to the the redshift exponent $\eta_{\mathrm{IA}}$.  This is as expected:  $\Omega_\mathrm{m}$, $\sigma_8$  and $A_{\mathrm{IA}}$ have confounding effects on the amplitude of the weak lensing signal, whereas the dark energy parameters are highly sensitive to redshift uncertainty.
 
We next consider redshift uncertainty and multiplicative bias, setting fixed priors of 0.1 on the two intrinsic alignment parameters. The prospects for self-calibration are shown in Fig.\ref{fig:multFoM2}.  Once again we indicate our criterion for self-calibration by horizontal grey lines. Vertical dashed lines indicate the \textit{Euclid} accuracy requirements. For redshift bin means, if the power spectrum only is used the self-calibration regime starts at a prior of around 0.1 for both FoMs. This means that the only way to improve cosmological parameter constraints is through narrow external priors.  In contrast, if the bispectrum is also used then the boundary of the self-calibration regime changes to about 0.001, within the \textit{Euclid} requirement.  A similar pattern is seen for the multiplicative bias parameters. When only the power spectrum is used the boundary of the self-calibration regime is at a value of about 0.3, again implying that tight external priors are needed to improve parameter constraints.  If the bispectrum is used as well self-calibration starts at around 0.005, just outside the  \textit{Euclid} requirement, except in the case of the $\Omega_\mathrm{m}$ -- $\sigma_8$ FoM where the self-calibration boundary is almost exactly at the \textit{Euclid} requirement.

In Fig. \ref{fig:contour1} we explore the joint effect of priors on redshift and multiplicative bias parameters, together with the effect of using the bispectrum as well as the power spectrum.  This figure shows the ratio between the FoM obtained with the combined power spectrum and bispectrum with the default prior of 0.1 imposed on the intrinsic alignment parameters but varying priors on the other parameters, and the FoM obtained with the power spectrum only and  priors of 0.1 for the intrinsic alignment parameters and 0.002 for the redshift and multiplicative bias parameters.   Thus the panels show, for each FoM, the improvement from using the bispectrum compared with the baseline \textit{Euclid} scenario with the power spectrum only and default priors.  The grey stars indicate the default values of the redshift and multiplicative bias priors.  At these points the  $\Omega_\mathrm{m}$ -- $\sigma_8$ FoM is around 65 times greater than the baseline \textit{Euclid} value and the $w_0$ -- $w_a$ FoM is around 13 times greater.  Thus including the bispectrum as well as the power spectrum produces a large gain compared with any further tightening of the priors with the power spectrum only.  This is true even when the redshift and multiplicative bias priors are greatly relaxed.   
Figure \ref{fig:contour1} also shows that there is little interaction between the redshift parameters on the one hand and the multiplicative bias parameters on the other.  Thus there is only limited opportunity for trade-offs between the accuracy of the two sets of parameters. 

\section{Conclusions}\label{sec:sysconc}
In the context of a \textit{Euclid}-like tomographic weak lensing survey we have considered three major sources of systematic uncertainty: contamination by intrinsic alignments which adds additional terms to the cosmic shear power spectrum and bispectrum; uncertainty in the mean redshifts of the tomographic bins due to the use of photometric redshift measurements; and multiplicative bias which affects the amplitude of the shear signal.  We modelled the effects of these systematics on the weak lensing bispectrum by extending existing methods which are well-tested for the power spectrum and which have been used to analyse data from current weak lensing surveys.  

We used figures of merit based on Fisher matrices to forecast the effect of these systematics on parameter constraints, focusing in particular on the large-scale structure parameters $\Omega_\mathrm{m}$ and $\sigma_8$ and the dark energy equation of state parameters $w_0$ and $w_a$.  Whether we consider the power spectrum only or the combined power spectrum and bispectrum, the presence of  systematic uncertainties causes an order of magnitude decrease in the figures of merit in both the \mbox{$\Omega_\mathrm{m}$ -- $\sigma_8$} and \mbox{$w_0$ -- $w_a$} planes.   

We compared two strategies for combatting this loss of information.  The first approach rests on using the power spectrum only and imposing tight priors on the systematic nuisance parameters, informed by external calibration data or simulations.  This is what is normally done in weak lensing analysis. In our analysis we assume that the external calibration meets requirements in the Euclid Definition Study Report   \citep{laureijs2011euclid}, where these exist. The second strategy involves analysing the bispectrum alongside the power spectrum.  We find that this greatly reduces the impact of systematic uncertainties, especially intrinsic alignments which, with our modelling assumptions, contribute at different levels to the power spectrum and bispectrum.  

Thus much more can be gained by using the bispectrum than by setting tight priors but using only the power spectrum. This is true even though our analysis is based on  a \lq cut down\rq \ bispectrum which depends only on equilateral triangles.  Using more triangle configurations could be expected to produce even greater gains.  Our results are also conservative because we used only a limited number of tomographic bins.  Increasing the number of bins would increase the constraining power from both the power spectrum and the combined bispectrum and power spectrum.  The relative gain from the bispectrum might be then smaller but we would still expect a substantial improvement.

In all the cases which we have considered, combining the bispectrum with the power spectrum improves self-calibration power: the self-calibration regime starts at a smaller prior value than with the power spectrum alone and there is less degradation in the figures of merit in the self-calibration regime.  For redshift and multiplicative bias uncertainties,  the self-calibration regime for a combined power spectrum-bispectrum analysis starts near or within \textit{Euclid} accuracy requirements. For intrinsic alignment parameters, where there are no specified \textit{Euclid} requirements, self-calibration starts close to our conservatively chosen default prior values.   It is important to recognize, however, that the added constraining power due to the bispectrum will lead to tighter accuracy requirements on all nuisance parameters if systematic errors are to be kept well below statistical errors.

These results are encouraging and we plan to explore several aspects further.
Firstly, we intend to validate our intrinsic alignment modelling by revisiting the analysis in \citet{semboloni2008sources} using state-of-the-art simulations such as the Euclid Flagship Mock Galaxy Catalogue\footnote{https://sci.esa.int/web/euclid/-/59348-euclid-flagship-mock-galaxy-catalogue}, or survey data such as the Dark Energy Survey Instrument Bright Galaxy Survey\footnote{https://www.desi.lbl.gov/the-desi-survey/} \citep{levi2019dark}.   
Secondly, as discussed in Sect. \ref{sec:cosmoparams}, we will  review the new  formula for the matter bispectrum derived by \cite{takahashi2020fitting} to improve our bispectrum modelling and investigate the potential for further self-calibration. 
Finally, we intend to explore the performance of weak lensing three-point statistics which are readily derived from real data, such as aperture mass statistics \citep{schneider1998new}.
Extending our work in these ways will help to confirm the practical value of using three-point statistics to control systematics in \textit{Euclid} and other next-generation weak lensing surveys.

\section*{Acknowledgements}
We are grateful to the anonymous reviewer for a helpful and positive report and to Hiranya Peiris for constructive comments on an earlier draft of this work. We would also like to thank Saroj Adhikari and Kwan Chuen Chan for providing information about their respective bispectrum calculations, and Stephen Stopyra for advice on using \textsc{Mathematica}.
\section*{Data Availability}
The data underlying this article will be shared on reasonable request to the corresponding author.


\typeout{}
\bibliographystyle{mnras}
\bibliography{refs} 
\appendix
\section{Origin of terms in the matter power spectrum and bispectrum covariance}\label{sec:covterms}
If  $\tilde{\delta}$ is the Fourier transform of the underlying density contrast, then an estimator for the power spectrum will involve the product of two Fourier modes $\tilde{\delta}\tilde{\delta}$. Similarly a bispectrum estimator involves $\tilde{\delta}\tilde{\delta}\tilde{\delta}$.
From this we can use Wick's theorem to understand the corresponding covariances. 

The power spectrum covariance has two terms.  Schematically, one term involves $\langle\tilde{\delta}\tilde{\delta}\rangle\langle\tilde{\delta}\tilde{\delta}\rangle$,
 the product of two power spectra, and the other involves the connected \mbox{four-point} function or trispectrum, $\langle\tilde{\delta}\tilde{\delta}\tilde{\delta}\tilde{\delta}\rangle_\mathrm{c}$.  

The bispectrum covariance has terms involving 
\begin{align}
\hspace{0.4cm}\langle\tilde{\delta}\tilde{\delta}\rangle\langle\tilde{\delta}\tilde{\delta}\rangle\langle\tilde{\delta}\tilde{\delta}\rangle\notag,\
\langle\tilde{\delta}\tilde{\delta}\tilde{\delta}\rangle_\mathrm{c}\langle\tilde{\delta}\tilde{\delta}\tilde{\delta}\rangle_\mathrm{c}\notag,\
\langle\tilde{\delta}\tilde{\delta}\tilde{\delta}\tilde{\delta}\rangle_\mathrm{c}\langle\tilde{\delta}\tilde{\delta}\rangle\notag \ \text{  and  }
\langle\tilde{\delta}\tilde{\delta}\tilde{\delta}\tilde{\delta}\tilde{\delta}\tilde{\delta}\rangle_\mathrm{c}\notag\ ,
\end{align}
and the power spectrum--bispectrum cross-covariance involves 
\begin{align}
\hspace{0.4cm}\langle\tilde{\delta}\tilde{\delta}\rangle\langle\tilde{\delta}\tilde{\delta}\tilde{\delta}\rangle_\mathrm{c} \text{  and  }\langle\tilde{\delta}\tilde{\delta}\tilde{\delta}\tilde{\delta}\tilde{\delta}\rangle_\mathrm{c}\ .\notag
\end{align}
Terms which depend only on the power spectrum  are referred to as Gaussian. If the underlying field was Gaussian then these would be the only non-zero parts of the covariance. The cross-covariance has no Gaussian terms.

The remaining, non-Gaussian, terms generate  non-zero off-diagonal elements. They arise from mode coupling either between small-scale modes within the survey window (in-survey covariance) or between in-survey modes and long-wavelength modes longer than the survey window dimension (supersample covariance).  Supersample covariance is generated by the \mbox{four-point} correlator in the power spectrum covariance and the \mbox{six-point} correlator in the bispectrum covariance.
\section{Matter power spectrum and bispectrum supersample covariance}\label{sec:SSC}
Background modes  which cause supersample covariance are essentially constant across the survey footprint, so their effect  can be equated to a change in the mean density within the survey region.   Supersample covariance can thus be  thought of as the response of the power spectrum and bispectrum to a long-wavelength mode $\delta_b$ \citep{hamilton2006measuring,rimes2006information,takada2013power,li2014super1,li2014super2,barreira2018complete,chan2018bispectrum,lacasa2018covariance}. 

In this view the power spectrum supersample covariance is \citep{takada2013power}
\begin{align}
\mathbfss{Cov}^{\mathrm{PP}}_\mathrm{SSC}&= \sigma_W^2\frac{\partial{P(k_1)}}{\partial{\delta_b}}\Bigg\vert_{\delta_b=0}\frac{\partial {P(k_2)}}{\partial{\delta_b}}\Bigg\vert_{\delta_b=0}\ ,\label{eq:PScov}
\end{align}
where $\sigma_W^2$ is the  variance of the long-wavelength background mode within the survey window, given by
\begin{align}
\sigma_W^2&= \frac{1}{V_W^2}\int\, \frac{\mathrm{d}q}{(2\mathrm{\pi})^3}\lvert \tilde{W}(\bm{q})\rvert^2P_\mathrm{L}(q) \ .\label{eq:sigmaw}
\end{align}
Here $V_W$ is the volume defined by the survey window,  $\tilde{W}$ is the Fourier transform of the survey window function and $P_\mathrm{L}$ is specifically the linear power spectrum because the long-wavelength mode is in the linear regime. The power spectrum $P$  may be in the linear or non-linear regime. 

Similarly the bispectrum supersample covariance is \citep{chan2018bispectrum}
\begin{align}
\mathbfss{Cov}^{\mathrm{BB}}_\mathrm{SSC}&=\sigma_W^2\frac{\partial {B(\bm{k}_1,\bm{k}_2,\bm{k}_3)}}{\partial {\delta_b}}\Bigg\vert_{\delta_b=0}\frac{\partial {B(\bm{k}_4,\bm{k}_5,\bm{k}_6)}}{\partial {\delta_b}}\Bigg\vert_{\delta_b=0}\ ,\label{eq:BScov}
\end{align} 
and the power spectrum--bispectrum supersample cross-covariance is
\begin{align}
\mathbfss{Cov}^{\mathrm{PB}}_\mathrm{SSC}&=\sigma_W^2\frac{\partial {P(k_1)}}{\partial {\delta_b}}\Bigg\vert_{\delta_b=0}\frac{\partial {B(\bm{k}_2,\bm{k}_3,\bm{k}_4)}}{\partial {\delta_b}}\Bigg\vert_{\delta_b=0}\ .\label{eq:PSBScov}
\end{align} 
Again, the bispectrum $B$ can be in the linear or non-linear regime.

The response functions in Eqs. (\ref{eq:PScov}), (\ref{eq:BScov}) and (\ref{eq:PSBScov}) can be derived using the halo model \citep{cooray2001power,cooray2002halo} together with perturbation theory \citep{bernardeau2002large}.  

The  halo model  is based round the integrals
\begin{align}
 I_\mu^\beta(k_1,k_2,\ldots,k_\mu)&\equiv
 \int \mathrm{d}Mf(M,z)\left(\frac{M}{\bar{\rho}}\right)^\mu b_\beta(M)\\
 &\hspace{2cm}\times\tilde u_M(k_1)\tilde u_M(k_2)\ldots\tilde u_M(k_\mu)\notag\ .
\end{align}
Here $M(z)$ is the halo mass, $n(M,z)$ is the number density of halos,  $f(M,z)\equiv\mathrm{d}n/\mathrm{d}M$ is the halo mass function, $\tilde u_M(k)$ is the Fourier transform of the halo density profile, $\mu$ is the number of points being correlated, and  $b_\beta(M)$ is the halo bias. We assume a Navarro-Frenk-White halo matter density profile \citep{navarro1997universal} and the mass-concentration relation given in \citet{duffy2008dark}. We use results from  \citet{tinker2008toward} to model the halo mass function and halo bias. 

The bias quantifies the $\beta$-th order response of the halo mass function to the long-wavelength mode $\delta_b$ \citep{mo1996analytic,schmidt2013peak}:
\begin{align}
b_\beta(M)&=\frac{1}{f(M,z)}\frac{\partial^{\;\beta} f(M,z)}{\partial\delta_b^{\;\beta}}\Bigg \vert_{\delta_b=0}\label{eq:bbeta}\ .
\end{align}
We assume linear bias so that $b_0=1$, $b_1=b(M)$ and $b_\beta=0$ for $\beta>1$.

 The halo model expression for the power spectrum is
\begin{align}
P_\mathrm{HM}(k)&=I_2^0(k,k)+[I_1^1(k)]^2P(k)\ ,
\end{align}
giving
\begin{align}
\frac{\partial{P_\mathrm{HM}(k)}}{\partial{\delta_b}}\Bigg \vert_{\delta_b=0}&=\frac{\partial{I_2^0(k,k)}}{\partial{\delta_b}}\Bigg \vert_{\delta_b=0} + \;[I_1^1(k)]^2\frac{\partial{P(k)}}{\partial{\delta_b}}\Bigg \vert_{\delta_b=0}\ ,\label{eq:dPddelta}
\end{align}
where we assume that the one-halo term $I_1^1(k)$ is not affected by the background  mode $\delta_b$ \citep{chiang2014position}.  We further assume that in the presence of $\delta_b$  the halo mass function changes from $f(M,z)$ to \mbox{$(1+\delta_b )f(M,z)$}  but the halo profile does not change \citep{schaan2014joint}, so that
 
 \begin{align}
\frac{\partial{I_2^0(k,k)}}{\partial{\delta_b}}\Bigg \vert_{\delta_b=0}&=  \int {\mathrm{d}M} \ \left(\frac{M}{\bar{\rho}}\right)^2 (1+\delta_b)
\frac{\partial{f}}{\partial{\delta_b}} \ [\tilde u_M(k)]^2\Bigg \vert_{\delta_b=0}\ .\label{eq:dI20}
\end{align}
From Eq. (\ref{eq:bbeta}) 
\begin{align}
\frac{\partial{f(M,z)}}{\partial{\delta_b}}\Bigg \vert_{\delta_b=0} &= f(M,z)\, b(M) \ .
\end{align}
 Substituting into Eq. (\ref{eq:dI20}) gives
\begin{align}
\frac{\partial{I_2^0(k,k)}}{\partial{\delta_b}}\Bigg \vert_{\delta_b=0}&=   \int {\mathrm{d}M}\frac{\mathrm{d}n}{\mathrm{d}M}\left(\frac{M}{\bar{\rho}}\right)^2 b(M)\ [\tilde u_M(k)]^2\\ 
&=I_2^1(k_1,k_2)\ ,
\end{align}
and Eq. (\ref{eq:dPddelta}) becomes
\begin{align}
\frac{\partial{P_\mathrm{HM}(k)}}{\partial{\delta_b}}\Bigg \vert_{\delta_b=0}&= I_2^1(k,k) + [I_1^1(k)]^2 \frac{\partial{P(k)}}{\partial{\delta_b}}\Bigg \vert_{\delta_b=0}\label{eq:dPddeltab}\ .
\end{align}

Thus we need the  response of the modulated power spectrum $P(k|\delta_b)$ to the background fluctuation $\delta_b$. This has been derived in several ways, including from perturbation theory and consistency relation arguments  \citep{takada2013power}; a separate universe approach \citep{li2014super1}; and from the position-dependent power spectrum and integrated bispectrum \citep{chiang2014position}.    The resulting expression is 
\begin{align}
\frac{\partial{P(k)}}{\partial{\delta_b}} \Bigg \vert_{\delta_b=0}&= \left(\frac{47}{21}-\frac{1}{3}\frac{\partial{\,   \ln}P(k)}{\partial{\, \ln k}}\right)P(k)\ .\label{eq:dPddeltab1}
\end{align}
Substituting into Eq. (\ref{eq:dPddeltab}) gives the halo model power spectrum response
\begin{align}
\frac{\partial{P_\mathrm{HM}(k)}}{\partial{\delta_b}}\Bigg \vert_{\delta_b=0}&= I_2^1(k,k) + [I_1^1(k)]^2\left(\frac{47}{21}-\frac{1}{3}\frac{\partial{\,   \ln}P(k)}{\partial{\, \ln k}}\right)P(k)\ .\label{eq:Presp}
\end{align}
 The bispectrum response can be derived in a similar way by expressing the bispectrum as the sum of \mbox{1-halo}, \mbox{2-halo} and \mbox{3-halo} terms. 
\begin{align}
&\frac{\partial {B_\mathrm{HM}(\bm{k}_1,\bm{k}_2,\bm{k}_3)}}{\partial{\delta_b}}\Bigg \vert_{\delta_b=0}=\frac{\partial {}}{\partial{\delta_b}}\left(B^{1h}+B^{2h}+B^{3h}\right)\Bigg \vert_{\delta_b=0}\\
&= \frac{\partial {}}{\partial{\delta_b}}
\bigg[ I_3^0(k_1,k_2,k_3)+ \left(I_1^1(k_1)I_2^1(k_2,k_3)P(k_1)+ \text{ 2 perms.}\right)\notag\\
&\hspace{1cm}+ I_1^1(k_1)I_1^1(k_2)I_1^1(k_3)B_{\mathrm{PT}}(\bm{k}_1,\bm{k}_2,\bm{k}_3)\bigg]\, \Bigg \vert_{\delta_b=0}\\
&=I_3^1(k_1,k_2,k_3)
+ \left(I_1^1(k_1)I_2^1(k_2,k_3) \frac{\partial {P(k_1)}}{\partial{\delta_b}}+ \text{ 2 perms.}\right)\Bigg \vert_{\delta_b=0}\\
&\hspace{1cm} +I_1^1(k_1)I_1^1(k_2)I_1^1(k_3)\frac{\partial {B_{\mathrm{PT}}(\bm{k}_{1},\bm{k}_{2},\bm{k}_{3}) }}{\partial{\delta_b}}\Bigg \vert_{\delta_b=0}\notag\\
&= I_3^1(k_1,k_2,k_3)\label{eq:dBddeltab}\\
&\hspace{1cm}+ I_1^1(k_1)I_2^1(k_2,k_3)P(k_1)\left(\frac{47}{21}-\frac{1}{3}\frac{\partial{\, \ln P(k_1)}}{\partial{\, \ln k_1}}\right)+ \text{ 2 perms.}\notag\\
&\hspace{1cm} +I_1^1(k_1)I_1^1(k_2)I_1^1(k_3)\frac{\partial {B_{\mathrm{PT}}(\bm{k}_{1},\bm{k}_{2},\bm{k}_{3}) }}{\partial{\delta_b}}\Bigg \vert_{\delta_b=0}\ .\notag
\end{align}
Here $B_{\mathrm{PT}}$ is the tree-level matter bispectrum given by
\begin{align}
B_{\mathrm{PT}}(\bm{k}_{1},\bm{k}_{2},\bm{k}_{3})&=2\big[F_2(\bm{k}_{1},\bm{k}_{2})P(k_1)P(k_2)\\
&\hspace{0.5cm}+F_2(\bm{k}_{2},\bm{k}_{3})P(k_2)P(k_3)\notag\\
&\hspace{0.5cm}+F_2(\bm{k}_{3},\bm{k}_{1})P(k_3)P(k_1)\big]\notag\ ,
\end{align}
where the symmetrised mode-coupling kernel $F_2$ is \citep{bernardeau2002large}
\begin{align}
F_2(\bm{k}_{1},\bm{k}_{2}) &= \frac{5}{7} 
+\frac{1}{2}\frac{\bm{k}_1\cdot\bm{k}_2}{k_1k_2}
\left(\frac{k_1}{k_2}+\frac{k_2}{k_1}\right)+\frac{2}{7}\frac{(\bm{k}_1\cdot\bm{k}_2)^2}{k_1^2k_2^2}\ .\label{eq:F2}
\end{align}
Thus we need the response of the tree-level bispectrum to the long mode.

\citet{chan2018bispectrum} used perturbation theory to obtain 
\begin{align}
\frac{\partial {B_\mathrm{PT}(\bm{k}_1,\bm{k}_2,\bm{k}_3 )}}{\partial{\delta_b}} \Bigg\vert_{\delta_b=0}&= \frac{433}{126} B_\mathrm{PT}(\bm{k}_1,\bm{k}_2,\bm{k}_3)+\frac{5}{126}B_{\mathrm{G}}(\bm{k}_1,\bm{k}_2,\bm{k}_3)\notag\\
&\hspace{1cm}-\frac{1}{3}\sum_{i=1}^3\frac{\partial {B_{\mathrm{PT}}(\bm{k}_1,\bm{k}_2,\bm{k}_3)}}{\partial{ \, \ln k_i}}\ ,\label{eq:Bresp}
 \end{align}
where  $B_{\mathrm{G}}$ is identical to $B_\mathrm{PT}$ but with the density coupling function  $F_2$ replaced by its velocity counterpart $G_2$:
\begin{align}
G_2(\bm{k}_{1},\bm{k}_{2}) = \frac{3}{7} 
+\frac{1}{2}\frac{\bm{k}_1\cdot\bm{k}_2}{k_1k_2}
\left(\frac{k_1}{k_2}+\frac{k_2}{k_1}\right)+\frac{4}{7}\frac{(\bm{k}_1\cdot\bm{k}_2)^2}{k_1^2k_2^2}\ .
\end{align}
For equilateral triangles we use the computer algebra package \textsc{Mathematica}\footnote{https://www.wolfram.com/mathematica} and Eq. (\ref{eq:Bresp}) to obtain
\begin{align}
\frac{\partial B_\mathrm{PT}^\mathrm{equi}}{\partial \delta_b}\Bigg \vert_{\delta_b=0}&= 
\left(\frac{2623}{98}-\frac{36}{7}\frac{\partial{\,   \ln}P(k)}{\partial{\, \ln k}}\right)P(k)^2\ .\label{eq:Brespequi}
 \end{align}
  
An alternative way to obtain the bispectrum response is to extend the concept of the position-dependent power spectrum  developed by \citet{chiang2014position} to three-point statistics \citep{adhikari2016constraining, munshi2017integrated}. We  use this method, for equilateral triangles only, as a check on the validity of Eq. (\ref{eq:Bresp}).
 
We  define the position dependent bispectrum  as $ \langle B(\bm{k}_1,\bm{k}_2,\bm{k}_3,\bm{r})\,\bar{\delta}(\bm{r})\rangle$, the correlation between the bispectrum measured in a sub-volume of the survey $V_L$, with length scale $L$ and centred at position $\bm{r}$,  and the mean density contrast at  $\bm{r}$.  It is equivalent to an integrated trispectrum and can be expressed as 
\begin{align}
\langle B(\bm{k}_1,&\bm{k}_2,\bm{k}_3,\bm{r})\,\bar{\delta}(\bm{r})\rangle\equiv iT(\bm{k}_1, \bm{k}_2,\bm{k}_3)\\
&= \frac{1}{V_L^2} \int \frac{\mathrm{d}^3\bm{q}_1}{(2\pi)^3}\int \frac{\mathrm{d}^3\bm{q}_2}{(2\pi)^3}\int \frac{\mathrm{d}^3\bm{q}_3}{(2\pi)^3}\label{eq:iT2}\\
&\hspace{0.8cm} \times T(\bm{k}_1-\bm{q}_1,\bm{k}_2-\bm{q}_2,\bm{k}_3 + \bm{q}_1+\bm{q}_2+\bm{q}_3,-\bm{q}_3)\notag\\
&\hspace{0.8cm} \times \tilde{W}_L(\bm{q}_2) \tilde{W}_L(-\bm{q}_1-\bm{q}_2-\bm{q}_3) \tilde{W}_L(\bm{q}_3)\notag\ ,\end{align}
where  $\tilde{W}_L(\bm{q})$ is the Fourier transform of the sub-volume window function which we take to be 1 within the sub-volume and 0 otherwise.

Following \citet{adhikari2016constraining} we  make the assumption that the trispectrum is dominated by the squeezed limit in which one wavevector is much smaller than the other three.  Eq. (\ref{eq:iT2}) can then be simplified through the bispectrum triangle condition  \mbox{$\bm{k}_1+\bm{k}_2+\bm{k}_3=0$}  and a change of variables to get
\begin{align}
iT(\bm{k}_1, \bm{k}_2)\approx 
\frac{1}{V_L^2} 
\int \frac{\mathrm{d}^3\bm{q}}{(2\pi)^3}|\tilde{W}_L(\bm{q})|^2 T(\bm{k}_1,\bm{k}_2,-\bm{k}_1+\bm{k}_2+\bm{q},-\bm{q})\ .
\end{align} 
Averaging  over the solid angles $\Omega_{12}$ and $\Omega_{13}$ between two pairs of wavevectors and fixing the direction of one $\bm{k}$ vector results in  
\begin{align}
	iT(k_1,k_2)&\equiv \int \frac{\mathrm{d}^2\Omega_{12}}{4\pi}\int \frac{\mathrm{d}^2\Omega_{13}}{4\pi}iT(\bm{k}_1, \bm{k}_2)\\
	&\approx\frac{1}{V_L^2}\int \frac{\mathrm{d}^3q}{(2\pi)^3}|\tilde{W}_L(\bm{q})|^2
 \int \frac{\mathrm{d}^2\Omega_{12}}{4\pi}T(\bm{k}_1,\bm{k}_2,-\bm{k}_{2}+\bm{q},-\bm{q})\ .\label{eq:iTav}
\end{align}
 This derivation depends critically on the assumption that the squeezed-limit configuration dominates the trispectrum. Strictly the result is only valid for trispectra which depend only on four items, say four sides of a quadrilateral or three sides and a diagonal  (see the discussion in \citealt{adhikari2016constraining}).   Nevertheless we take the squeezed-limit assumption it to be a reasonable approximation.
 
We proceed to  show that in the equilateral case Eq. (\ref{eq:iTav}) has the form 
 \begin{align}
iT(k,k)&\approx \sigma^2_L\, f(k)P(k)^2,\label{eq:iTkf}
\end{align}
for some function $f(k)$, where $ \sigma^2_L$ is the variance of the density field on the sub-volume scale, given by
\begin{align}
\sigma_L^2&=\frac{1}{V_L^2}\int \frac{\mathrm{d}^3 q}{(2\pi)^3}|\tilde{W}_L(\bm{q)}|^2 P(q)\ .
\end{align}
It then follows that  $iT(k,k)$ measures how the equilateral bispectrum responds to a large-scale density fluctuation with variance $ \sigma^2_L$.

 To evaluate the trispectrum on the right-hand side of Eq. (\ref{eq:iTav}) we use perturbation theory. The general trispectrum $T_{\mathrm{PT}}(\bm{k}_1,\bm{k}_2,\bm{k}_3,\bm{k}_4)$ can be expressed as \citep{bernardeau2002large,pielorz2010fitting}
\begin{align}
T_{\mathrm{PT}}&=4T_\mathrm{a}+6T_\mathrm{b}\ , \label{eq:TPT}
\end{align}
 where $T_\mathrm{a}$ is the sum of 12 terms like
\begin{align}
F_2(\bm{k}_{1}, -\bm{k}_1-\bm{k}_3)F_2(\bm{k}_{2},\bm{k}_1+\bm{k}_3)P(k_{1})P(k_2)P(|\bm{k}_1+\bm{k}_3|) \ , \notag
\end{align}
and $T_\mathrm{b}$ is the sum of four terms of the form  $F_3(\bm{k}_1,\bm{k}_2,\bm{k}_3)P(k_1)P(k_2)P(k_3)$. Here $F_3$ is the symmetrised third-order coupling kernel: 
 \begin{align}
 F_3(\bm{k}_1, \bm{k}_2, \bm{k}_3 )&=
\frac{7}{54}\left [\alpha(\bm{k}_1,\bm{k}_{23})F_2(\bm{k}_2, \bm{k}_3) + \text{2 perms}\right]\\
&\hspace{0.5cm}+\frac{4}{54}\left[\beta(\bm{k}_1,\bm{k}_{23})G_2(\bm{k}_2, \bm{k}_3)+ \text{2 perms}\right]\notag\\
&\hspace{0.5cm}+\frac{7}{54}\left[\alpha(\bm{k}_{12},\bm{k}_3)G_2(\bm{k}_1, \bm{k}_2)+ \text{2 perms}\right]\ ,\notag\\
\alpha(\bm{k}_1,\bm{k}_2) &= \frac{\bm{k}_{12}\cdot\bm{k}_1}{k_1^2}\ ,\hspace{0.5cm}
\beta(\bm{k}_1,\bm{k}_2) = \frac{k_{12}^2(\bm{k}_1\cdot\bm{k}_2)}{2k_1^2k_2^2}\ ,
\end{align}
with $\bm{k}_{ij} = |\bm{k}_i+\bm{k}_j|$.

We use this formulation and the computer algebra package \textsc{Mathematica} to derive an approximation for the squeezed trispectrum with three equal short modes and one long mode $\bm{q}$.
  In this configuration three terms of $T_\mathrm{a}$  are zero in the limit $\bm{q} \to 0$ because  \mbox{$F_2(\bm{k},-\bm{k})=0$} and one term of $T_\mathrm{b}$ is zero because \mbox{$F_3(\bm{k}_1,\bm{k}_2, \bm{k}_3) = 0$}  if $\bm{k}_1+\bm{k}_2+ \bm{k}_3 = 0$.
  
We set $|\bm{k_i}|=k$ for $i=1,2,3$ and $\bm{k}_i\cdot\bm{k}_j=k^2/2$ for $i\ne j$, write $P^\prime(k)=\partial{P(k)}/\partial{k}$, 
 and Taylor-expand all terms in $T_\mathrm{a}$ and $T_\mathrm{b}$ to first order in $q/k$.
This leads to
\begin{align}
T_\mathrm{a}^{\mathrm{equi}}&= \frac{9P(k)^2P(q)}{98k^2q^2}\left[17(\bm{k}\cdot\bm{q})^2+42 k^2\bm{k}\cdot\bm{q} + 32 k^2q^2\right]\\\notag
&\quad- \frac{9P(k)P^\prime(k) P(q) }{98k^3q^2}\left[(\bm{k}\cdot\bm{q})^3+ 14k^2(\bm{k}\cdot\bm{q})^2+6k^2q^2\right]\ , \\
T_\mathrm{b}^{\mathrm{equi}}&= \frac{P(k)^2P(q)}{84k^2q^2}\left[106(\bm{k}\cdot\bm{q})^2-216 k^2\bm{k}\cdot\bm{q} + 53 k^2q^2\right]\\\notag
&\quad+ \frac{P(k)P^\prime(k) P(q) }{126k^3q^2}\left[70(\bm{k}\cdot\bm{q})^3-216 k^2(\bm{k}\cdot\bm{q})^2+93k^2q^2\right]\ ,\\
T_{\mathrm{PT}}^{\mathrm{equi}}&= \frac{P(k)^2P(q)}{98k^2q^2}\left[1354(\bm{k}\cdot\bm{q})^2+ 2055k^2q^2\right]\\\notag
&- \frac{P(k)P^\prime(k) P(q) }{147k^3q^2}\left[436(\bm{k}\cdot\bm{q})^3-2268k^2(\bm{k}\cdot\bm{q})^2+327k^2q^2\right]\ .
\end{align}

 \begin{figure}
\captionsetup[subfigure]{font={large},justification=centering}
\begin{subfigure}[b]{1\linewidth}
\includegraphics[width=9cm]{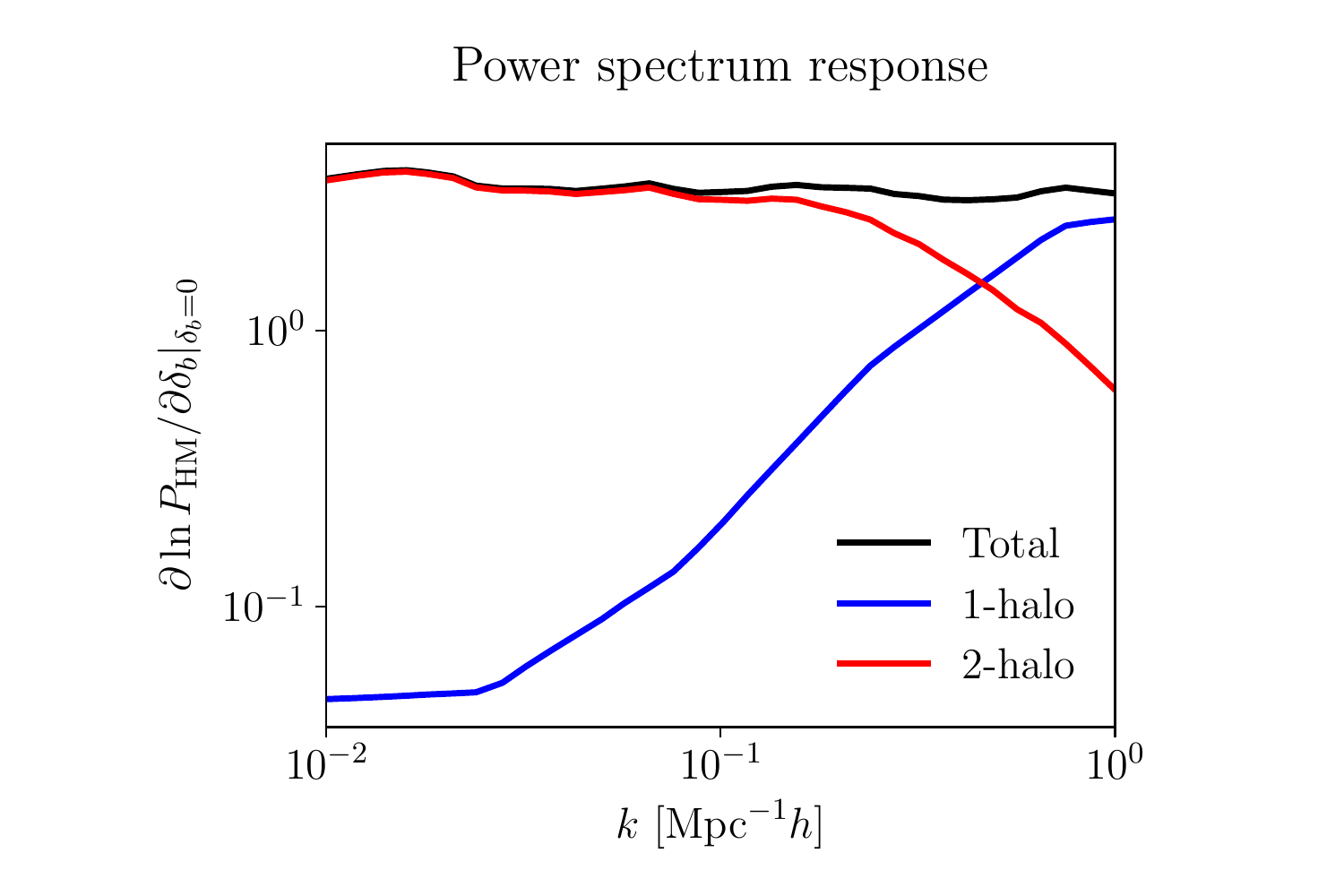}
\vspace{0.5cm}
\end{subfigure}

\begin{subfigure}[b]{1\linewidth}
\includegraphics[width=9cm]{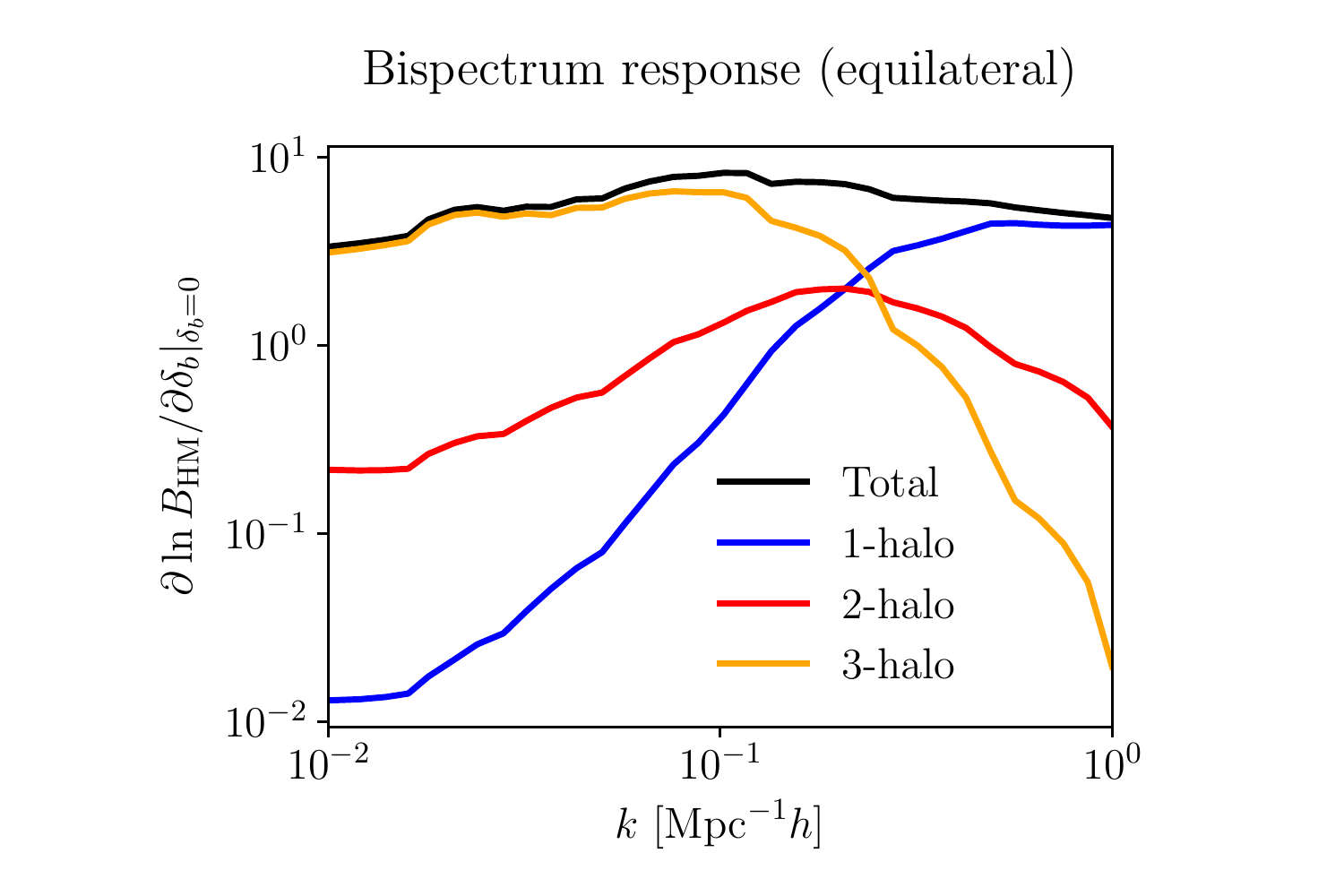}
\end{subfigure}
\caption{Responses of the halo model matter power spectrum (top) and equilateral bispectrum (bottom) to a long-wavelength super-survey mode $\delta_b$.  }\label{fig:respfn} 
\end{figure}

Although $T_\mathrm{a}^{\mathrm{equi}}$ and $T_\mathrm{b}^{\mathrm{equi}}$ include terms in $\bm{k}\cdot\bm{q}/q^2$ which are divergent as $q\to0$, these cancel out in the final expression for $T_{\mathrm{PT}}^{\mathrm{equi}}$.

Finally, ignoring terms which vanish as $q\to0$, we get
\begin{align}
T_{\mathrm{PT}}^{\mathrm{equi}}&= P(q)P(k)\Bigg[\left(\frac{2055}{98}+\frac{1354(\bm{k}\cdot\bm{q})^2}{98k^2q^2}\right)P(k)\\\notag
&\hspace{1.5cm}+\left(\frac{327}{147k} -\frac{2268(\bm{k}\cdot\bm{q})^2}{147k^3q^2}\right)\frac{ \partial P(k)}{ \partial k}\Bigg]\ .
\end{align}
We substitute this into Eq.  (\ref{eq:iTav})
and take the average over the solid angle $\Omega$  between $\bm{k}$ and $\bm{q}$. The angular average of \mbox{$(\bm{k}\cdot\bm{q})^2/k^2q^2$} is 1/3 and so we get
\begin{align}
iT^{\mathrm{equi}}
&= \sigma_L^2P(k)^2\left(\frac{7519}{294} -\frac{143}{49}\frac{\partial \, \ln P(k)}{\partial \,\ln k}\right)\ ,\label{eq:angavT}
\end{align}
and 
\begin{align}
\frac{\partial {B^\mathrm{equi}_\mathrm{PT}} }{\partial{\delta_b}}\Bigg \vert_{\delta_b=0}&=
\left(\frac{7519}{294} -\frac{143}{49}\frac{\partial \, \ln P(k)}{\partial \, \ln k}\right)P(k)^2\ .
\end{align}

This is close to but not identical with the result for equilateral triangles obtained by  \citet{chan2018bispectrum}, Eq. (\ref{eq:Brespequi}). We would not  expect exact agreement given the approximations we have made and the fact that our results are only correct to first order.  Nevertheless our final expression broadly confirms Eq. (\ref{eq:Bresp}) for equilateral bispectra.  We therefore use Eq. (\ref{eq:Bresp}) in our work since it is more complete and general, substituting into the halo model expression  Eq. (\ref{eq:dBddeltab}).

We note however that both results are quite different from the expression obtained by  \citet{adhikari2016constraining}  who  derived 
\begin{align}
  \left\langle T^{\mathrm{equi}}(k)\right\rangle_\mathrm{ang-av} &=P(q) P(k)^2 \left(\frac{579}{98}-\frac{8}{7}\frac{\partial \, \ln P(k)}{\partial \, \ln k}\right)\ .\label{eq:Adhikari}
\end{align}
It is difficult to determine the source of the discrepancies since all three derivations rely on \textsc{Mathematica} (private communications) and intermediate steps are not transparent.

  Figure \ref{fig:respfn} illustrates our calculated matter response functions evaluated
  using  modelling assumptions from Sect. \ref{sec:cosmoparams}.  This figure shows the individual halo model terms and the total response but excludes the dilation terms  -- the $- \! 1/3$ terms in Eqs. (\ref{eq:dPddeltab1}) and (\ref{eq:Bresp}) -- which are consistently small as also found by \citet{li2014super1} and \citet{chan2018bispectrum}.
 
To obtain the full halo model expressions for the matter supersample covariance  we substitute  Eqs. (\ref{eq:Presp}) and (\ref{eq:dBddeltab}) into Eqs. (\ref{eq:PScov}) and (\ref{eq:BScov}).
\section{Weak lensing covariance}\label{sec:WLcov}
In this appendix we give expressions for the components of the convergence power spectrum and bispectrum covariance for a single redshift bin. The power spectrum--bispectrum cross-covariance can easily be derived in a similar way \citep{kayo2013cosmological,rizzato2019tomographic}.  Further details and derivations can be found in \citet{takada2009impact}, \citet{kayo2013cosmological} and \citet{sato2013impact}. \citet{rizzato2019tomographic} give all permutations of terms in these covariances for a tomographic survey. 

We assume a survey with area $\Omega_\mathrm{s}$ in steradians and consider angular bins of width $\Delta\ell_i$ centred on the values $\ell_i$. Thus \mbox{$l_i -\Delta\ell/2 \le |\boldsymbol{\ell}_i| \le l_i +\Delta\ell/2$}. 
For simplicity  we omit noise terms in this appendix but in a real survey the intrinsic ellipticity of galaxies will always induce shape noise.   In the main part of this paper we always implicitly include shape noise in the Gaussian terms of the covariances (both the power spectrum and the bispectrum). We assume this noise is Gaussian, so the observed power spectrum between redshift bins \mbox{$i$ and $j$} has the form
\begin{align}
C^{(ij)}_\mathrm{obs}(\ell)&= C^{(ij)}(\ell) + \frac{\sigma_\epsilon^2}{2\bar{n}_i}\delta_{ij}^\mathrm{K}\ ,
\end{align}
where $\delta_{ij}^\mathrm{K}$ is the Kronecker delta,  $\sigma_\epsilon$ is the total intrinsic ellipticity dispersion, which we take to be 0.35,  and $\bar{n}_i$ is the galaxy number density in \mbox{redshift bin $i$}. 
\subsection*{Gaussian covariance}
The Gaussian part of the convergence power spectrum covariance is
\begin{align}
	\mathrm{Cov}[C(\ell_1),C(\ell_2)]_\mathrm{G}
	&=\frac{2\delta^\mathrm{K}_{\ell_1\ell_2}}{N_\mathrm{pairs}(\ell_1)}
	C(\ell_1)C(\ell_2)\ ,
\end{align}
where  $\delta^\mathrm{K}_{\ell_1\ell_2}$ is the Kronecker delta  which is 1 if  $\ell_1=\ell_2$ and $\ell_1$ is within the bin width $\Delta \ell_1$, and zero otherwise.  $N_{\mathrm{pairs}}(\ell_1)$ is the number of independent pairs of modes within the bin width.

The Gaussian part of the convergence bispectrum covariance is
\begin{align}
	\mathrm{Cov}[B(\boldsymbol{\ell}_1,\boldsymbol{\ell}_2, \boldsymbol{\ell}_1)&,B(\boldsymbol{\ell}_4,\boldsymbol{\ell}_5, \boldsymbol{\ell}_6)]_\mathrm{G}=\\ \notag
	\frac{\Omega_\mathrm{s}}{N_\mathrm{trip}(\ell_1,\ell_2,\ell_3)}&C(\ell_1)C(\ell_2)C(\ell_3)\\\notag
\times \ 	[\delta^\mathrm{K}_{\ell_1\ell_4}\delta^\mathrm{K}_{\ell_2\ell_5}\delta^\mathrm{K}&_{\ell_3\ell_6} + \delta^\mathrm{K}_{\ell_1\ell_4}
	\delta^\mathrm{K}_{\ell_2\ell_6}\delta^\mathrm{K}_{\ell_3\ell_5}+\delta^\mathrm{K}_{\ell_1\ell_5}
	\delta^\mathrm{K}_{\ell_2\ell_4}\delta^\mathrm{K}_{\ell_3\ell_6} 	
+ \text{3 perms}]\ ,
\end{align}
where $N_\mathrm{trip}(\ell_1,\ell_2,\ell_3)$ is the number of triplets of modes which form triangles of side lengths $\ell_1$, $\ell_2$ and $\ell_3$  within the specified bin widths $\Delta\ell_i$.

 $N_{\mathrm{pairs}}$ and $N_{\mathrm{trip}}$ can be approximated by  \citep{takada2007probing,joachimi2009bispectrum}
\begin{align}
 	N_{\mathrm{pairs}}(\ell) &= \frac{\Omega_\mathrm{s}\ell\Delta \ell}{2\mathrm{\pi}}\ ,\\
	N_{\mathrm{trip}}(\ell_1,\ell_2,\ell_3) &= \frac{\Omega_\mathrm{s}^2\ell_1\ell_2\ell_3\Delta \ell_1\Delta \ell_2\Delta \ell_3}{2\mathrm{\pi}^2\sqrt{2\ell_1^2\ell_2^2 +2\ell_2^2\ell_3^2+2\ell_3^2\ell_1^2 -\ell_1^4-\ell_2^4-\ell_3^4}}\ .
\end{align}
\subsection*{In-survey non-Gaussian covariance}
 The in-survey non-Gaussian part of the convergence power spectrum covariance is
\begin{align}
	\mathrm{Cov}[C(\ell_1),C(\ell_2)]_\mathrm{NG}
	&=\frac{2\mathrm{\pi}}{\Omega_\mathrm{s}}\int _{|\boldsymbol{\ell}|\in \ell_1}\frac{\mathrm{d}^2 \boldsymbol{\ell}}{\ell_1\Delta\ell_1}\int _{|\boldsymbol{\ell}^\prime|\in \ell_2}\frac{\mathrm{d}^2 \boldsymbol{\ell}^\prime}{\ell_2\Delta\ell_2}\notag\\
	&\hspace{3cm}\times T(\boldsymbol{\ell},-\boldsymbol{\ell},\boldsymbol{\ell}^\prime,-\boldsymbol{\ell}^\prime)\ ,
\end{align}
 where $T$ is the convergence trispectrum. The integrals are over all wavevectors which are within the bin width $\Delta\ell$ around \mbox{$\boldsymbol{\ell}$ or $\boldsymbol{\ell}^\prime$}.
 
The in-survey non-Gaussian part of the convergence bispectrum covariance involves similar integrals over the angular bins. However it can be considerably simplified by making use of triangle conditions and other reasonable assumptions such as that the trispectrum does not vary much within the bin width. Full details can be found in Appendix A of \citet{kayo2013cosmological}.  The resulting simplified expression is
\begin{align}
	\mathrm{Cov}&[B(\boldsymbol{\ell}_1,\boldsymbol{\ell}_2, \boldsymbol{\ell}_3),
	B(\boldsymbol{\ell}_4,\boldsymbol{\ell}_5, \boldsymbol{\ell}_6)]_\mathrm{NG}=	\\\notag
	&\frac{2\mathrm{\pi}}{\Omega_\mathrm{s}}B(\ell_1,\ell_2, \ell_3)B(\ell_4,\ell_5, \ell_6)\left[\frac{\delta^\mathrm{K}_{\ell_1\ell_4}}{\ell_1\Delta\ell_1} + \frac{\delta^\mathrm{K}_{\ell_1\ell_5}}{\ell_1\Delta\ell_1}+\text{7 perms}\right]\\\notag
	&+\frac{2\mathrm{\pi}}{\Omega_\mathrm{s}}\Bigg[ \frac{\delta^\mathrm{K}_{\ell_1\ell_4}}{\ell_1\Delta\ell_1}C(\ell_1)T(\boldsymbol{\ell}_2,\boldsymbol{\ell}_3,\boldsymbol{\ell}_5,\boldsymbol{\ell}_6)	\notag\\
	&\hspace{2.cm}+ \frac{\delta^\mathrm{K}_{\ell_1\ell_5}}{\ell_1\Delta\ell_1}
	C(\ell_1)T(\boldsymbol{\ell}_2,\boldsymbol{\ell}_3,\boldsymbol{\ell}_4,\boldsymbol{\ell}_6)+\text{7 perms}\Bigg] \notag\\
	&+ \frac{1}{\Omega_\mathrm{s}}\int_0^{2\mathrm{\pi}}\frac{\mathrm{d}\psi}{2\mathrm{\pi}}P_6(\boldsymbol{\ell}_1,\boldsymbol{\ell}_2,\boldsymbol{\ell}_3,\boldsymbol{\ell}_4,\boldsymbol{\ell}_5,\boldsymbol{\ell}_6;\psi)\ ,
\end{align}
where $P_6$ is the six-point function or pentaspectrum.   The triangle conditions \mbox{$\boldsymbol{\ell}_1+\boldsymbol{\ell}_2+ \boldsymbol{\ell}_3=0$} and \mbox{$\boldsymbol{\ell}_4+\boldsymbol{\ell}_5+ \boldsymbol{\ell}_6=0$} mean that  the $P_6$ term depends automatically on two triangles. The only remaining freedom is the angle $\psi$ between any two wavevectors, one in each of these triangles, so we can replace the integrals over $\boldsymbol{\ell}_i$ with integrals over $\psi$ \citep{kayo2013cosmological}.
\subsection*{Supersample covariance}
Using the standard Limber approximation and assuming a spatially flat universe, the weak lensing power spectrum supersample covariance for a single redshift bin is \citep{takada2013power} 
\begin{align}
&\mathrm{Cov}[C(\ell_1),C(\ell_2)]_\mathrm{SSC}\notag\\
&\qquad=\frac{1}{\Omega_\mathrm{s}}\int_0^{\chi_\mathrm{lim}} \mathrm{d}\chi \ q^4(\chi)\,\chi^{-6}\sigma_W^2(\chi)\frac{ \partial{P_\delta(\ell_1/\chi;\chi)}}{\partial{\delta_b}}\frac{\partial{P_\delta(\ell_2/\chi;\chi)}}{\partial{\delta_b}}\notag\\
\end{align}
where $\chi_\mathrm{lim}$ is the maximum comoving distance  of the survey, $q(\chi)$ is the lensing weight function  given by Eq. (\ref{eq:q}),  and $\sigma_W^2(\chi)$ is defined in Eq. (\ref{eq:sigmaw}).

Similarly the weak lensing bispectrum supersample covariance is \citep{kayo2013cosmological}
\begin{align}
&\mathrm{Cov}[B(\boldsymbol{\ell}_1,\boldsymbol{\ell}_2, \boldsymbol{\ell}_3),
	B(\boldsymbol{\ell}_4,\boldsymbol{\ell}_5, \boldsymbol{\ell}_6)]_\mathrm{SSC}=	\\\notag
	&\quad\frac{1}{\Omega_\mathrm{s}}\int_0^{\chi_\mathrm{lim}} \mathrm{d}\chi \ q^6(\chi)\,\chi^{-10} \sigma_W^2(\chi)\frac{\partial{B_\delta( \boldsymbol{\ell}_1/\chi, \boldsymbol{\ell}_2/\chi, \boldsymbol{\ell}_3/\chi;\chi)}}{\partial{\delta_b}}\\\notag
	&\hspace{4.15cm}\times\frac{\partial{B_\delta( \boldsymbol{\ell}_4/\chi, \boldsymbol{\ell}_5/\chi, \boldsymbol{\ell}_6/\chi;\chi)}}{\partial{\delta_b}}\ .	
\end{align}
\onecolumn
\subsection*{Magnitudes of covariance terms}
\begin{figure*}
\captionsetup[subfigure]{justification=centering}
\hspace{-1.1cm}
\begin{subfigure}[b]{0.3\linewidth}
 \includegraphics[width=7.5cm]{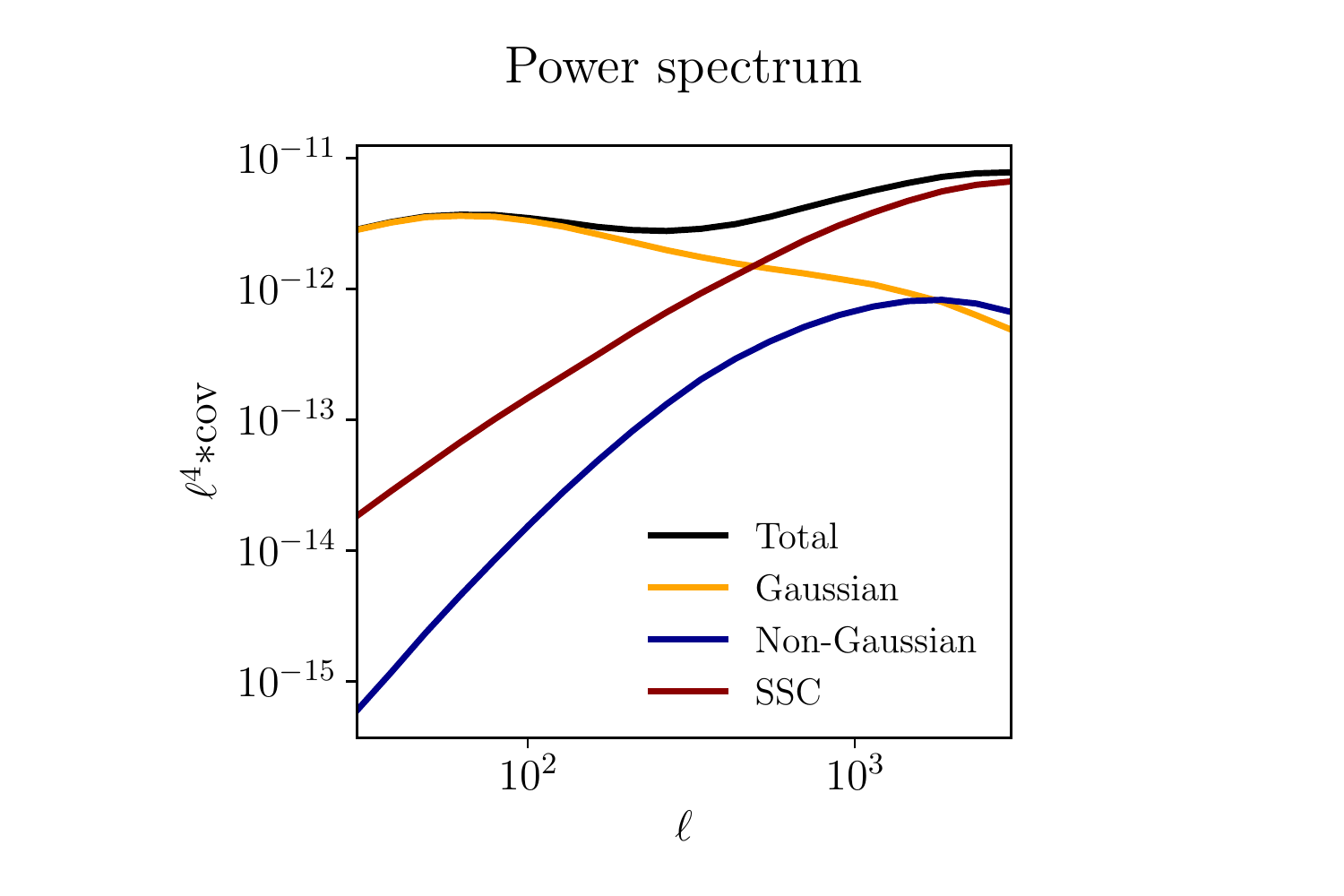}
\end{subfigure}
\hspace{0.4cm}
\begin{subfigure}[b]{0.3\linewidth}
 \includegraphics[width=7.5cm]{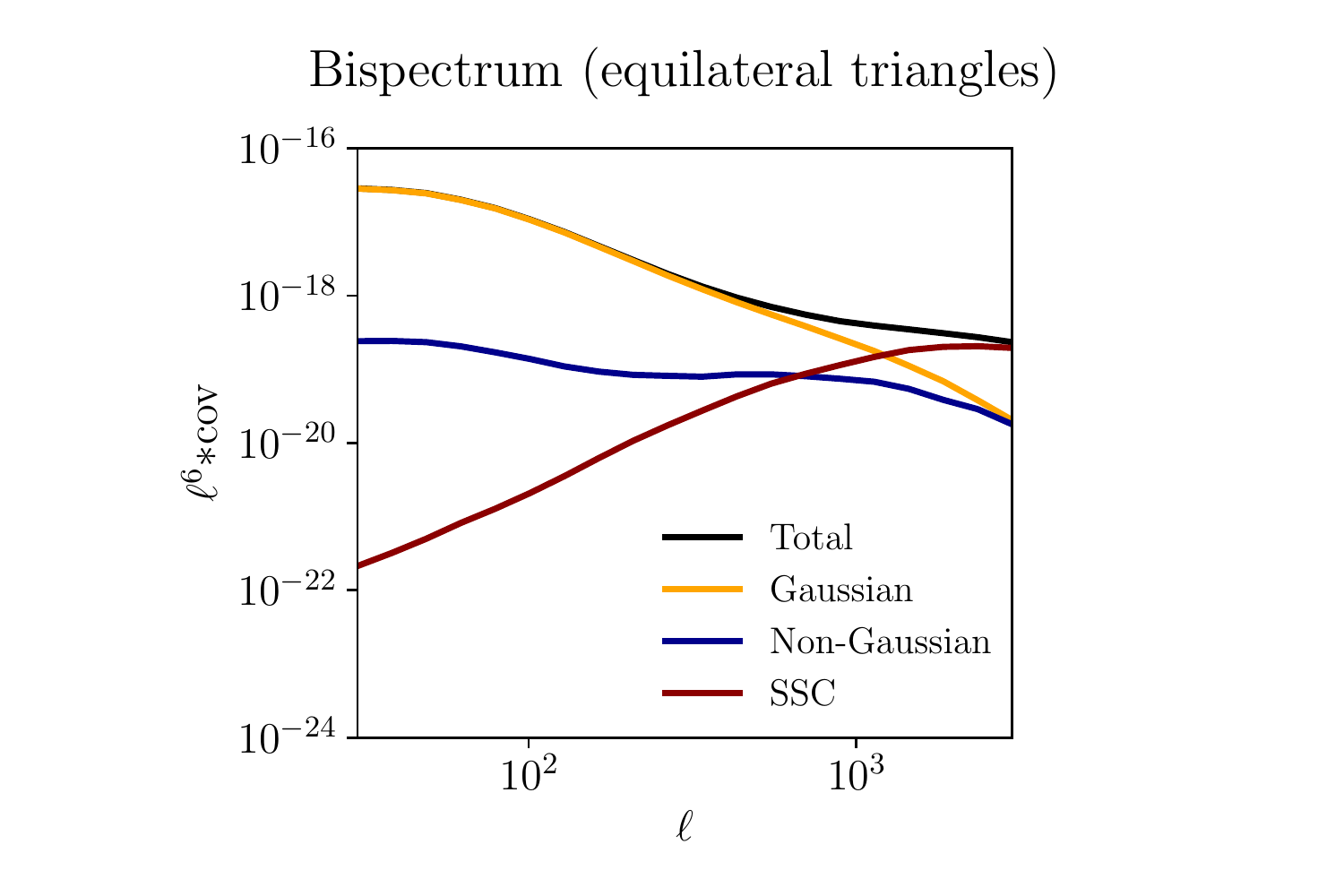}
\end{subfigure}
\hspace{0.4cm}
\begin{subfigure}[b]{0.3\linewidth}
 \includegraphics[width=7.5cm]{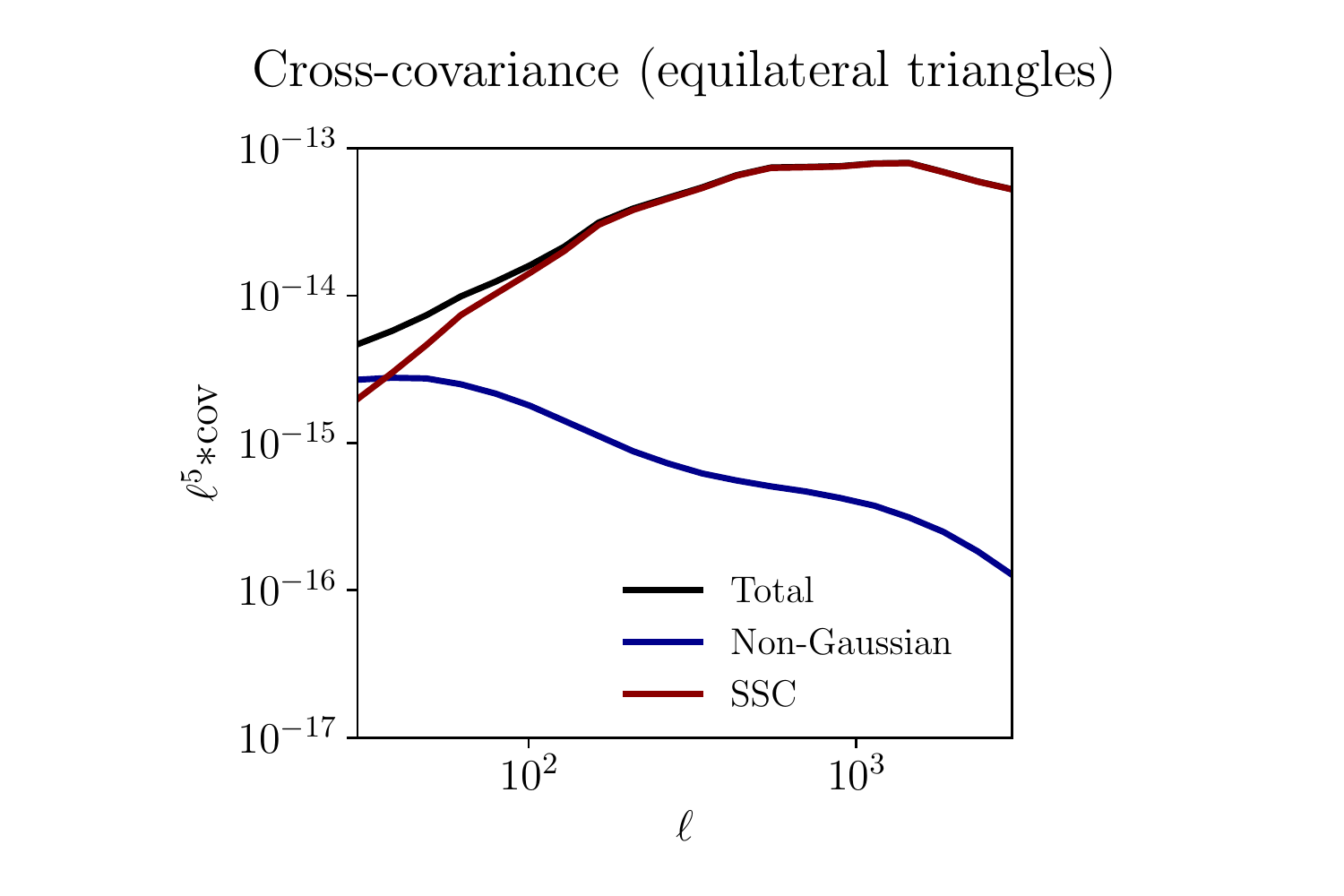}
\end{subfigure}
\vspace{-10pt}
\caption{Diagonal terms of the weak lensing power spectrum and bispectrum covariance matrices and their cross-covariance, calculated for a single redshift bin at $z=0.2$. The cross-covariance does not have a Gaussian term. }\label{fig:covel} 
\end{figure*}

Figure \ref{fig:covel} shows the diagonal Gaussian, in-survey non-Gaussian and supersample terms of the weak lensing power spectrum and bispectrum covariance and their cross-covariance across a range of angular scales, calculated for a single redshift bin for a 15\,000 deg$^2$ survey.  For the bispectrum and cross-covariance we show results for equilateral configurations only. For the power spectrum and bispectrum Gaussian terms dominate except at small scales where the supersample terms are large. The supersample term dominates the cross-covariance, which has no Gaussian term.  The results are consistent with other similar numerical calculations for \textit{Euclid}-like surveys \citep{barreira2018complete, rizzato2019tomographic}.

\begin{figure*}
\hspace{-0.1cm}
\begin{subfigure}[b]{0.5\linewidth}
    \centering
     \includegraphics[width=9cm]{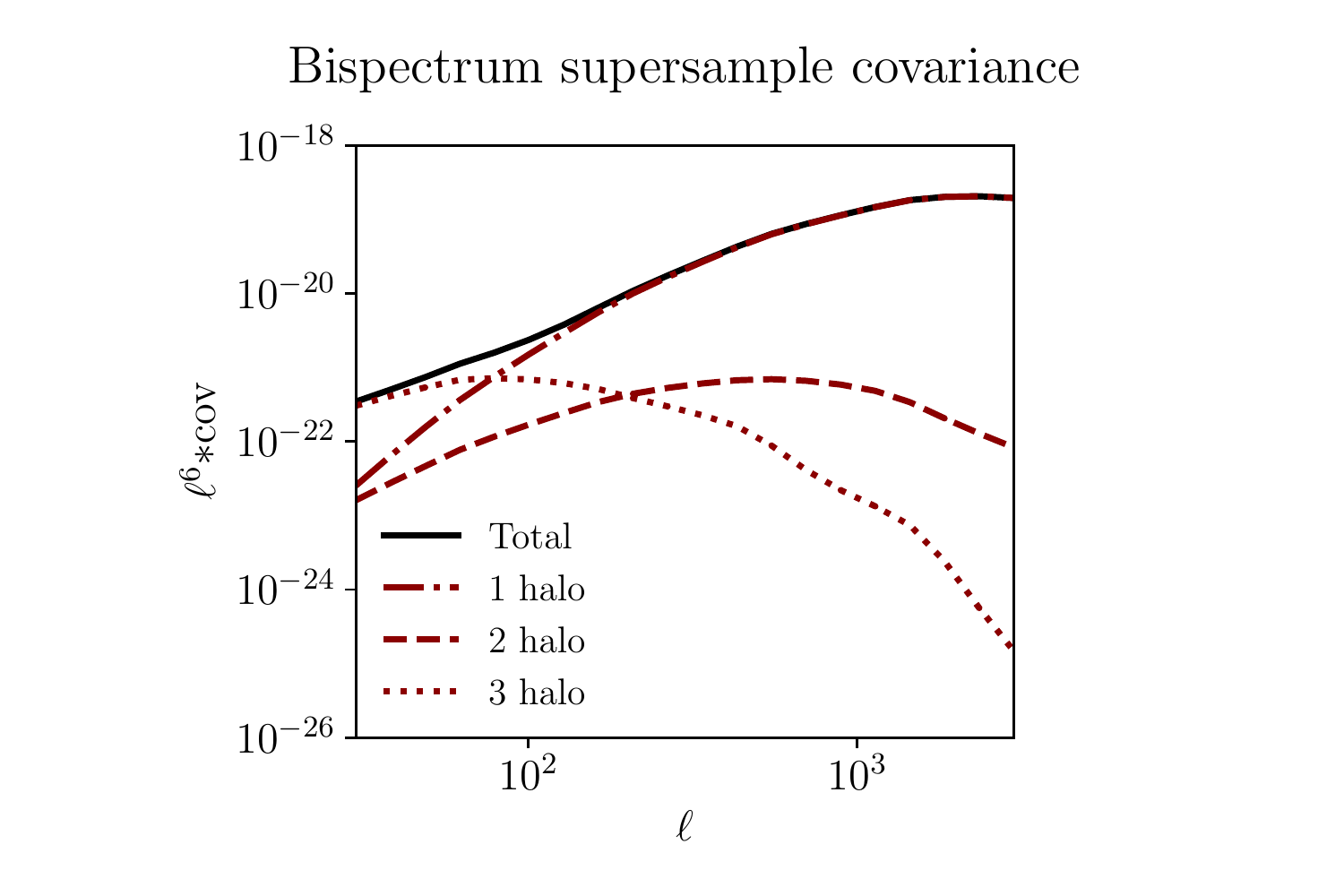}
\end{subfigure}
\hspace{-0.3cm}
\begin{subfigure}[b]{0.5\linewidth}
     \includegraphics[width=9cm]{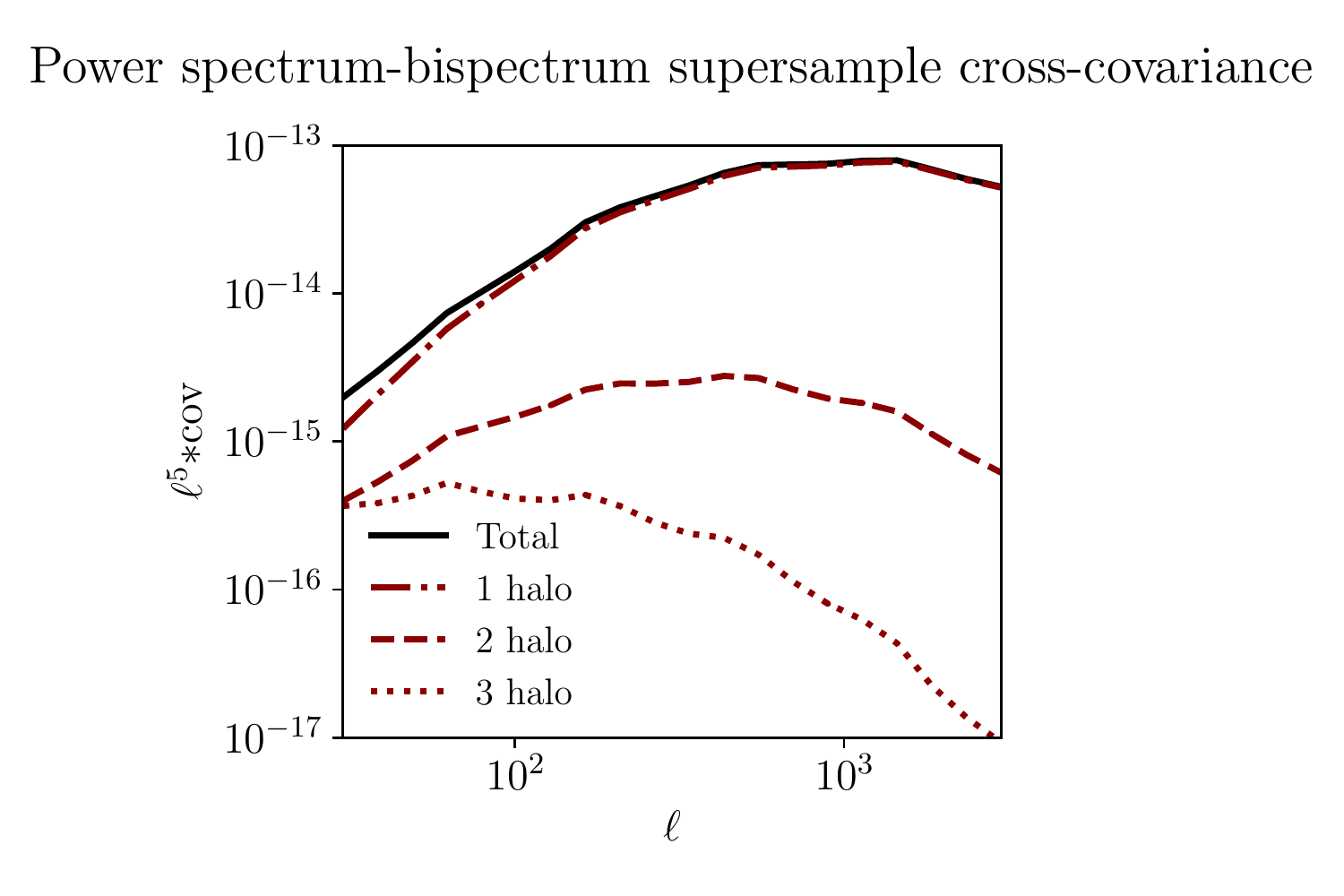}
\end{subfigure}
\vspace{-0.5cm}
\caption{Individual halo model terms of the weak lensing bispectrum supersample covariance and power spectrum-bispectrum supersample cross-covariance, calculated for a single redshift bin at $z=0.2$. 
 The power spectrum includes both the one-halo and two-halo terms.}\label{fig:SSCcovel}
\end{figure*}

Figure \ref{fig:SSCcovel} splits the supersample terms of the equilateral-triangle bispectrum covariance and the power spectrum-bispectrum cross-covariance  into their one-halo, two-halo and three-halo components. 
The one-halo term is dominant for  $\ell>50$ in the bispectrum covariance and at all scales in the cross-covariance. At larger scales the three-halo 
term dominates the bispectrum covariance. However this is inconsequential since the signal-to-noise ratio in this regime is very small.

\clearpage
\section{Self-calibration -- supplementary plots}\label{sec:additional_plots}

\begin{figure}
\hspace{0.5cm}
     \begin{subfigure}[b]{0.5\linewidth}     
      \caption*{Power spectrum  derivatives, first redshift bin}     
                 \includegraphics[width=8.cm]{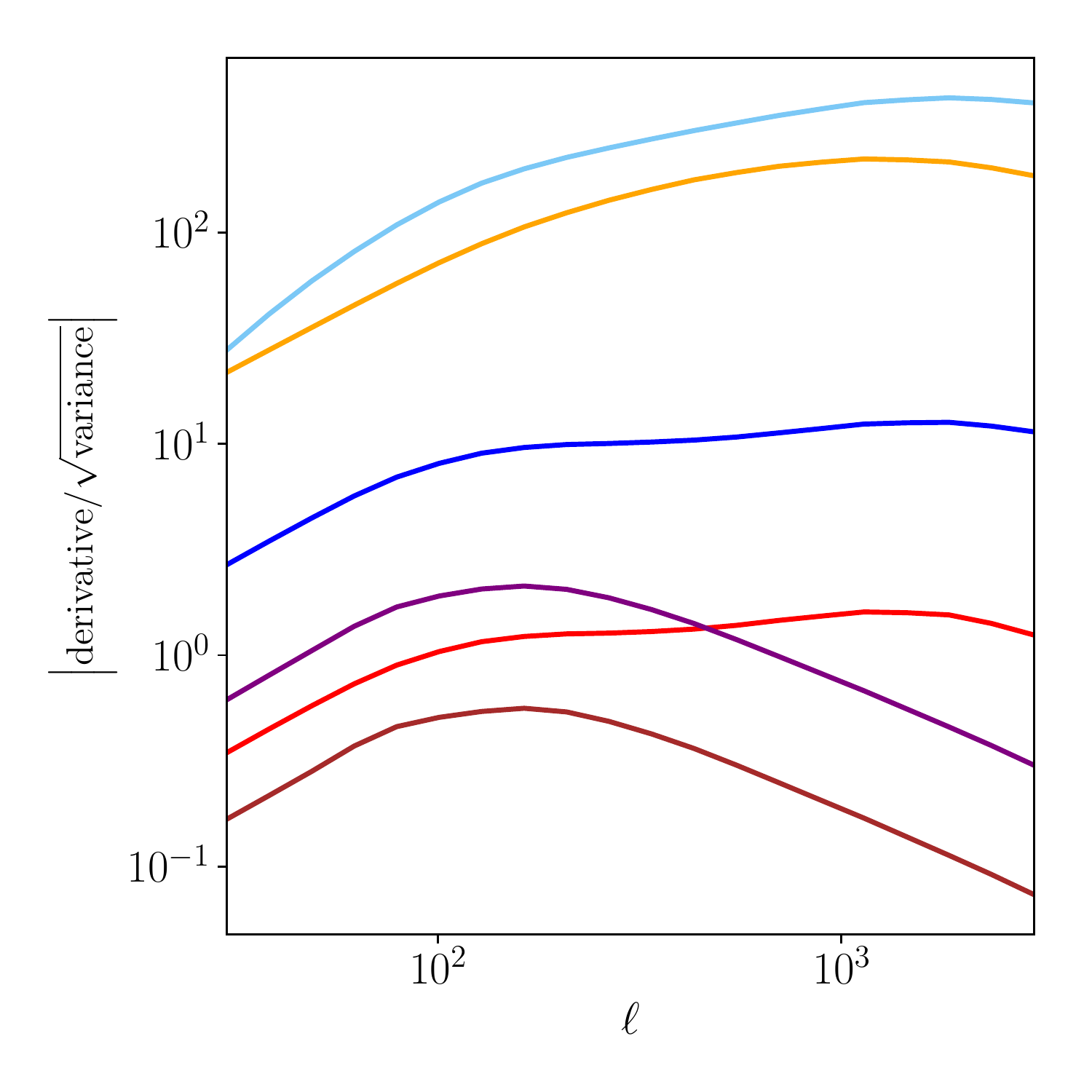}                        
    \end{subfigure}  
\hspace{0.5cm}
\begin{subfigure}[b]{0.5\linewidth}     
 \caption*{Bispectrum  derivatives, first redshift bin}  
                 \includegraphics[width=8.cm]{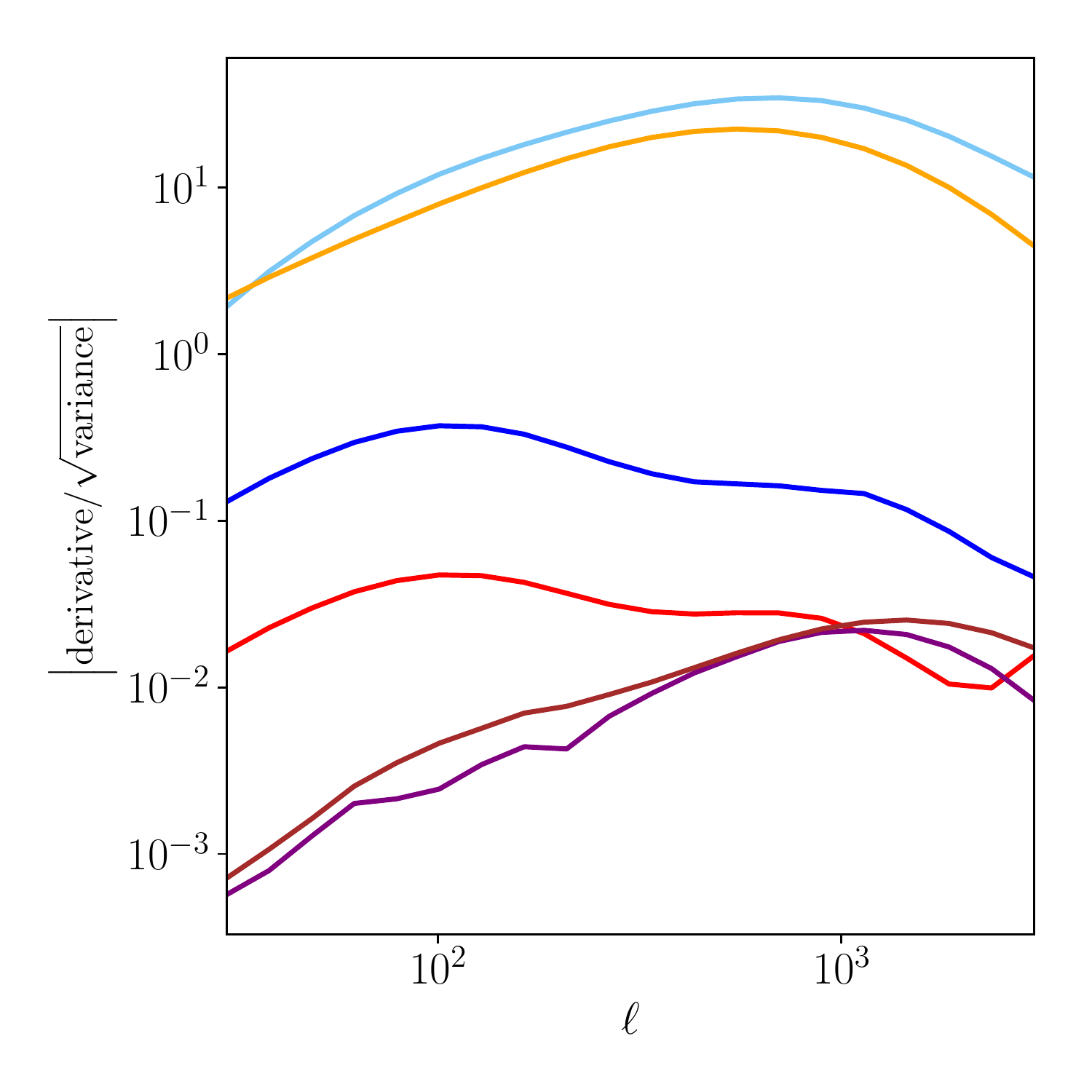}                                 
    \end{subfigure}
    
    \hspace{2cm}
 \begin{subfigure}[b]{1.\linewidth}     
                 \includegraphics[width=15cm]{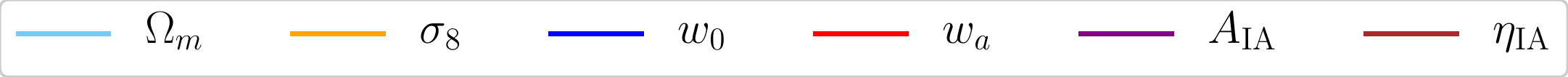}                    
    \end{subfigure}
    \caption{The ratio of the derivatives of the power spectrum and bispectrum with respect to the cosmological and intrinsic alignment parameters and the square root of the data variance.} \label{fig:derivs}
\end{figure}

To demonstrate how the bispectrum contributes to the breaking of degeneracies, Fig. \ref{fig:derivs} compares the derivatives of the power spectrum and bispectrum with respect to the cosmological and intrinsic alignment parameters with the square root of the data variance.  The results are for the first redshift bin only.  The power spectrum and bispectrum results clearly differ in amplitude and shape.\\

\begin{figure*}
 \hspace{-3cm}
      \begin{subfigure}[b]{0.25\linewidth}     
                  \includegraphics[width=8cm]{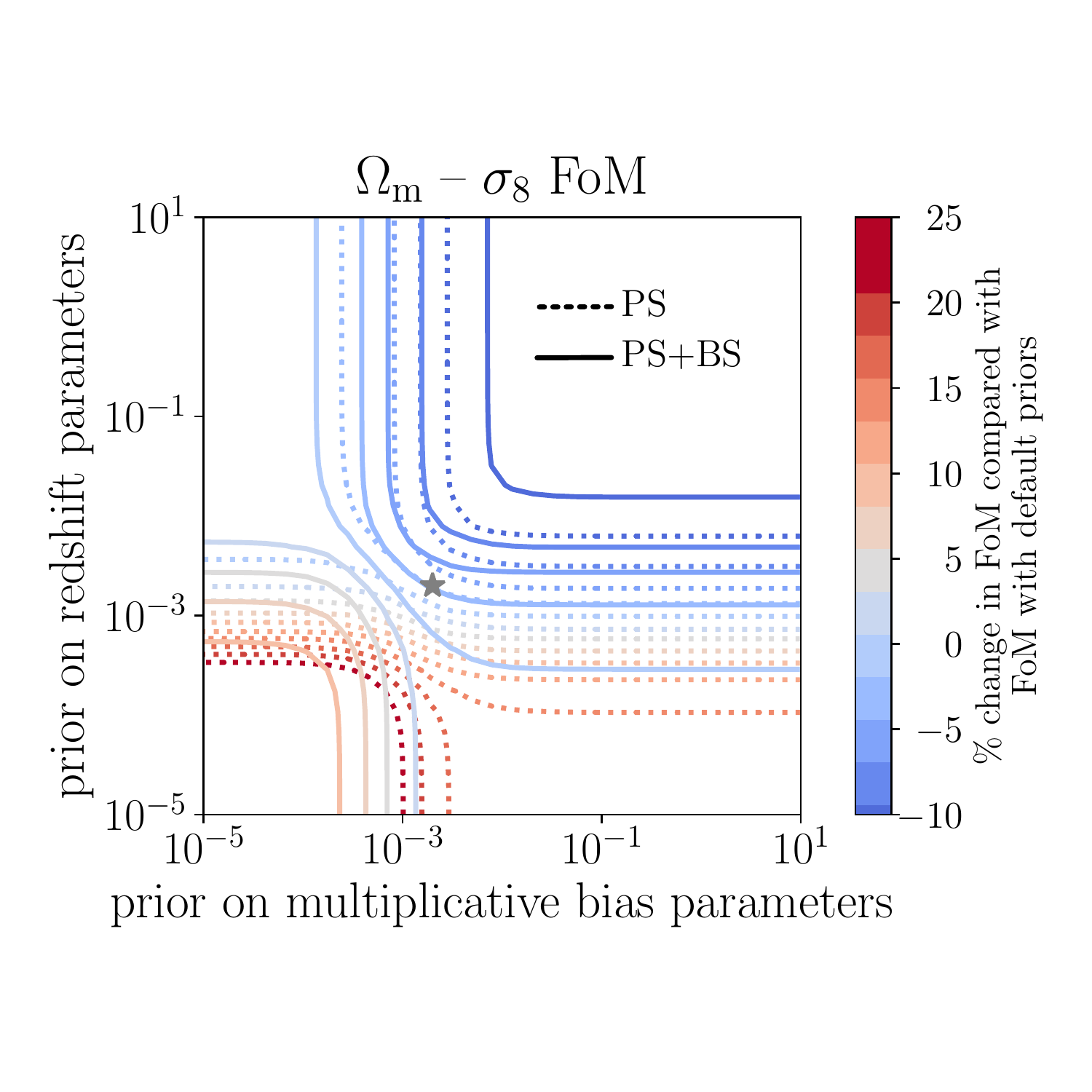}                    
    \end{subfigure}
\hspace{4cm}
    \begin{subfigure}[b]{0.25\linewidth}     
                  \includegraphics[width=8cm]{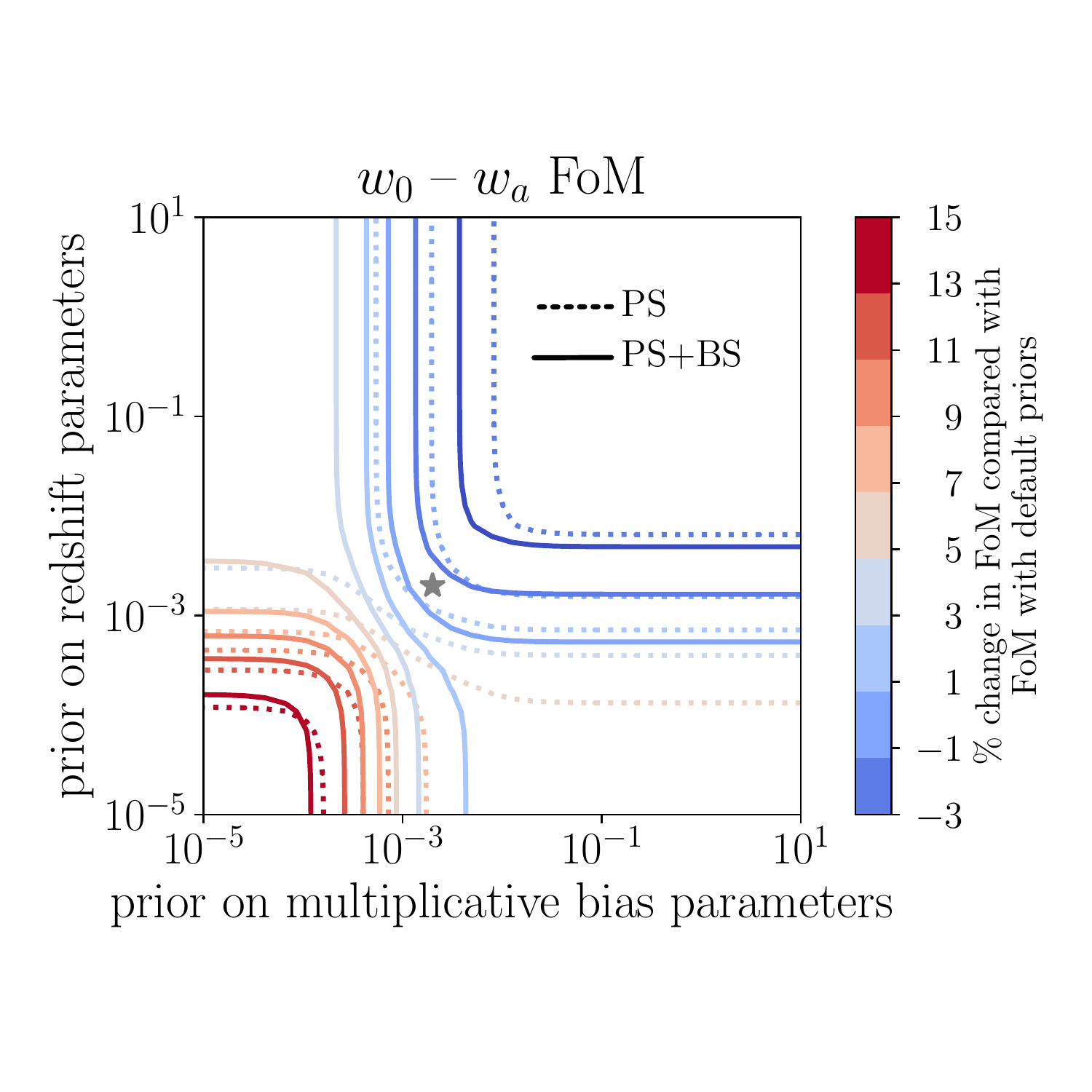}                    
    \end{subfigure}
  \vspace{-1cm}
 \caption{  Percentage change in the figures of merit compared with the figure of merit obtained with the default prior of 0.1 on the intrinsic alignment parameters and priors equal to the  \textit{Euclid} accuracy requirement  0.002 for redshift and multiplicative bias  parameters \citep{laureijs2011euclid}. Priors on the intrinsic alignment parameters are set to their default values - of 0.1.The grey stars indicate the default prior values for redshift bin mean and multiplicative bias parameters.    \textit{Left}: $\Omega_\mathrm{m}$ -- $\sigma_8$ FoM.   \textit{Right}: $w_0$ -- $w_a$ FoM. Note different scales in the two panels.  } 
  \label{fig:contour}
\end{figure*}

\begin{figure*}            \hspace{-2.5cm}
            \begin{subfigure}[b]{1.\linewidth}
            \hspace{8cm}
              \large{\qquad Power spectrum only}
              \end{subfigure}
    \begin{subfigure}[b]{0.25\linewidth}
    \hspace{-1.5cm}
     \includegraphics[width=7.5cm]{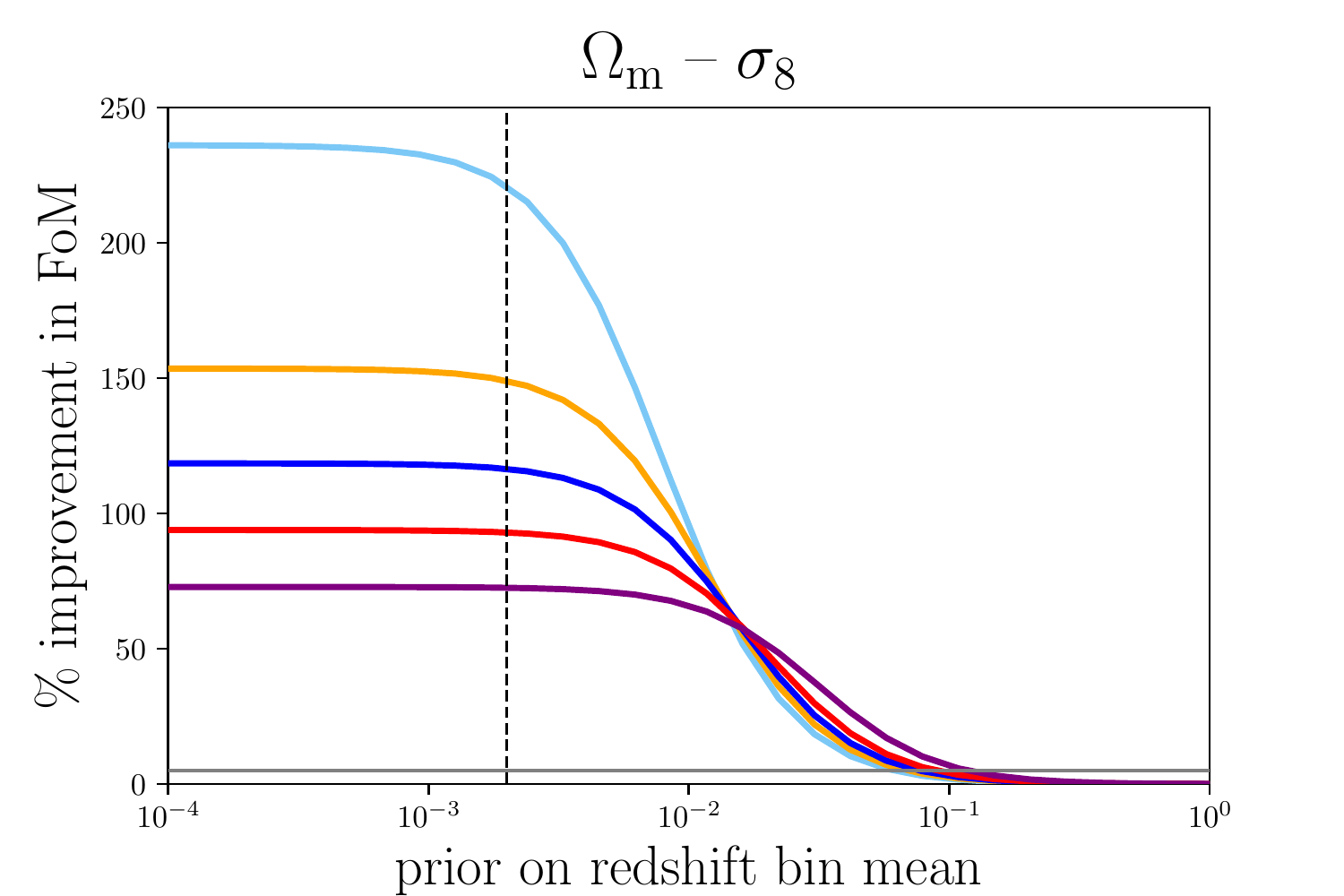}      
    \end{subfigure}
  \hspace{2cm} 
    \begin{subfigure}[b]{0.25\linewidth}
        \includegraphics[width=7.5cm]{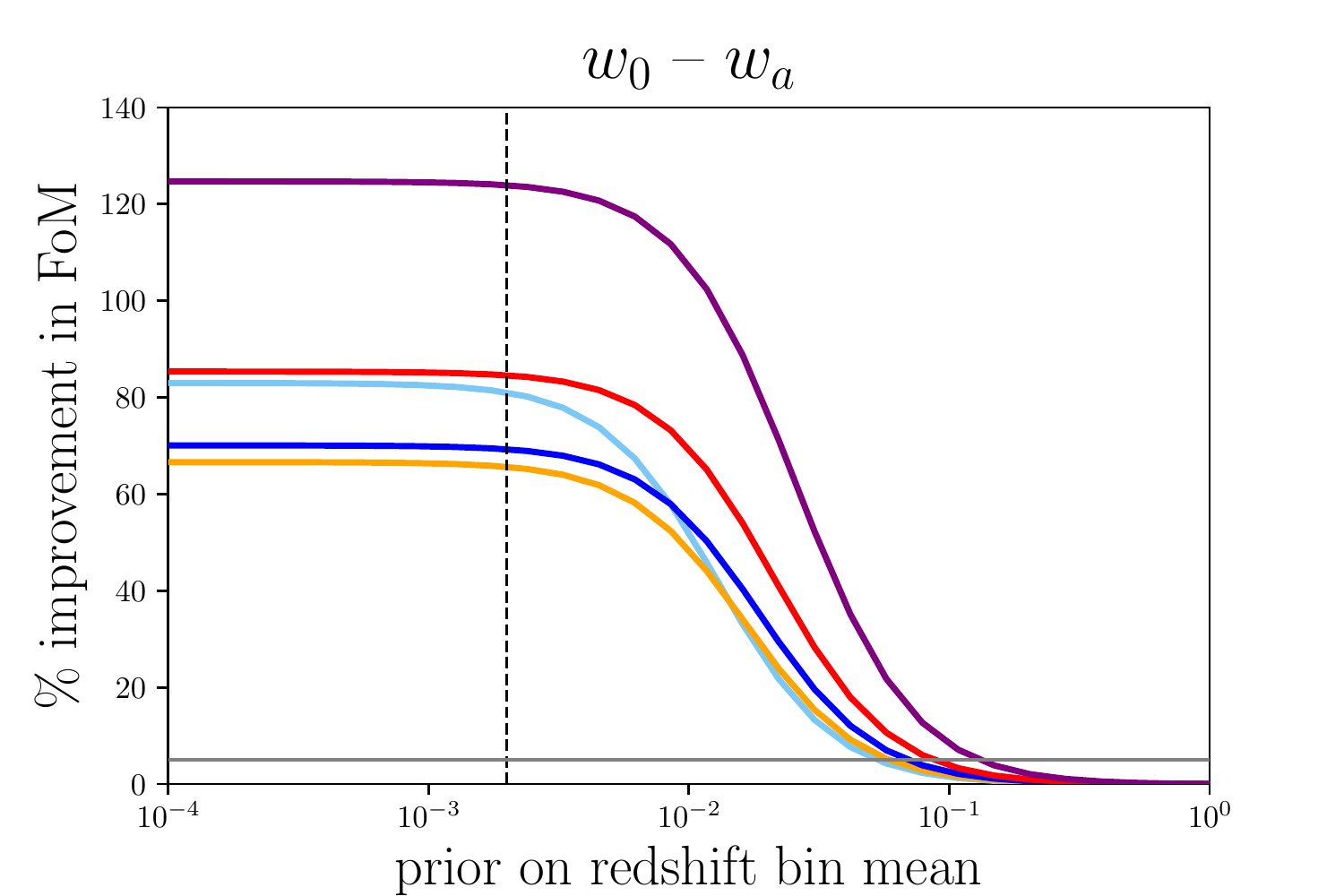}   
    \end{subfigure}
    \vspace{20pt}
    
     \hspace{-2.5cm}
                 \begin{subfigure}[b]{1.\linewidth}
            \hspace{8cm}
              \large{Power spectrum  plus bispectrum}
              \end{subfigure}
    \begin{subfigure}[b]{0.25\linewidth} 
    \hspace{-1.5cm}
           \includegraphics[width=7.5cm]{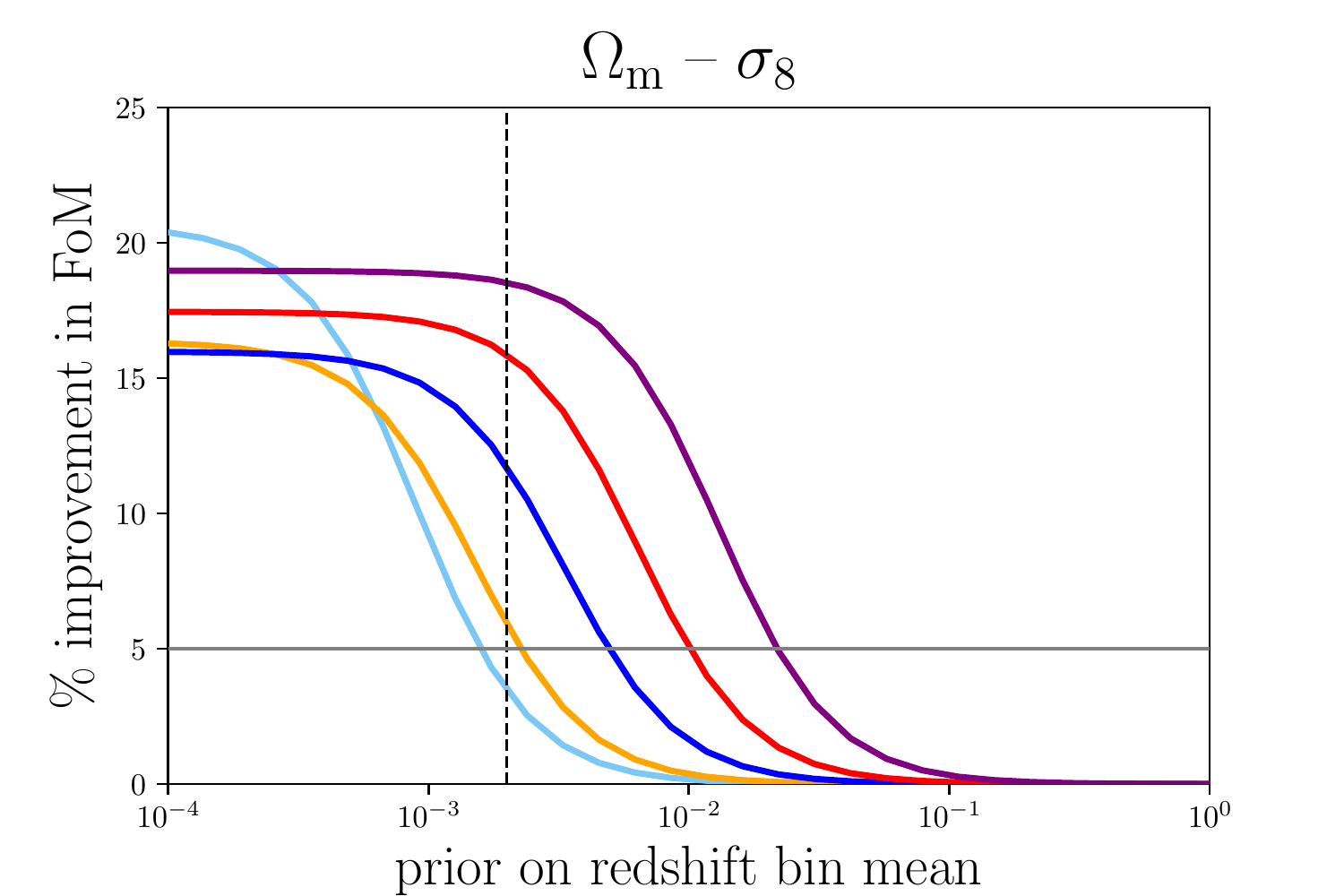}     
    \end{subfigure}
     \hspace{2cm} 
    \begin{subfigure}[b]{0.25\linewidth}
           \includegraphics[width=7.5cm]{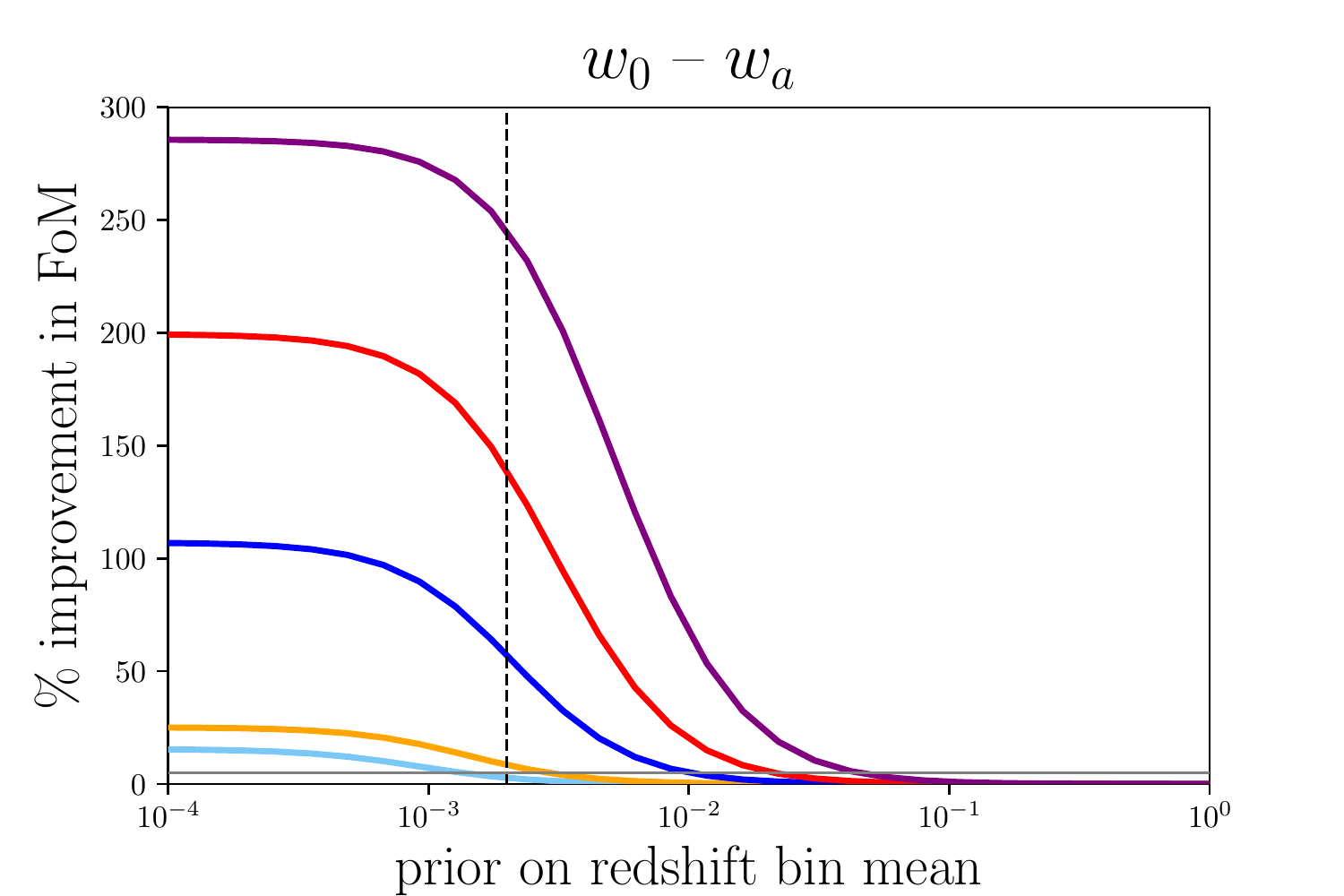}    
    \end{subfigure}  
    
     \centering
     \vspace{0.5cm}
    \begin{subfigure}[b]{1.0\linewidth} 
    \hspace{1.8cm}
       \includegraphics[width=15.0cm]{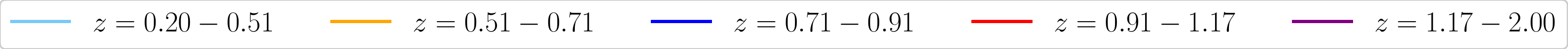} 
     \end{subfigure}
\vspace{1pt}
    \caption{ Percentage increase in figures of merit when the  prior on the mean of each redshift bin is tightened individually, compared to a wide prior of 10.  Priors on all other nuisance parameters are set to their default values - 0.1 for the intrinsic alignment parameters and 0.002 for multiplicative bias parameters. \textit{Top}: Power spectrum. \textit{Bottom}: Power spectrum and bispectrum combined.  \textit{Left}: $\Omega_\mathrm{m}$ -- $\sigma_8$ figure of merit. \textit{Right}: \mbox{$w_0$ -- $w_a$} figure of merit. The vertical dashed lines indicate the \textit{Euclid} redshift accuracy requirement \citep{laureijs2011euclid}. The horizontal grey lines indicate a 5\% improvement in the FoM. An improvement less than this is our criterion for self-calibration. Note different vertical scales in each panel.   
 }\label{fig:zbinFoM1}  
\end{figure*}

\noindent
Figure \ref{fig:contour} compares the FoMs obtained with the combined bispectrum and power spectrum  with those obtained with the power spectrum only, in each case imposing default priors of 0.1 on the intrinsic alignment parameters and 0.002 on all other nuisance parameters. \\
Figure \ref{fig:zbinFoM1} shows the effect of varying priors on redshift bin means individually, for the the power spectrum only and the combined power spectrum and bispectrum. The vertical dashed lines indicate the redshift accuracy requirement from the Euclid Definition Study Report  \citep{laureijs2011euclid}. The horizontal grey lines indicate a 5\% improvement in the FoM; an improvement less than this is our criterion for self-calibration.  With only the power spectrum, the self-calibration regime starts at a prior value of around 0.1 for every bin. The bispectrum is more sensitive to redshift and when it is also used the self-calibration point increases with redshift, from roughly 0.001 for the nearest bin to 0.1 for the furthest. Nevertheless  self-calibration mainly starts at smaller values  than for the power spectrum only.  In all cases the height of the step between the self-calibration regime and the region where the FoM is fixed by the prior is less when the bispectrum is combined with the power spectrum, reducing reliance on external priors.





\bsp	
\label{lastpage}
\end{document}